\documentclass{JHEP3}

\usepackage{amsfonts}
\usepackage{amsmath}
\usepackage{amssymb}
\usepackage{latexsym}
\usepackage{graphicx}
\numberwithin{equation}{section}

\def\hR{\textbf R}
\def\hA{\textbf A}
\def\ze{{\zeta_1,\zeta_2,\zeta_3,\zeta_4}}

\def\sg{\hat\sigma}

\def\ZZ{{\mathbb Z}}

\def\IR{{\mathbb R}}

\def\ap{{\alpha'}}
\newcommand{\be}{\begin{equation}}
\newcommand{\ee}{\end{equation}}
\newcommand{\bea}{\begin{eqnarray}}
\newcommand{\eea}{\end{eqnarray}}
\def\p{\partial }

\def\a{\alpha }

\def\s{\sigma }
\def\sg{\sigma }

\def\threeh{{\scriptstyle {3 \over 2}}}
\def\fiveh{{\scriptstyle {5 \over 2}}}
\def\non{\nonumber }

\def\calE{{\cal E}}
\def\calI{{\cal I}}

\def\0chi{{0_\chi}}
\def\Vlambda{{V^\Lambda}}
\def\calT{{\cal T}}

\def\calR{{\cal R}}
\def\calF{{\cal F}}
\def\calS{{\cal S}}
\def\calD{{\cal D}}

\def\calV{{\cal V}}

\def\nn{\nonumber}
\def\half{{\scriptstyle {1 \over 2}}}
\def\third{{\scriptstyle {1 \over 3}}}

\def\i{h}

\preprint{
DAMTP-2008-54,
IPhT-T-08-100,
UB-ECM-PF-08/13}

\title{Modular properties of two-loop maximal supergravity
and connections with string theory}
\author{Michael B. Green\\
 Department of Applied Mathematics and
Theoretical Physics\\
Wilberforce Road, Cambridge CB3 0WA, UK\\
\email{\tt M.B.Green@damtp.cam.ac.uk}}
\author{Jorge G. Russo\\
Instituci\' o Catalana de Recerca i Estudis Avan\c{c}ats (ICREA)\\
Department ECM,  Facultat de Fisica, University of Barcelona\\
 Av. Diagonal, 647,  Barcelona 08028 SPAIN\\
\email{\tt jrusso@ub.edu}}
\author{Pierre Vanhove\\
Institut de Physique Th\'eorique,\\
CEA, IPhT, F-91191 Gif-sur-Yvette, France\\
CNRS, URA 2306, F-91191 Gif-sur-Yvette, France\\
and\\
Niels Bohr Institute, University of Copenhagen,\\
  Blegdamsvej 17, DK--2100 Copenhagen \O, Denmark\\
\email{pierre.vanhove@cea.fr}}

\abstract{  The low-momentum expansion  of the  two-loop four-graviton
  scattering amplitude in eleven-dimensional supergravity compactified
  on  a circle  and a  two-torus is  considered up  to terms  of order
  $S^6\, \calR^4$ (where $S$ is  a Mandelstam invariant and $\calR$ is
  the  linearized  Weyl  curvature).   In  the case  of  the  toroidal
  compactification  the coefficient  of each  term in  the  low energy
  expansion is generically a  sum of a number of $SL(2,\ZZ)$-invariant
  functions of the complex structure of the torus.  Each such function
  satisfies  a   separate  Poisson  equation  on   moduli  space  with
  particular source  terms that are bilinear in  coefficients of lower
  order  terms,   consistent  with  qualitative   arguments  based  on
  supersymmetry.  Comparison is made  with the low-energy expansion of
  type II  string theories in  ten and nine dimensions.   Although the
  detailed behaviour of the string amplitude is not generally expected
  to be reproduced by  supergravity perturbation theory to all orders,
  for the terms considered here  we find agreement with direct results
  from  string   perturbation  theory.   These  results  point   to  a
  fascinating pattern  of interrelated  Poisson equations for  the IIB
  coefficients at  higher orders in  the momentum expansion  which may
  have a significance beyond the particular methods by which they were
  motivated.  }

\date{\Date}

\newpage
\begin{document}

%%%%%%%%%%%%%%%%%%%%%%%%%%%%%%%%%%%%%%%%%%%%%%%%%%%%%%%%%%%%%
\section{Introduction}\label{sec:intro}
%%%%%%%%%%%%%%%%%%%%%%%%%%%%%%%%%%%%%%%%%%%%%%%%%%%%%%%%%%%%%

The rich network of string theory dualities  provides powerful
constraints on the structure of M-theory.  These are particularly
restrictive for maximally supersymmetric backgrounds although the full
power of maximal supersymmetry has proved difficult to exploit.
The purpose of this paper is to further investigate the small corner of string
theory associated with the low-energy expansion of the
four-graviton scattering in nine or ten dimensions and its
connection to eleven-dimensional supergravity.
In terms of an effective action, this corresponds to an investigation of terms
 involving derivatives acting
on four powers of the linearized Riemann curvature.

More precisely, our aim is to further develop the connections between
multi-loop eleven-dimensional supergravity compactified on $\calS^1$ and $\calT^2$  and the type II
superstring  theories, making use of the conjectured relationships
between M-theory and type IIA  and IIB superstring theories
\cite{Hull:1994ys,Witten:1995ex,Schwarz:1995jq,Aspinwall:1995fw}.   In earlier
work, a number of terms in the low-energy expansion of the type II string theory amplitudes
were determined from the compactified one-loop and two-loop supergravity amplitudes \cite{ggv:oneloop,
Russo:1997mk,gkv:twoloop,gv:D6R4}.  The fact that the full, nonperturbative moduli dependence of
the string amplitudes was reproduced is presumably a consequence of the constraints of maximal
supersymmetry on `protected' terms.
In the absence of a complete understanding of which terms are protected it is of interest to
pursue the connections with quantum supergravity further.
Here we will develop the low-energy expansion of the two-loop
supergravity amplitude
in a more systematic fashion and determine several orders beyond those considered previously.
We will, furthermore, investigate the extent to which this makes contact with the type II string theories in
nine and ten dimensions.
We will find that the scalar-field dependent coefficients of the higher-derivative terms
in the expansion satisfy a suggestive pattern of differential equations on moduli space.
Comparing these coefficients with known `data' from the low-energy expansion of tree-level and
genus-one perturbative string theory in nine and ten dimensions
\cite{gv:stringloop,grv:oneloop} shows a surprising degree of agreement.
Although it is obvious that there is far more to M-theory than perturbative supergravity, these results
suggest patterns that could persist to all orders in the low-energy expansion.

\subsection{Overview of low orders in the momentum expansion}

In ten dimensions
there is a clear distinction between type IIA and type IIB  superstring theories even though it is known that they
have identical four-graviton amplitudes at least up to, and including,
 genus-four in string perturbation theory \cite{Berkovits2006}.
  The IIA theory has a single real modulus, and at strong coupling this is
 identified  with  the  radius   of  a  single  compact  dimension  in
 eleven-dimensional supergravity \cite{Witten:1995ex}.
 The ten-dimensional IIB theory has a complex modulus (a complex scalar coupling constant) that is identified with the
 complex structure of the torus in the $\calT^2$ compactification of eleven-dimensional
 supergravity in the limit in which the torus volume vanishes \cite{Schwarz:1995dk,Aspinwall:1995fw}.
 Invariance of M-theory under large diffeomorphisms of $\calT^2$ implies that the IIB theory
 possesses a $SL(2,\ZZ)$ duality symmetry  \cite{Hull:1994ys} that relates strong and weak coupling in a
 manner that involves both the  perturbative and
 non-perturbative ($D$-instanton) interactions.
 After compactification to nine dimensions
on a circle the two string theories are identified by the action of T-duality, which inverts
the  radius  of  the  compact  dimension and  transforms  the  dilaton
appropriately.
The nine-dimensional  duality group is  $SL(2,\ZZ)\otimes \IR^+$.

Although the explicit calculations in this paper concern the four-graviton amplitude, maximal supersymmetry ensures that 
the conclusions apply equally to the scattering of any four states in the supermultiplet.  In fact, maximal supersymmetry guarantees that 
the general type IIA or IIB amplitude has the structure\footnote{We are grateful to Nathan Berkovits for emphasizing the generality
of this structure.}
\be
\hA_\ze= F(s,t,u)\,\hR_\ze^4\,,
\label{eLoopa}
\ee
where we have labeled each external massless particle by its superhelicity $\zeta_r$, which takes 256 values (the dimensionality 
of the maximal supergravity
multiplet) and its momentum $p_r$ ($r=1,2,3,4$), where $p_r^2 =0$.   
$F(s,t,u)$ is a function of the Mandelstam invariants\footnote{
The (dimensionless) Mandelstam  invariants, $s=-\ap\,(p_1+p_2)^2$,
$t=-\ap\,(p_1+p_4)^2$ and $u=-\ap\,(p_1+p_3)^2$,
are subject to the mass-shell condition $s+t+u=0$ and
 $\sqrt{\alpha'}=l_s$ is the string length scale.}
 $s$, $t$, $u$.
The kinematical factor in (\ref{eLoopa})
is given by (see  (7.4.57) of \cite{Green:1987mn})
  \begin{equation}
  \hR_\ze^4(p_{1},p_{2},p_{3},p_{4})= \zeta_{1}^{AA'} \zeta_{2}^{BB'}
  \zeta_{3}^{CC'}\zeta_{4}^{DD'}\, K_{ABCD}\,\tilde K_{A'B'C'D'}\,,
  \label{kinfact}
 \end{equation}
where the indices $A, B$ on the polarization tensors $\zeta_r^{AB}$  run over both
 vector and spinor values (for example, the graviton polarization is $\zeta^{\mu\nu}$,
 where $\mu,\nu = 0,1,\dots,9$) and the tensor $K\, \tilde K$ is defined in
 \cite{Green:1987mn}.   For the purposes of this paper we will consider the case of external gravitons
for which  $\hR$ reduces to the the momentum-space form of the linearized Weyl tensor,
\be
\label{riemdef}
\calR_{\mu\nu\rho\sigma} = -4\, p_{[\mu} \zeta_{\nu][\sigma} p_{\rho]}\, \, ,
\ee
where the symmetric traceless polarization tensor satisfies
$p^\mu  \zeta_{\mu\nu}=0$.   
 The kinematic factor $\hR_\ze^4$ in (\ref{eLoopa}) becomes  $\calR^4$, which denotes
 the product of four Weyl curvatures
 contracted into each other by a well-known  sixteen-index   tensor (often denoted $t_{8}t_{8}$).

The low-energy expansion of the four-particle amplitude requires the expansion of the function $F(s,t,u)$ in (\ref{eLoopa}) for 
small $s,t,u$.  This can be expressed as a complicated mixture of terms that are analytic and
nonanalytic functions  of the Mandelstam invariants.
The analytic terms may be expanded as power series' in integer powers of $s$, $t$ and $u$
in a straightforward manner.  The lowest-order terms contain the poles and contact terms characteristic of the supergravity tree diagrams.
A great deal is also known about higher-order  analytic terms up to order ${\alpha'}^6$.
The nonanalytic terms contain massless threshold
singularities whose form is determined by unitarity and depends on the number of
noncompact space-time dimensions.
 Generically, there
are fractional powers or logarithmic branch points, giving rise to
non-integer powers of $s$ or $\log s$ factors.

In what follows we shall separate the low-energy  expansion of the  ten-dimensional amplitude in either type II theory
into the sum of an analytic part and a non-analytic part,
\be
A_{II} = i{\alpha'}^4\, ( A_{II}^{an} + A_{II}^{nonan})\,,
\label{annonan}
\ee
where $A_{II}$ has been normalized to be dimensionless.
In the IIB theory the coefficients
in the series in the analytic term $A_{II}^{an}$ in (\ref{annonan}) are $SL(2,\ZZ)$-invariant functions of the 
complex coupling and the series has the form
\be
A_{IIB}^{an} = \sum_{p\ge 0,q\geq -1} g_B^{p+\threeh q-\half}
\,\calE_{(p,q)}(\Omega)\,
\hat\sigma_2^p\hat\sigma_3^q\, \calR^4\,,
\label{gtensum}
\ee
where
\be
\label{sigmadefs}
\hat \sigma_n = \frac{s^n + t^n +u^n}{4^n}\, .
\ee
The factors $\hat \sigma_2^p\,
\hat\sigma_3^q$ are the most general scalars that are symmetric monomials in
$s$, $t$, $u$ of order $2p+3q$.  The functions $\calE_{(p,q)}$'s are modular functions of
the complex scalar, $\Omega=\Omega_1 + i \Omega_2$, where
\be
\Omega_1 = C^{(0)}\,,\qquad \Omega_2 = e^{-\phi_B} = g_B^{-1}\,,
\label{omdef}
\ee
and $C^{(0)}$ is the Ramond--Ramond scalar, $\phi_B$ is the type IIB dilaton and $g_B$ is the
type IIB coupling constant.  The expression (\ref{gtensum}) includes the Born term with its
poles and the coefficient $\calE_{(0,-1)} = 1$.   The nonanalytic contribution
is  a series  that  contains  multi-particle thresholds  of
symbolic form
\be
A_{IIB}^{nonan} = \left( s\log (-s) +
g_B^{\threeh}\,  \calF_4(\Omega)\, s^4 \log(-s) +
g_B^2\, \calF_5(\Omega)\,s^5 \log(-s) + \dots
  \right)\, \calR^4\,,
\label{nonansum}
\ee
where the $\calF_r(\Omega)$'s are modular functions of
$\Omega$, which begin with terms that are genus-one or higher.

The coefficients in the expansion are known up to terms of order $\hat\sigma_3\, \calR^4$,
\bea
A_{\rm IIB}&=&  e^{-2\phi}\frac{1}{3\hat \sigma_3}\, \calR^4 +
 e^{-\phi_B/2}\, E_{3\over 2} (\Omega)\, \calR^4 +
 e^{\phi_B/2}{1\over 2}\, E_{5\over 2}(\Omega)\, \hat \sigma_2\, \calR^4\nn\\
&& \qquad +
 e^{\phi_B}{1\over 6}\, \calE_{(3/2,3/2)} (\Omega)\,\hat\sigma_3\, \calR^4 +\cdots
\,,
\label{unzy}
\eea
The terms in (\ref{unzy}) are analytic in $s,t$ and $u$ and  translate into
local higher-derivative interactions in a
$SL(2,\ZZ)$-invariant effective action,
\bea
S_{\rm IIB}&=& {1\over {\alpha'}^4 \, 2^7\pi^6}\,\int d^{10}x \sqrt{-g}\, \bigg[
e^{-2\phi}\, R^{(10)}+ {\alpha'}^3 \,e^{-\phi_B/2}\, E_{\threeh} (\Omega)\, \calR^4 +
 {{\alpha'}^5\over 2} e^{\phi_B/2}\, E_{\fiveh}(\Omega )\,  D^{4}\calR^4\nn\\
&& \qquad +{{\alpha'}^6\over6}
 e^{\phi_B}\,  \calE_{(\threeh,\threeh)} (\Omega )\,D^6\calR^4\bigg]+\cdots \, ,
\label{unzyact}
\eea
where $R^{(10)}$ is the curvature scalar, $g$ the ten-dimensional type IIB string metric and
the coefficients $E_s(\Omega)$ and $E_{(\threeh,\threeh)}(\Omega)$ will be described below.
The derivatives in (\ref{unzyact})  are
contracted so  that the four-point amplitude contributions  arise in a
manner that is defined by the pattern of Mandelstam invariants in (\ref{unzy}).
From (\ref{unzy}) it follows that the coefficients in (\ref{gtensum}) are given by
\bea
  \calE_{(0,0)}(\Omega)=E_{\threeh}(\Omega),   \qquad
\calE_{(1,0)} (\Omega) ={1\over 2}\,E_{\fiveh}(\Omega)\,, \qquad
\calE_{(0,1)}(\Omega)= {1\over 6}\calE_{(\threeh,\threeh)}(\Omega)\,.
\eea

The quantities $E_s$ in (\ref{unzy}) and  (\ref{unzyact}) are
Eisenstein series that  solve the Laplace eigenvalue equations
on the fundamental domain of $SL(2,\ZZ)$,
\be
\Delta_{\Omega} E_s\equiv \Omega_2^2 \,\left( {\partial^2\over \partial \Omega_1^2}
+ {\partial^2\over \partial \Omega_2^2}\right)\, E_s = s(s-1) E_s\, .
\label{lapeig}
\ee
Given the fact that the $E_s$ is a $SL(2,\ZZ)$ function that can have
no worse than power growth as $\Omega_{2}\to \infty$ (which is required for consistency
with string perturbation theory at weak coupling) the solution of this equation is uniquely
given by
\be
E_s= \sum_{(m,n)\ne (0,0)}\frac{\Omega_2^s}{|m+n\Omega|^{2s}}\,,
\label{esdef}
\ee
which can be expanded at weak coupling in the form
\bea
\label{zexpand}
 E_s (\Omega) & =& 2
\zeta(2s) \Omega_2^{s}  + 2\sqrt \pi  \Omega_2^{1 - s }
{\Gamma(s-\half)\zeta(2s -1)\over \Gamma(s)}\nn\\
&&  + {2\pi^{s}\over \Gamma(s)}  \sum_{k\ne
0} \mu(k,s)
 e^{-2\pi(|k|\Omega_2 - i k \Omega_1)}
 |k|^{s-1}
 \left(1+{s(s-1) \over 4\pi |k| \Omega_2} +\dots\right)\, .
\eea
The two power-behaved terms in this expansion correspond to the tree-level and
genus-$(s-1/2)$ contributions in string theory\footnote{In order to avoid confusion, we will refer to the
number of `loops' (denoted by $L$) in the context of the supergravity Feynman rules, and the `genus'
(denoted by $h$)  in the
context of the string theory perturbative expansion.}, as can be seen by taking into account
the powers of $e^{\phi_B}$ in (\ref{unzy}) and identifying $\Omega_2^{-1}$ with the IIB string
coupling, $g_B$.  The exponential terms correspond to the infinite set of $D$-instanton
contributions.

The fact that $\calE_{(0,0)}(\Omega) = E_{3/2}(\Omega )$ is the coefficient
associated with the $\calR^4$ term in (\ref{unzy})
was initially deduced via indirect arguments \cite{gg:dinstanton,ggv:oneloop}.
One of these made use of properties of loop amplitudes
of eleven-dimensional supergravity compactified on a circle or on a two-torus, combined
with dualities that relate M-theory to type II string theory in nine
dimensions.  In this way the function $E_{3/2}(\Omega)$ describes the
dependence of the low-energy limit of the one-loop ($L=1$) four-graviton scattering amplitude on the modulus
of the compactification torus \cite{ggv:oneloop}.   The ultraviolet divergence, which behaves as
$\Lambda^3 \calR^4$, where $\Lambda$ is a momentum cutoff,  is independent of $\Omega$
and can be subtracted by a local counterterm.   The coefficient of this counterterm
is fixed by requiring the IIA and IIB amplitudes to be equal, as they are known to be.
The modular function $E_{3/2}$ can also be derived as a consequence of supersymmetry
combined with $SL(2,\ZZ)$-duality \cite{Green:1998by}.
Although it is suspected that the other modular functions appearing in
higher derivative terms (at least up to the order shown in
(\ref{unzy})) should also be determined by supersymmetry combined with non-perturbative
dualities, there is no systematic procedure for doing this (a sketchy outline is given in section~\ref{sec:modular}
of this paper).

Expanding the $L=1$ supergravity
 amplitude in powers\footnote{
The dimensionless Mandelstam invariants of eleven-dimensional supergravity are denoted by upper case
letters     $S=-l_{11}^2\,(p_1+p_2)^2$,    $T=-l_{11}^2\,(p_1+p_4)^2$,
$U=-l_{11}^2\,(p_1+p_3)^2$, where $l_{11}$ is the
eleven-dimensional Planck length, and related to the invariants in the ten-dimensional string frame
by $S=R_{11}\,s\, \dots$, where $R_{11}$ is the radius of the eleventh
dimension.}  of $S$, $T$ and $U$
 leads to higher-order terms in the derivative
expansion of the form \cite{Russo:1997mk,gkv:twoloop}. This results in an infinite set 
of analytic terms that are interpreted in IIB string coordinates
as modular invariant coefficients multiplying powers of order $r_B^{1-2k}\, s^k$,
\be
A_{L=1}= r_B\,\big(g_B^{-{1\over
      2}}E_{3\over 2}(\Omega)\calR^4+
\sum_{k=2}^\infty h_k \, r_B^{-2k}g_B^{k-{1\over 2}}
E_{k-{1\over 2}}(\Omega)\, \calS^{(k)}\, \calR^4\big) +\cdots\, ,
\label{loopone}
\ee
where the ellipsis stand for the non-analytic contributions~\cite{gkv:twoloop} and $h_k$ are simple constants and $\calS^{(k)}$ is a polynomial in
$\hat\sigma_2$ and $\hat\sigma_3$  of order $k=2p+3q$ in the
Mandelstam invariants.
All contributions with $k\geq 2$ vanish in the ten-dimensional type
IIB limit where the two-torus volume, $\calV_2$, vanishes.
So we see that in the ten-dimensional limit the compactified one-loop ($L=1$) eleven-dimensional supergravity amplitude
contributes only at order $\calR^4$.
In order to obtain higher-derivative interactions
one has to consider eleven-dimensional supergravity at higher loops ($L>1$).
The coefficient
 $\calE_{(1,0)}= E_{5/2}(\Omega)/2$ of the ten-dimensional IIB theory
 indeed arises from a one-loop subdivergence of the
low-energy limit of the  two-loop amplitude of eleven-dimensional supergravity compactified
on a two-torus in the limit in the limit $\calV_2\to 0$
\cite{gkv:twoloop,gv:D6R4}.

 The function
$\calE_{(0,1)}\, (\Omega )$ in (\ref{gtensum}) is obtained by expanding the two-loop
supergravity amplitude  to the next order in $S$, $T$, $U$ and compactifying on a two-torus
\cite{gv:D6R4}.  It
satisfies the Poisson equation
\be
\Delta_\Omega \calE_{(0,1)} =
12 \calE_{(0,1)} -E_\threeh\, E_\threeh\, ,
\label{poiss}
\ee
in which the source term on the right-hand side is quadratic in the $O({\alpha'}^3)$
modular function $E_{3/2}$.   We will denote the solution to this equation by
$\calE_{(0,1)}=\calE_{(3/2,3/2)}/6$,  as in \cite{gv:D6R4}.
This source term makes the equation quite different from the Laplace
eigenfunction equation (\ref{lapeig}).  Its
structure was argued in \cite{gv:D6R4} to follow, at
least qualitatively, from the constraints of supersymmetry.
The solution of (\ref{poiss}) is complicated, but the zero-mode of $\calE_{(0,1)}$, which contains the
perturbative terms, is found to have the form
\be
\int_{-\half}^\half \Omega_2^{-1}\, \calE_{(0,1)}\, d\Omega_1 = {2\over 3}\zeta(3)^2 \Omega_2^{2} +{4\over 3} \zeta(2)\zeta(3) + {8\over 5}\zeta (2)^2  \Omega_2^{-2} +
{4\over 27}\zeta(6) \Omega_2^{-4} +O(\exp(-4\pi\Omega_2))\, ,
\label{ethreehthreeh}
\ee
so it contains tree-level, genus-one, genus-two
and genus-three perturbative string theory terms as well an infinite series of
$D$-instanton -- anti-$D$-instanton pairs. The tree-level and genus-one terms agree precisely with direct string theory
calculations, while the genus-two term has not yet been extracted directly from string theory.  The
genus-three term cannot yet be computed in string perturbation theory but it is gratifying that the
value of its coefficient agrees, as it should,
 with that of the genus-three term in the IIA theory that is predicted
from one loop in eleven-dimensional supergravity compactified on $\calS^1$.  This agreement is
striking since extracting the coefficient in the IIB theory from the $L=2$ amplitude involves the use
of a Ramanujan identity (see  appendix~\ref{sec:betaterm}), whereas the coefficient in the IIA theory
obtained from the $L=1$ amplitude
arises from a simple integral.
Although there is no proof that $\calE_{(0,1)}(\Omega)$ is the exact modular function, these
agreements strongly suggest that it is.  It is notable that the terms in the expression (\ref{ethreehthreeh})
are not of uniform transcendental weight.  Whereas, there is a correlation of the power of $\Omega_2$ and the
weight of the $\zeta$ values for the first three terms, this breaks down for the genus-three term.  We will see
an analogous lack of transcendentality in many of the examples to be described later in this paper.

The first nonanalytic term beyond the Born (pole) term arises at order ${\alpha'}^4$
and comes from the ten-dimensional supergravity one-loop
diagrams.
It has the symbolic form given by the first term on the right-hand side of
(\ref{nonansum}).  Its precise expression, reviewed in \cite{grv:oneloop},
has a much more complicated threshold structure but it
has the notable property that the scale of
the logarithm cancels, using $s+t+u=0$.

Obviously  the  analysis  of  Feynman diagrams  of  eleven-dimensional
supergravity  has limited  use  since  it does  not  capture the  full
content  of  quantum  string  theory,  or M-theory.   To  begin  with,
eleven-dimensional supergravity is  not renormalizable.  Our procedure
is to  regulate the ultraviolet divergences by  introducing a momentum
cutoff and  subtracting the divergences with  counterterms. The result
is finite but the  counterterms contribute arbitrary coefficients that
parameterize our ignorance of the short-distance physics.  However, at
low orders  the values of some  of these coefficients are  known to be
determined  by supersymmetry   if  we also  assume  the result
should be in accord with string dualities.  One of the aims here is to
investigate the extent to which this continues at higher orders.

A related issue is that the Feynman diagrams describe a semi-classical approximation to the theory in a
particular classical background space-time.  This can only be motivated
in the limit in which the radii of the compact dimensions are much larger than
the eleven-dimensional Planck length.  This means $R_{11} \gg 1$ for the $\calS^1$ compactification
(where $R_{11}$ is the dimensionless radius of the eleventh dimension in Planck units).
This is the limit of large IIA string coupling, $g_A = R_{11}^{3/2}
\gg1$.  Bearing in mind that this is
far from the regime of string perturbation theory, we will see to what
extent there
is agreement between the compactified two-loop Feynman diagrams and corresponding
perturbative string theory results.  For
the $\calT^2$ compactification the analogous condition is $\calV_2=R_{10}R_{11} \gg 1$ (where $\calV_2$ is the
dimensionless volume  of  $\calT^2$  in Planck  units).  In IIA string theory compactified to nine dimensions
this is the limit in which
$r_A =1/r_B \gg  g_A^{-\third}$, where $r_A$ is the radius of the compact dimension (in string units).
Nevertheless,  the
coefficients of the $\calR^4$, $\calD^4\, \calR^4$ and $\calD^6\, \calR^4$ terms reviewed above
give the correct values in the $r_B \to \infty$ limit -- presumably the extrapolation from small $r_B$
to large $r_B$ works because these terms are protected by supersymmetry.
The low energy
limit we are considering is one in which $S\, R_{11}^2\ll 1$. Since the supergravity loop diagrams are
ultraviolet divergent we will also introduce a dimensionless momentum cutoff
$\Lambda  \gg R_{11}^{-1}  \gg 1$  measured in  units of  $l_{11}$ the
eleven-dimensional Planck length.
We will see that the low energy expansion of
the Feynman diagrams possesses a very rich structure.  In particular, the coefficients that depend
on the scalar fields satisfy a series of mathematically intriguing Poisson equations that are nontrivial
extensions of (\ref{poiss}) satisfied by $\calE_{(0,1)}$, as we will see.

\subsection{Outline of paper}

In this paper we will consider the higher-order terms in the
low-energy expansion of the four-graviton amplitude that are
obtained by expanding the two-loop amplitude of
eleven-dimensional supergravity, compactified to ten dimensions on
$\calS^1$ and nine dimensions on $\calT^2$ to several higher orders in the Mandelstam invariants.

The four-graviton amplitude (\ref{annonan})
 at two loops ($L=2$) in maximal supergravity has the form \cite{BernDunbar}
\be
A_{sugra}^{an} + A_{sugra}^{nonan} =  \! i{\kappa_{11}^6\over 2\,(2\pi)^{22}\,l_{11}^{12}}\; \calR^4\, \calI(S,T,U)\,,
\label{eberndun}
\ee
where the scalar function $\calI(S,T,U)$ has the structure
\be
\calI(S,T,U) =S^2 I^{(S)}(S;T,U) + T^2 I^{(T)}(T;U,S) + U^2 I^{(U)}(U;S,T)\,.
\label{calidef}
\ee
The terms in brackets are sums of $\varphi^3$ scalar field theory two-loop planar and non-planar ladder diagrams,
\be
I^{(S)} (S;T,U)=I^P(S;T,U) + I^P(S;U,T)+ I^{NP}(S;T,U) +I^{NP}(S;U,T)\, ,
\label{sumall}
\ee
with analogous expressions for $I^{(T)}$ and $I^{(U)}$.
The  expression (\ref{eberndun}) has an overall prefactor of $\calR^4$,
which has eight powers of the external  momenta,  together with
four more powers from the factors of $S^2$, $T^2$ or $U^2$.  This means
that the loop integrals, $I^{(P)}$ and $I^{(NP)}$, are much less divergent than they would
naively appear.
We  will  be  interested   in  the  compactified  amplitude,  so  that
$\calI(S,T,U)$ is a function of the moduli of the compact space.
 Ignoring for the moment the nonanalytic pieces, we shall expand
the analytic part of $\calI(S,T,U)$ in a power series,
\be
\calI^{an}(S,T,U) = \sum_{p,q\geq0}\,n_{(p,q)} \sigma_2^p\, \sigma_3^q\,I_{(p,q)}\,,
\label{ipqdef}
\ee
where $I_{(p,q)}$ is a function of the moduli that will be defined by the integral (\ref{e:DefIn}) and the constant
coefficients $n_{(p,q)}$ can be read off from~(\ref{eberz}).  Note that $I_{(0,0)}=0$ since the
well-known $\calR^4$  term
only arises at one loop.  The dependence on the Mandelstam invariants in (\ref{ipqdef}) is contained in the
$\sigma_2$ and $\sigma_3$, which are defined by
 \be \sigma_n= S^n+T^n+U^n
\label{msigdef}
\ee
(whereas in the string variables  we used the symbol $\hat \sigma_n$ in (\ref{sigmadefs})).
The coefficients $I_{(p,q)}$ in (\ref{ipqdef}) depend on the $\calT^n$ moduli in a
manner to be determined.
The infrared massless threshold effects give rise to nonanalytic terms that we will also need to discuss.

In section~\ref{sec:expand}  we will show how the expression for $\calI(S,T,U)$ can be
reexpressed in a useful form  that would also arise naturally in a world-line functional integral
describing the two-loop process.
This involves attaching vertex operators for external
states of momentum $p_r$ ($r=1,2,3,4$) to points $t_r$ on the three world-lines, of length $L_k$ ($k=1,2,3$),
 of the two-loop vacuum diagram.   The amplitude involves the
the usual factor of $\exp( -\sum_{r,s}\, p_r\cdot p_s\, G_{rs})$,
where $G_{rs}$ is the
Green function connecting pairs of points on these world-lines, as discussed in \cite{Dai:2006vj}.
This provides a very compact expression for the sum of all diagrams as an integral over all insertion points
$t_r$  and over the lengths $L_k$ of the three world-lines,  with an appropriate measure.
The low energy expansion is obtained, formally, by expanding the integrand in powers of the Green
function
\be
\exp(-\sum_{r,s=1}^4\, p_r \cdot p_s\, G_{rs}) = \sum_{N=0}^\infty {1\over
  N!}\,(-\sum_{r,s=1}^4\, p_r \cdot p_s\, G_{rs})^N\,,
\label{gexpand}
\ee
which are to be integrated over the positions $t_r$ with a specific measure.

We will discuss a `hidden' modular
invariance that acts on the three Schwinger parameters, $L_k$.
This symmetry is particularly useful in evaluating the compactification of the
amplitude on a spatial $n$-torus and was used in
\cite{gkv:twoloop,gv:D6R4} in evaluating terms of order $\calD^4\, \calR^4$ and $\calD^6\, \calR^4$.
This becomes more explicit after a change of variables from the Schwinger parameters, $L_k$, to
variables $\tau_1$, $\tau_2$ and $V$.  The quantity $\tau=\tau_1+i\tau_2$ enters in a manner analogous to
the modulus of a world-sheet torus embedded in the target space in genus-one string theory.
  After the above redefinition of variables we will see that
the coefficient $I_{(p,q)}$ in (\ref{ipqdef}) has the schematic form
(the precise coefficients will be included later)
\be
I_{(p,q)}= \int dV V^{5-2p-3q}\, \int {d^2\tau\over \tau_2^2}\,
B_{(p,q)} (\tau)\, \Gamma_{(n,n)}(G_{IJ}; V,\tau)\,,
\label{caliform}
\ee
where $\Gamma_{(n,n)}$  is  a lattice  factor  that contains  the
information about  the compactified target space  with metric $G_{IJ}$
($I,J=1,\cdots,n$).
 It will be important that the integrand is invariant under $SL(2,\ZZ)$, when
suitably extended outside the fundamental domain.
 This integral has ultraviolet and infrared divergences, depending
on the values  of $p$ and $q$.   These will require a careful treatment of the
 integration  limits,  which  will  be  discussed in detail in
 section~\ref{sec:limits}.

An important  property of  the coefficients, $B_{(p,q)}(\tau)$  in the
integrand    is   that   they    can   be    written   as    sums   of
components $b_{(p,q)}^r(\tau)$,
\be
B_{(p,q)}(\tau) = \sum_{i=0}^{\lceil 3N/2\rceil} b_{(p,q)}^{3N-2i}(\tau)
\label{sumsbi}
\ee
where $N=2p+3q-2$ and the components satisfy Green function equations
in $\tau$  of the form
\be
(\Delta_\tau - r(r+1))\, b_{(p,q)}^r = \tau_2\, c_{(p,q)}^r(\tau_2)\, \delta(\tau_1)\,,
\label{genpoisson}
\ee
where   $\Delta_\tau  =\tau_2^2\,(\partial_{\tau_1}^2+\partial_{\tau_1}^2)$,   $c_{(p,q)}^r$    is   a
polynomial in $\tau_2+\tau_2^{-1}$ of degree $N-1$
  (see appendix~\ref{sec:Modular}  for details)\footnote{We  would  like to  thank  Don  Zagier  for
  explaining the mathematical significance of this decomposition}. This  property will  be used
extensively to determine $I_{(p,q)}$.

 The $\calS^1$ compactification to ten dimensions
will be described in  section~\ref{sec:sone}, together with appendix~\ref{sec:s1compact}.
This will lead to coefficients for higher-momentum terms in the type
IIA theory up to order $S^6\, \calR^4$.
Although this reproduces the terms considered in earlier work,
important new issues are encountered
at order $S^4\,  \calR^4$ ($k=4$) where further
non-analytic terms arise.  Such nonanalytic behaviour arises from infrared threshold effects that are not
captured by the power series expansion (\ref{gexpand}), so we will be
careful to regulate the infrared limit of the integrals.
In ten dimensions  unitarity implies that such thresholds are logarithmic and arise
at this order in $\alpha'$ at genus-one and genus-two.   Further logarithmic singularities arise
 at genus-two at  order $s^5\,  \calR^4$, and at
genus-one and genus-three at order $s^6\,  \calR^4$, with a complicated
pattern of thresholds at all orders in $\alpha'$ thereafter.
Unlike in the case of the lowest-order nonanalytic term (\ref{nonansum}),
the scales of the logarithms, which we will not evaluate, do not cancel.
The translation of these supergravity results into the language of type IIA superstring theory
is summarized in section~\ref{connectten}.

Compactification to nine dimensions on a two-torus
will be considered in section~\ref{sec:ttwo}. The coefficients in the expansion
now have a richer structure since they depend on the three moduli of
$\calT^2$, or the complex coupling, $\Omega$, and
the radius of the compact dimension, $r_B$,  in the type IIB string
theory language.  Each term with a distinct kinematic structure must have a
coefficient that is an independent function that is invariant under the nine-dimensional
duality group, $SL(2,\ZZ)\times \IR^+$.
We will determine certain analytic  terms in the double
expansion of the amplitude up to order  $S^6\, \calR^4$ that are associated with particular inverse powers
of $r_B$.  In order for the Feynman diagram approximation to have a chance of being a
sensible approximation it is necessary that $r_B\ll 1$, or $r_A \gg 1$.
The
coefficients will be modular functions of $\Omega$.
In fact, we will see that each coefficient is generally
a sum of a number of modular functions
that satisfy independent Poisson equations analogous to
(\ref{poiss}).
The structure of these equations, which generalizes
(\ref{lapeig}), is summarized by (\ref{ipoisso}), which is
one of the most intriguing results of this paper.

In nine dimensions almost all
the low-order nonanalytic terms have branch points that are non-integer
powers of the Mandelstam invariants rather than logarithms, and so they can
be separated from the analytic part unambiguously -- the exception is the term of order $S^5 \,\log(-S)$,
which is the contribution from nine-dimensional supergravity
and can be obtained by dimensional
regularization, as summarized in appendix~\ref{sec:ninereg}.
However, there are terms that are power-behaved in $r$ as well as terms containing,
 factors such as $\log r^2$, which is nonanalytic in $r$,  and exponentially suppressed terms
of the form $e^{-cr}$.  A series of terms that are power behaved in $r_B$ was
seen to arise from the expansion of the $L=1$ supergravity amplitude in (\ref{loopone}).
Similarly, we will find that the momentum expansion of the $L=2$
amplitude gives a sum of higher-momentum modular invariant terms,
\be
A_{L=2}^{an} = \sum_{q\geq1}\sum_{p\geq 0} \sum_l r_B^{1-l}
 g_B^{\half N +\half +{l\over 4}}\calE_{(p,q)}^{(l)}
(\Omega)\,
\hat\sigma_2^p\hat\sigma_3^q\, \calR^4\,,
\label{analytr}
\ee
for various values of $l$ that will be specified later.
Terms proportional to $r$ reproduce the $d=10$ expansion, so that $\calE_{(p,q)}^{(0)}\equiv \calE_{(p,q)}$.
 All contributions with $l\geq 0$
vanish in the ten-dimensional type IIB limit,
but they give rise to well defined modular functions in
nine dimensions. In addition to terms that are power-behaved in the radius $r_A$ or $r_B$,
there are also terms proportional to $\log r_A$ or $\log r_B$.  Such terms arise
explicitly at genus-one in nine-dimensional string theory \cite{grv:oneloop}.
For example, there is a term of the  form $r \log r\times\, s^4 \, \calR^4$, which is intimately related to the
presence  of  the  genus-one  $s^4  \log  s\,  \calR^4$  term  in  ten
dimensions determined in~\cite{grv:oneloop}.
We will see in the following that this dependence on $r$ can also be seen from the $\calT^2$
reduction of two-loop ($L=2$) eleven-dimensional supergravity.
Terms of the form $e^{-cr_B}$ that arise in string theory when $2p+2q\ge 4$ are not reproduced
by Feynman diagrams at any number of loops.

Perturbative contributions to
the string amplitude are obtained from the weak-coupling
expansion of these modular functions (making use of the methods described in
appendix~\ref{sec:Diff}).  Each term in the momentum expansion  derived in this manner
is accompanied by a particular inverse power of the radius $r_B$ and the new terms do
not contribute in the large-$r_B$ limit.  However, after T-duality to the IIA theory, we
are able to compare a number of coefficients with those derived explicitly from genus-one
in string theory compactified on a circle \cite{grv:oneloop} and find precise agreement.
Special issues concerning the terms that contain $\log r$ factors will also be discussed.
The issue of the pattern of logarithms is intimately related to the threshold behaviour
in maximal  supergravity in various dimensions.
In appendix~\ref{sec:DimReg} we will evaluate the supergravity amplitude in nine, ten and
eleven dimensions, making use of dimensional regularization.  These expressions are of
relevance to various pieces of the argument in the body of the paper.
For example, in ten dimensions the pole
term gives rise to a term of order $S^5\, \log S\, \calR^4$ that is identified with a genus-two
contribution to $s^5\, \log s\,\calR^4$ in ten-dimensional string theory.
In section~\ref{sec:modular} we will sketch the way in which
supersymmetry constrains higher derivative terms and argue that
the structure of the Poisson equations satisfied by the coefficients
of the terms in the derivative expansion of the
nine-dimensional IIB theory can be motivated by supersymmetry.

%%%%%%%%%%%%%%%%%%%%%%%%%%%%%%%%%%%%%%%%%%%%%%%%%%%%%%%%
\section{ Properties of the two-loop supergravity amplitude}
\label{sec:expand}
%%%%%%%%%%%%%%%%%%%%%%%%%%%%%%%%%%%%%%%%%%%%%%%%%%%%%%%%

\begin{figure}[ht]
\centering
\includegraphics[width=10cm]{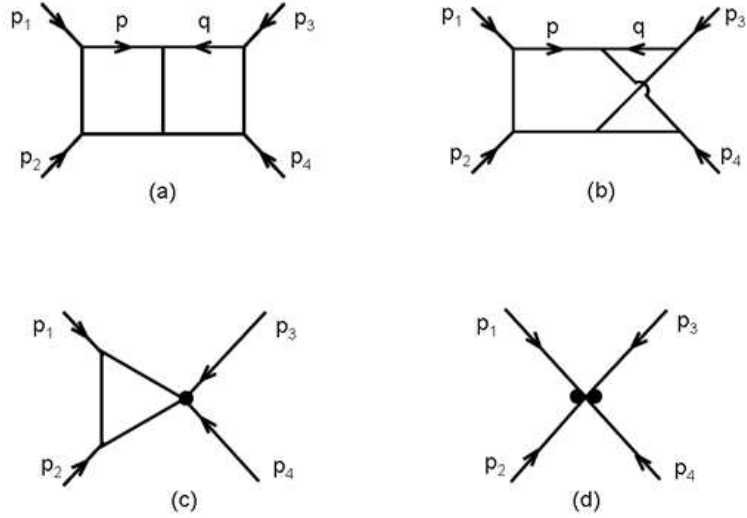}
\caption{The two-loop four-graviton amplitude in eleven dimensions.
(a) The $S$-channel planar diagram  reduces to
$S^2\, \calR^4$ multiplying a scalar field theory double-box diagram.
(b) The $S$-channel nonplanar diagram  reduces to
$S^2\, \calR^4$ multiplying a nonplanar scalar field theory two-loop diagram.
(c)  The triangle diagram with a one-loop counterterm at one vertex that subtracts
a sub-divergence.
(d)  A new two-loop primitive divergence.}
\label{fig:TwoBoxDiagram}\end{figure}

It has been known for a long time that the sum of one-loop Feynman diagrams that
contribute to four-graviton scattering in maximal supergravity in any dimension has the
form of a box diagram of $\varphi^3$ scalar field theory multiplying
$\calR^4$, where
$\calR$ is the linearized Weyl curvature, as discussed in the introduction. Similarly,
the sum of all two-loop diagrams,
$A(S,T,U)$,  is very economically expressed in terms of
two  particular diagrams of $\varphi^3$ scalar field theory
\cite{BernDunbar}.  These are the planar double-box
diagram, $I^P(S,T)$ of figure~\ref{fig:TwoBoxDiagram}(a), and the non-planar double box
diagram, $I^{NP}(S,T)$ of figure~\ref{fig:TwoBoxDiagram}(b), together
with the other diagrams obtained by permuting the external particles. In
addition, one must include the one-loop triangle
diagram of figure~\ref{fig:TwoBoxDiagram}(c)
containing a one-loop counterterm
at one vertex (indicated by the blob),
which subtracts the one-loop sub-divergences from the two-loop diagrams. In addition there are
two-loop primitive divergences (that are indicated by the double-blob in
figure~\ref{fig:TwoBoxDiagram}(d)).

The two-loop integrals  appearing in the amplitude are  sums of planar
and   non-planar  pieces,  (\ref{sumall}).    We  are   interested  in
compactifying these expressions on  the $n$-torus $\calT^n$ with $n=1$
or $2$.  After manipulations  that are given in \cite{gkv:twoloop} the
loop  integrals can  be expressed  as integrals  over  seven Schwinger
parameters,  one  for each  propagator.   The  integrations over  loop
momenta  in the  compact  directions  are replaced  by  sums over  the
Kaluza--Klein integers in each loop $m_I$ and $n_I$, where $I=1,\dots,
n$.    After   performing   the   integration  over   the   continuous
$(11-n)$-dimensional loop momenta,  the planar and non-planar diagrams
reduce to
  \bea
 I^P(S;T,U) &=&  {\pi^{11-n}\over \calV_n^2}\,
\int_0^\infty  d  L_1  d  L_2 dL_3\,  \Gamma_{(n,n)} \,  \int_0^{L_3}\!\!  dt_4\int_0^{t_4}\!\!
dt_3\int_0^{L_1}dt_2 \int_0^{t_2}dt_1
\, \Delta^{n-11\over 2}
\, e^{h_P} \,,
\label{iplanar}
\eea
and\footnote{ In  this  section  we  ignore  the  ultraviolet  and
  infrared divergences. A treatment of these divergences and a proper
definition of the integration limits of the integrals will
  be discussed in section~\ref{sec:limits}.}
\bea
I^{NP}(S;T,U) &=& {\pi^{11-n}\over \calV_n^2}\,
\int_0^\infty\!\! d L_1 d L_2 dL_3 \, \Gamma_{(n,n)}
\int_0^{L_3} \!\! dt_3 \int_0^{L_2}\!\! dt_4 \int_0^{L_1}\!\! dt_2\int^{t_2}_0\!dt_1 \Delta^{n-11\over 2}
\,  e^{h_{NP}} \,,
\label{inonplanar}
\eea
where
\be
\Delta=   L_1 L_2+  L_3 L_1+ L_2  L_3\,.
\label{deltadef}
\ee
The lattice factor $\Gamma_{(n,n)}$ is defined by
\be
\Gamma_{(n,n)}(G^{IJ};\{L_k\}) = \sum_{(m_I,n_I)\in \ZZ^{2n}} e^{-\pi G^{IJ} \left( L_1 m_Im_J + L_3 n_I n_J +
L_2 (m+n)_I (m+n)_J \right)}\,.
\label{latdef}
\ee
where $G^{IJ}$ is the inverse metric on $\calT^n$ and $\calV_n=\sqrt{\det G_{IJ}}$ is its
volume.  The quantities $h_P$ and $h_{NP}$ are given by\footnote{The variables in this
section are related to those of \cite{gkv:twoloop}
by $L_1 =\lambda$,
$L_2=\rho$, $L_3=\sigma$, $t_1=L_1 w_1$, $t_2=L_1 w_2$, $t_3 = L_3 v_1$ and in the planar
case, $t_4 = L_3 v_2$, while in the non-planar case, $t_4= L_2 u_1$.}
\bea
h_P &=& T{ L_2\over
 \Delta}(t_4-t_3)(t_2-t_1)\\
&+&S\, \left[{L_2\over L_1  L_3 \Delta}
 (L_1 t_3- L_3 t_1)(L_1 t_4- L_3 t_2)+ {1\over L_3} t_3(L_3-t_4)+ {1\over L_1} t_1(L_1-t_2)
 \right]\nn\\
 &=& {1\over \Delta} \left(-S (t_1t_2  (L_2+L_3) + t_3t_4(L_2+L_1))
 +T (t_2t_4+t_1 t_3) L_2
 +U  (t_1 t_4 + t_2 t_3)L_2\right)\nn\\
 \nn &&\qquad + S\, t_3+ S\, t_1
 \,,
\label{hpdef}
\eea
and
\bea
h_{NP}&=& T{ 1\over
 \Delta}(L_2 t_3- L_3 t_4)(t_2-t_1)\\
&+&S\, \left({ 1\over L_1\Delta}
 (L_1 t_4- L_2t_1)(L_1 t_3- L_3t_2)+  {1\over L_1} t_1(L_1-t_2) \right)\nn\\
 &=&{1\over \Delta} \left(S (-t_1t_2 (L_2+L_3) + t_3t_4 L_1)
 + T(t_1t_4 L_3 + t_2t_3 L_2) +U(t_1t_3 L_2 + t_2t_4 L_3) \right)
  +  S\, t_1   \,.\nn
\label{hnpdef}
\eea
In writing these expressions we have ignored the ultraviolet divergences, which are manifested
as divergences at the $L_k=0$ endpoints ($k=1,2,3$) that will be regulated by a cutoff
in subsection~\ref{sec:counter} (as in \cite{gkv:twoloop}).
The complete expression, $\calI(S,T,U)$ in (\ref{calidef}) is obtained by summing the $S$-channel,
$T$-channel and $U$-channel diagrams.

\subsection{World-line presentation of the two-loop amplitude}
\label{sec:worldpres}

The above
structure of the two-loop amplitude can, in principle, be deduced by considering the quantum mechanics
functional integral associated with the world-lines for the internal propagators in the two-loop
diagrams. This has a structure that bears a close resemblance to the world-sheet description of
the genus-two string theory amplitude (although that is formulated in ten-dimensional space-time).
We will here rewrite the expressions for the two-loop Feynman diagrams of the previous subsection in
order to make this explicit.
The advantage of this description is that it naturally packages together
the planar and nonplanar diagrams of the $S$, $T$ and $U$ channels.

\begin{figure}[ht]
\centering
\includegraphics[width=10cm]{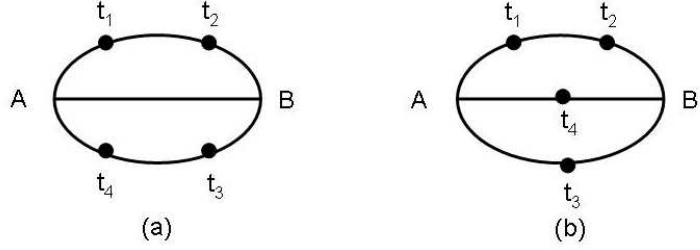}
\caption{(a) A planar diagram is represented by the
skeleton with a pair of external states connected to
 each of two internal lines.
(b)  The nonplanar configuration in which one pair of external states is attached to a single line
and the other states are each attached to separate lines.  Integrating
the positions of the four states over the whole network generates
the sum of all Feynman diagrams.
}
\label{fig:skeleton}\end{figure}
The `skeleton', or vacuum diagram, has
three scalar propagators joining the junction $A$ to junction $B$
in figure~\ref{fig:skeleton}.
The lengths of these lines, $L_k$
($k=1,2,3$), are moduli that are to be integrated between $0$ and $\infty$.
The scattering particles with momenta $p_r^\mu$ ($r=1,2,3,4$) are
associated with plane-wave vertex operators that are inserted at
positions $t_r$ on any of the three lines of the skeleton, as
shown in figure~\ref{fig:skeleton}.  These positions are then to be
integrated over the whole network. Since there are four vertex operators and only three lines,
 at any point in the integration
domain one pair of vertex operators  is
attached to one line, say line $1$, while the other two  may both be
attached to one of the other two lines (line $2$ or $3$), which is the planar situation, or else the
other two lines may have only one vertex operator attached, which is the non-planar situation.
The labelling of the positions $t_r$ of the vertex operators is
arbitrary, but it is convenient to choose coordinates $t_r^{(k_r)}$
for particle $r$ on line $k_r$ such that
\be
t_r = t_r^{(k_r)}\,,
\label{coordpart}
\ee
where $0\le t_r^{(k_r)}\le L_{k_r}$,  and $t_1^{(k_1)}=0$, $t_2^{(k_2)}=0$, $t_3^{(k_3)}=0$ and
$t_4^{(k_4)}=0$ coincide at the junction $A$.
In other words, the integral
over the whole network decomposes into sectors labeled by $\{k_r\}$,
\be
\oint \prod_{r=1}^4 dt_r \equiv \sum_{\{k_r\}}\int_0^{L_{k_1}} dt_1^{(k_1)}
\cdots \int_0^{L_{k_4}} dt_4^{(k_4)}
\label{measureone}
\ee

The expression for the Feynman diagrams can be written in a compact form in terms of the Green function,
$G_{rs}$, between two vertices at points $t_r$ and $t_s$ on the skeleton diagram.  Following
\cite{Dai:2006vj} this is written in terms of two-vectors
\be
{\bf v^{(k_r)}} = t_r^{(k_r)}\, {\bf u}^{(k_r)}\, \quad {\rm or,\ in\ components,} \quad  v_I^{(k_r)} = t_r^{(k_r)}\, u_I^{(k_r)}\,,
\label{vdefs}
\ee
where $I=1,2$ labels the loop and $u_I^{(k)}$ are constant vectors
\be
{\bf u^{(1)}}= \begin{pmatrix} 1\\ 0 \end{pmatrix}\,\qquad {\bf u^{(2)}}=
\begin{pmatrix} -1\\ 1 \end{pmatrix}\,,\qquad {\bf u^{(3)}}= \begin{pmatrix} 0\\ -1 \end{pmatrix}
\label{udefs}
\ee

With this notation the sum of all two-loop contributions to the amplitude
defined in (\ref{eberndun}) and (\ref{calidef}) is given by
\bea
\calI(S,T,U)&=& {\pi^{11-n}\over \calV_n^2}\, \int_0^\infty dL_1\, dL_2\, dL_3\,
\,\Gamma_{(n,n)} \, \oint \prod_{r=1}^4 dt_r\, W^2\, \Delta^{\frac{n-11}{2}}\,
\, e^{-\sum_{r,s=1}^4 p_r \cdot p_s
\, G_{rs}}\,  ,
\label{funcint}
\eea
where $G_{rs}$ is the one-dimensional Green function for the
Laplace operator evaluated  between the points $t_r$ and  $t_s$ on the
skeleton diagram, to be discussed below. The lattice factor is defined
in (\ref{latdef}).

The  function  $W$ appearing  in  the  measure  in (\ref{funcint})  is
defined by\footnote{The world-line formulation of the
  two-loop four-graviton amplitude in ten dimensions would arise from a field theory limit of four-graviton genus two
  amplitude in type~II superstring theory. The function $W$, used here,
  is the field theory limit of the function $\mathcal{Y}_S$ that enters the string amplitude  derived in~\cite{D'Hoker:2005jc}.}
\bea
     3W     &=&     (T-U)\,
\Delta_{12}\Delta_{34} + (S-T)\, \Delta_{13}\Delta_{24} +(U-S)\,
\Delta_{14}\Delta_{32}  \nn\\
&=& \ \ \, S( u_1^{(k_1)} u_1^{(k_2)} u_2^{(k_3)} u_2^{(k_4)}
+u_2^{(k_1)} u_2^{(k_2)} u_1^{(k_3)} u_1^{(k_4)})\nn\\
&& +T( u_1^{(k_1)} u_2^{(k_2)} u_2^{(k_3)} u_1^{(k_4)}
+u_2^{(k_1)} u_1^{(k_2)} u_1^{(k_3)} u_2^{(k_4)})\nn\\
&&+U(  u_1^{(k_1)}  u_2^{(k_2)}  u_1^{(k_3)} u_2^{(k_4)}  +u_2^{(k_1)}
u_1^{(k_2)} u_2^{(k_3)} u_1^{(k_4)})\,
\label{ydef}
\eea
where
\be
\Delta_{rs} = \epsilon^{IJ}\, u_I^{(k_r)}\, u_J^{(k_s)}\,.
\label{wdef}
\ee
Note, in particular, that  $\Delta_{rs} =0$ if $k_r= k_s$ (i.e., $t_r$ and $t_s$ are on the same line).
Furthermore  $W=0$ if three of the vertices on the same line (using $S+T+U=0$),
so that the only non-zero contributions
come from the planar and non-planar diagrams of figure~\ref{fig:TwoBoxDiagram}.
It is easy to see that in any region in which $t_r$ and $t_s$ are on the same line,
$W = -6 k_r\cdot k_s$, so that
\begin{itemize}
\item{ $W^2= S^2$\ \ \ if $k_1=k_2$ and/or $k_3=k_4$\,;}
\item{$W^2=  T^2$\ \ \ if $k_1=k_4$ and/or $k_2=k_3$\,;}
\item{$W^2=  U^2$\ \ \ if $k_1=k_3$ and/or $k_2=k_4$\, .}
\end{itemize}

This setup makes contact with the discussion in \cite{Dai:2006vj}, where the Green function for an
arbitrary Feynman diagram of $\varphi^3$ scalar field theory was described.  Our case differs only due to
the presence of a measure factor $W^2$ in (\ref{funcint}) which encodes the fact that we are
discussing maximal supergravity.  However, the exponential factor involves the same Green function
as in \cite{Dai:2006vj}, which has the form
\be
G_{rs} = -\frac{1}{2} d_{t^{(k_r)}_r  t^{(k_s)}_s} +
\frac{1}{2} ({\bf v^{(k_r)\,T}- v^{(k_s)\, T}})\, K^{-1} \, ({\bf v^{(k_r)}- v^{(k_s)}})\,,
\label{genform}
\ee
where  $d_{t^{(k_r)}_r t^{(k_s)}_s}$  is the  modulus of  the distance
between $t_r$ on line $k_r$ and $t_s$ on line $k_s$. If $k_r=k_s$ then
$d_{t_r^{(k_r)} t_s^{(k_r)}}  = |t_r^{(k_r)}-t_s^{(k_r)}|$, if $k_r\ne
k_s$         then        $d_{t_r^{(k_r)}         t_s^{(k_s)}}        =
(t_r^{(k_r)}+t_s^{(k_s)})$.
The matrix $K^{-1}$ (analogous to the inverse of the imaginary part of
the period matrix  in the genus-two string calculation)  is defined by
\be K^{-1}=\frac{1}{\Delta}
\begin{pmatrix}
  L_3+L_2 & L_2\\
  L_2 & L_1+L_2
\end{pmatrix}\, ,
\label{periodinv}
\ee
where $\Delta$ is defined in (\ref{deltadef}).
The function $G_{rs}$ is constructed to be the Green function of the one-dimensional Laplace operator
that satisfies
\be
\frac{d^2}{dt_r^2}\, G_{rs}= -\delta(t_r-t_s) + \rho\,,
\label{greeneq}
\ee
and $G_{r,r}=0$,
where
\be
\oint \rho dt \equiv \sum_{k=1}^3 \int_0^{L_k} \rho^{(k)}\, dt^{(k)} = 1\,,
\label{rhocon}
\ee
and $\rho^{(k)}$ is a constant on line $k$.  The presence of $\rho$ in (\ref{greeneq})
ensures $\oint \ddot{G}_{rs} \, dt_s=0$, which is required by Gauss' law on the compact one-dimensional
network.

The Green function (\ref{genform})
satisfying the conditions (\ref{greeneq}) and (\ref{rhocon}) has a functional form
that depends on whether the points $t_r$ and $t_s$ are on the same line or on different lines.
If $t_r$ and $t_s$ are on the same line ($k_r=k_s$)
\be
G_{rs} = -\frac{1}{2}\, |t^{(k_r)}_r-t^{(k_r)}_s| +\frac{1}{2\Delta}\,
(L_l+L_m)\, (t_r^{(k_r)}- t_s^{(k_r)})^2\,,
\label{greensame}
\ee
where $l\ne m\ne k_r=k_s$. If they are on different lines ($k_r\ne k_s$)
the Green function is given by
\bea
G_{rs} &=& -\frac{1}{2}\, (t^{(k_r)}_r+t^{(k_s)}_s)  + \frac{  \left(
(L_l + L_{k_s})\, (t^{(k_r)}_r)^2+ (L_l+L_{k_r})\, (t^{(k_s)}_s)^2 + 2t_r^{(k_r)}\, t_s^{(k_s)}\,
L_l\right)}{2\Delta}\,,
\label{greensameb}
\eea
where $l\ne k_r\ne k_s$.  In verifying the conditions (\ref{greeneq}) and (\ref{rhocon})
we find that   $\rho^{(k)} = (L_l+L_m)/\Delta$, where $k\ne l\ne m$.
The terms quadratic in a single $t_r$ or $t_s$ in $G_{rs}$ do not contribute to the exponent in
(\ref{funcint}) due to the condition $S+T+U=0$ so the exponential factor ends up being extremely
simple.

The integral over the vertex operator positions, $t_r^{(k_r)}$, separates into the two distinct classes
described above, namely:
(a)  Planar configurations in which one pair is attached to one of the three internal lines, and the other
pair is attached to one of the other lines;
(b) Non-planar configurations in which one pair is attached to one of the
internal lines while the other vertices are each attached to the other two internal lines.
It is straightforward to see that these contributions are identical to those given by
the integrals (\ref{iplanar}) and (\ref{inonplanar}).

The complete integral over the $t_r$'s in (\ref{funcint})
automatically adds contributions that permute the lines and the
positions of the four states attached to them.
Using these expressions for $G_{rs}$ and $W$ the expression (\ref{funcint})
reproduces the sum of terms inside the square bracket in the last line of  (\ref{eberndun}),
which is the sum of planar and nonplanar diagrams in the $S$, $T$ and $U$ channels.
The expression (\ref{funcint}) has an obvious discrete symmetry under the shift $k_r \to k_r+1$,
fixing all $k_s$ with
$s\ne r$ (and with the identification $u_I^{(4)} \equiv u_I^{(1)}$), which
moves a vertex operator from one line of the skeleton to the next.  This can be thought of as a discrete
remnant of the reparametrization invariance of the world-line functional integral that corresponds to
cutting two of the lines of the skeleton to produce four endpoints, and regluing the endpoints in
a different order.
One  important insight one gains from this symmetry is that the planar and nonplanar diagrams in all
channels are required and their relative normalizations are fixed.

It is important to exploit the symmetries of the complete integral (\ref{funcint}),
which automatically combines
the planar and nonplanar diagrams and symmetrizes
(\ref{iplanar}) and (\ref{inonplanar}) over permutations
of $L_1$, $L_2$ and $L_3$.

Formally, if we ignore divergences, the low energy expansion of
(\ref{funcint}) can be written as
a power series in symmetric monomials of the Mandelstam invariants, as in
(\ref{ipqdef}), by expanding the factor of $e^{-\sum p_r\cdot p_s G_{rs}}$.
The resulting coefficients in (\ref{ipqdef}) may be written as
 \be
\tilde I_{(p,q)}={\pi^{11-n}\over N!}\,
\int \prod_{k=1}^3 d L_{k}\, \Delta^{-{1\over2} + p + \threeh q}\,
\tilde B_{(p,q)}(L_2/L_1,L_3/L_1)\, \Gamma_{(n,n)}\,,
\label{ibdef}
\ee
where $\tilde B_{(p,q)}$ is the coefficient of $\sigma_2^p\sigma_3^q$
(which is of order $N+2$ in the Mandelstam
invariants, with $N=2p+3q-2$) in the expansion of the exponential in the integrand of
(\ref{funcint}) and is given by
\be
\sum_{2p+3q=N+2} \sigma_2^p\sigma_3^q\,  \tilde B_{(p,q)}(L_2/L_1,L_3/L_1) = \Delta^{-2 - \half N} \,
\oint \prod_{r=1}^4 d t_r\,  W^2\,
\left(-\sum_{r,s=1}^4 p_r \cdot p_s\, G_{rs}\right)^N\, .
\label{bndefs}
\ee
We will need to evaluate the coefficients $\tilde B_{(p,q)}$ in order to evaluate $\calI(S,T,U)$ in
(\ref{ipqdef}).  The first two
cases are known.  The zeroth order term has
$N=0$ ($p=1, q=0$) and is given by
\be
\tilde B_{(1,0)} = \Delta^{-2}\,\sum_{\{k_r\}\in\{L_r\}}
\int_0^{L_{k_1}}dt_1^{(k_1)}\dots \int_0^{L_{k_4}}\, dt_4^{(k_4)} = 1\,,
\label{bzerodef}
\ee
which agrees with \cite{gkv:twoloop}.
For $N=1$ ($p=0$ and $q=1$), substituting the expression
for $G_{rs}$ leads to
\bea
\sigma_3\,\tilde B_{(0,1)}&=&
{\Delta^{-\fiveh}\over3}\, \sum_{\{k_r\}\in\{L_r\}}\left(\prod_{r=1}^4\int_0^{L_{k_r}}
d t_r^{(k_r)}\right)\, W^2 \, \left(S(G_{12} + G_{34})+T(G_{14}+G_{23}) +
U(G_{13}+G_{24})\right)
\nn\\
&=& \frac{\sigma_3}{12}\, \left({L_1+ L_2 + L_3\over\Delta^\half} +
5{L_1 L_2L_3\over \Delta^{\threeh}}\right)\,,
\label{bonedef}
\eea
in agreement with \cite{gv:D6R4} (allowing for the extra normalization factor of $1/12$).
In evaluating the integrals over $L_k$ in (\ref{ibdef}) care must be taken to subtract the
ultraviolet divergent parts, as we will review later.  There are also singular infrared
effects associated with the occurrence of massless particle thresholds, giving
 rise to nonanalytic behaviour that is not captured by the
expansion (\ref{ipqdef}), as will also be seen later.

First we will describe a change of integration variables that is  very useful for evaluating the
integral.

\subsection{Redefinition of the integration parameters}

\begin{figure}[h]
\centering
\includegraphics[width=11cm]{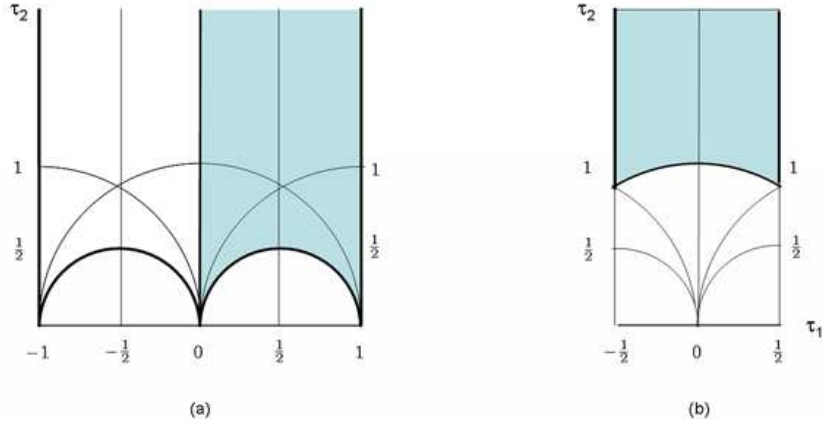}
\caption{(a) The region of integration of $\tau=\tau_1+i\tau_2$ is equivalent to a fundamental
domain of $\Gamma_0(2)$.  Ultraviolet divergences arise on the boundary of
this region.  (b)  The integrand can be mapped into a
threefold cover of the
fundamental domain, $\calF$,  of $SL(2,\ZZ)$.  The ultraviolet divergences arise from the $\tau_1=0$
axis.}
\label{fig:ModularRegions}\end{figure}

As in \cite{gv:D6R4} it is very useful to redefine the Schwinger parameters by replacing
$L_1$, $L_2$, $L_3$,  by the variables $V$ and $\tau = \tau_1+i\tau_2$,
defined by
\be
\tau_1={L_1\over L_2+ L_1 }\ ,\qquad \tau_2={\sqrt{\Delta}\over  L_2+ L_1 }\ ,\qquad
V= \Delta^{-\half}\,.
\label{paramdefs}
\ee
The integration measure transforms as
\be
d L_1\, d L_2\, d L_3 = 2 {dV\over V^4}\, {d^2\tau\over \tau_2^2}\,.
\label{meastrans}
\ee
The ranges of the new variables are
\be
0\le \tau_1 \le 1\, , \qquad |\tau-\half| \ge \half\,,\qquad 0\le V \le \infty\,,
\label{newrange}
\ee
which is the shaded region shown in figure~\ref{fig:ModularRegions}(a).
 This is a fundamental domain of the
group $\Gamma_0(2)$.
 The three segments of the boundary of this region are:
$\tau_1 = 0$ that comes from $L_1\to 0$ with $
L_2,L_3$ fixed;
$\tau_1 =1$ that comes from $ L_2\to 0$  with $
L_1, L_3$ fixed and $|\tau|^2 -\tau_1=0$ that comes from
$ L_3\to 0$. with $ L_2, L_3$ fixed.  It follows that
that the ultraviolet divergences, arise at the
boundary of this region. From its construction it is evident that the sum of two-loop integrals is invariant under the action of the
symmetric group, $\mathfrak{S}^3$ on the three parameters $L_1$, $L_2$ and $L_3$, which maps the six regions in
the shaded domain in figure~\ref{fig:ModularRegions}(a) into each other.
The action of the two-cycles in $\mathfrak{S}^3$  on the Schwinger parameters is given by the following
actions on $\tau$,
\be
L_1 \leftrightarrow L_2\ :\quad \tau\to 1-\tau^*\,,\quad
L_1 \leftrightarrow L_3\ :\quad \tau\to \frac{1}{\tau^*}\,,\quad
L_2 \leftrightarrow L_3\ :\quad \tau\to \frac{\tau^*}{\tau^*-1}\,.
\label{symmthree}
\ee
where  $\tau^*=\tau_1-i\tau_2$  is  the  complex conjugate  of  $\tau$

It is also easy to see that the integrand is invariant under the $T$ and $S$ transformations defined by
\be
T\, : \qquad\quad L_1\to 2L_1+ L_2\, , \ \quad\qquad L_2\to -L_1\,,
\qquad  L_3\to 2 L_1 +  L_3\,,
\label{tdef}
\ee
and
\be
S\, : \qquad\quad   L_1\to  -L_1\, ,
\qquad L_2\to 2L_1+L_3\,, \qquad L_3\to  2L_1+  L_2\,.
\label{sdef}
\ee
Invariance under the these transformations depends on the invariance of the lattice factor
$\Gamma_{(n,n)}$.
For example, consider the $\calS^1$ compactification, where the lattice factor
$\Gamma_{(1,1)}(\{L_r\})$ is given as a sum over $m$ and $n$ in (\ref{latdef}).
The $T$ transformation is an invariance when accompanied by the shift $m \to m+n$.
Likewise, the $S$ transformation is an invariance when
accompanied by the transformation $m\to -n$, $n\to m$.
In terms of the new variables these transformations become
\be
T\,  : \quad \tau\to \tau+1\,,\qquad
S\, : \quad  \tau \to -\frac{1}{\tau}\,.
\label{staudef}
\ee
Note that the cyclic permutation of the Schwinger parameters is generated by
\be
T\, S\,:\qquad\quad  L_1 \to   L_3\,,\qquad L_2\to L_1\, , \qquad L_3\to L_2\,,
\label{sttrans}
\ee

 We may now use the $S$ and $T$ transformations to map the shaded domain in
figure~\ref{fig:ModularRegions}(a) into a three-fold cover of the shaded area in figure~\ref{fig:ModularRegions}(b).
This is $\calF$, the fundamental domain of $SL(2,\ZZ)$, which is defined by $\{-\half\le \tau_1\le \half$,
$|\tau > 1\}$.  The boundaries at $\tau_1=0$, $\tau_1=1$ and $|\tau| =\half$ in
 figure~\ref{fig:ModularRegions}(a) map into the
 line $\tau_1=0$ in $\calF$.   In other words the integral over the domain (\ref{newrange}) is
three times the integral over  $\calF$.

In terms of $\tau$ and $V$ the matrix $K^{-1}$ in (\ref{periodinv}) takes the
$SL(2,\ZZ)$-covariant form
\be
K^{-1}= \frac{V}{\tau_2}
\begin{pmatrix}
|\tau|^2 & \tau_1\\
\tau_1 & 1
\end{pmatrix}
\, ,
\label{invtau}
\ee
while  (\ref{genform}) becomes
\be
G_{rs} = -\frac{1}{2} d_{ t_r t_s} +\frac{V}{2\tau_2}|
 v^{(k_r)}_2-  v^{(k_s)}_2 + \tau (v^{(k_r)}_1- v^{(k_s)}_1)|^2\,.
\label{genress}
\ee

For much of what follows it will be useful to perform Poisson resummations on the Kaluza--Klein
modes $(m_I, n_I)$  to express the lattice factor  in terms of winding numbers $(\hat m^I,\hat n^I)$,
just as in \cite{gkv:twoloop}\footnote{Recall that a Poisson resummation that
replaces a sum over a Kaluza--Klein
charge $m$ by a sum over a winding number $\hat m$ is expressed by the identity
$\sum_{m=-\infty}^\infty e^{-\pi A m^2 + 2m\pi A s} = A^{-\half} e^{\pi A s^2}
\sum_{\hat m=-\infty}^\infty e^{-\pi A^{-1}
\hat m^2 - 2i\pi \hat m s}$.},
\be
\Gamma_{(n,n)}(G_{IJ};V,\tau) = \calV_n^2\, \Delta^{-{n\over 2}}\, \hat\Gamma_{(n,n)}(G_{IJ};V,\tau)\,,
\label{latticetrans}
\ee
so that the $n$-dimensional lattice factor in (\ref{iplanar}),  (\ref{inonplanar})
and (\ref{latdef}) is given in terms of sums over winding numbers by
\bea
 \hat\Gamma_{(n,n)}(G_{IJ};V,\tau)= \sum_{( \hat m^I, \hat n^I)\in\ZZ^{2n}}
 e^{-\pi \,  {G_{IJ}\over  \tau_2\, V}
( \hat m^I+  \hat n^I\tau)(\hat m^J+ \hat n^J\bar\tau)}=
\sum_{( \hat m^I, \hat n^I)\in\ZZ^{2n}} e^{-\pi \hat E}\,,
\label{latticefact}
\eea
where we have defined
\be
\hat E(G_{IJ};V,\tau) = {G_{IJ}\over  \tau_2\, V}
( \hat m^I+  \hat n^I\tau)(\hat m^J+ \hat n^J\bar\tau)\,.
\label{edhatdef}
\ee
This expression is familiar in string theory as the partition function for the
mapping of a world-sheet torus with complex structure $\tau$
 into a target space torus with metric $G_{IJ}$.
This will prove to be important in evaluating the integrals (as it was in \cite{gkv:twoloop,gv:D6R4}).
Note that the factor $\calV_n^{-2}\Delta^{n/2}$
in the measure of $I^P$ and $I^{NP}$ cancels in the winding number basis.
In terms of the new variables the expression (\ref{funcint}) has the form
\be
\calI(S,T,U)= \int_0^\infty dV V^3\,\int_\calF \frac{d^2\tau}{\tau_2^2}
\, \oint \prod_{r=1}^4 dt_r \, W^2 \,e^{-p_r\cdot p_s\, G_{rs}}\, \hat\Gamma_{(n,n)} \,.
\label{funcintnew}
\ee
In section~\ref{sec:counter} we will discuss how the ultraviolet cutoff on
the loop momentum can be imposed by suitable choice of limits on these
integrals.

In order to evaluate the low energy expansion of the amplitude we need
to  evaluate  the coefficients  $I_{(p,q)}$  in (\ref{ipqdef}),  which
arise  as  the coefficients  of  the terms  in  the  expansion of  the
integrand  in (\ref{funcintnew}).  Since we want to express the integral in terms of the
new variables $\tau$ and $V$,  we need to change variables in the functions $\tilde B_{(p,q)}(L_2/L_1,L_3/L_1)$,
defined in (\ref{ibdef}).  In the process of changing variables from $(L_1,L_2,L_3)$ to $(\tau_1,\tau_2,V)$
it will turn out to be  convenient to redefine the normalization of
$\tilde B_{(p,q)}$ by a multiplicative constant
factor in order to arrive at final equations with simple coefficients.  We therefore define the
rescaled coefficients $B_{(p,q)}(\tau)$ by
\be
B_{(p,q)}(\tau)=d_{(p,q)}\, \tilde B_{(p,q)}(L_2/L_1,L_3/L_1)\,,
\label{ddef}
\ee
where, up to the order considered in this paper, the integer coefficients $d_{(p,q)}$ are arbitrarily chosen
to be
\bea
d_{(1,0)} &=&1\,,\quad d_{(0,1)} = 12\,,\quad d_{(2,0)} =  144\,,\nn\\
d_{(1,1)} &=&  15120\,,\quad
d_{(3,0)} = 302400\,,\quad
d_{(0,2)} =  {3\over 4}\, 302400\,.
\label{bbtildef}
\eea
The simplest examples of the $B_{(p,q)}$'s are
$B_{(1,0)}=1$, obtained in (\ref{bzerodef}) and \cite{gkv:twoloop}, and
$B_{(0,1)}$ obtained in (\ref{bonedef})
and \cite{gv:D6R4}, which is given in terms of $\tau_1$ and $\tau_2$ by
\be
 B_{(0,1)}(\tau) = {1\over \tau_2} (|\tau|^2 - |\tau_1 |+1)
+{5\over\tau_2^3} (\tau_1^2-|\tau_1|)(|\tau|^2-|\tau_1|)
\,.
\label{bonedefb}
\ee
A most important feature of $B_{(0,1)}$ is that it
satisfies the Poisson equation
\be
(\Delta_\tau-12)B_{(0,1)} (\tau)=-12\tau_2\delta(\tau_1)\, ,
\label{laplaceA}
\ee
where the laplacian can be written in terms of the original Schwinger parameters as\footnote{
The symbol $\Delta = L_1L_2+L_2L_3+L_1L_2$ should not be confused with $\Delta_\tau$.}
\be
\Delta_\tau \equiv \tau_2^2 (\partial^2_{\tau_1} + \partial^2_{\tau_2}) =
\Delta\, \partial_{L_k}\,\partial_{L_k} -2L_k\, \partial_{L_k}\, .
\ee
This  property  is  very  useful  for determining  properties  of  the
coefficients $I_{(p,q)} (S,T,U)$,
in the low energy expansion in (\ref{ibdef}).

As we will show in appendix~\ref{sec:Modular}, the higher-order  coefficients satisfy
generalizations of this Poisson equation.  In general these coefficients are sums of the form
$B_{(p,q)}(\tau) = \sum_{i} b_{(p,q)}^i(\tau)$,
where the components $b_{(p,q)}^i$ individually satisfy Poisson equations that generalize (\ref{laplaceA}).
The functions $B_{(2,0)}\,,  B_{(1,1)}\,, B_{(3,0)}\,, B_{(0,2)}$ are described in detail
in appendix~\ref{sec:Modular} together with the detailed Poisson equations of the form (\ref{poissoeq})
satisfied by the $b_{(p,q)}^i$'s.
These  Poisson  equations  are  again  the key  to  understanding  the
structure of the coefficients $I_{(p,q)}$.

%%%%%%%%%%%%%%%%%%%%%%%%%%%%%%%%%%%%%
\subsection{Integration limits}\label{sec:limits}
%%%%%%%%%%%%%%%%%%%%%%%%%%%%%%%%%%%%%%

Up to now we have ignored the fact that the Feynman integrals are ultraviolet divergent and therefore need to be
regulated in a systematic manner, such as by introducing a momentum cutoff.
We will implement this by introducing a lower cutoff on the Schwinger
 parameters, $L_k$, that are conjugate to the loop momentum \cite{ggv:oneloop}, as will be discussed
 in the nest two subsections.  Although this is feasible
 at one and two loops ($L=1$ and $L=2$) it is
unlikely to be convenient, and may not even be
consistent, for $L>2$.  In addition, although the loops are not infrared
divergent, they contain infrared singularities due to the presence of massless intermediate states.
Such nonanalytic terms cannot be expanded in a power series in $S$, $T$ and $U$ and need separate consideration,
as described in subsection~\ref{sec:infracut}.

%%%%%%%%%%%%%%%%%%%%%
\subsubsection{Ultraviolet divergences and counter-terms}
%%%%%%%%%%%%%%%%%%%%%%%%%
\label{sec:counter}

The nonrenormalizable divergences of eleven-dimensional supergravity
may be subtracted by the addition of local counterterms but this results in an increasing number of
apparently arbitrary coefficients as the number of loops is increased.  However, some of these
coefficients are fixed by imposing extra conditions implied by the correspondence of the compactified
theory with string theory.  Investigating the extent
to which coefficients can be determined in this manner is one of the
main motivations of this work\footnote{
We cannot make use of dimensional regularization since this cannot be
consistently applied to a nonrenormalizable theory.  In particular, standard
dimensional regularization would set to zero all the dimensional terms
associated with power divergences that we will need to keep.
 Nevertheless it is extremely useful
  for evaluating a subset of the nonanalytic terms, as we will see
in  appendix~\ref{sec:DimReg}}.

The  most  basic example  arises  at one  loop.   The  sum of  Feynman
diagrams  in  that case  has  the form  of  a  prefactor of  $\calR^4$
multiplying  a  $\varphi^3$  scalar  field  theory  box  diagram  with
eleven-dimensional loop  momentum $p$.  The presence  of eight factors
of momentum  in the  prefactor implies that  the loop amplitude  has a
$\Lambda^3$  ultraviolet divergence  in the  presence of  a  cutoff at
$|p|=\Lambda$.  Since the loop  momentum enters the integral over with
a  factor $e^{-p^2  L}$, where  $L$  is the  Schwinger parameter,  the
momentum cutoff  may implemented in  a gaussian manner  by introducing
the cutoff  $L\ge \Lambda^{-2}$, instead  of a step  function momentum
cutoff.  Upon compactification,  this short-distance divergence arises
entirely in the  sector with zero winding number.   This dependence on
$\Lambda$  can  be  subtracted  by introducing  a  local  counterterm,
$\delta_1 A = c_1\, \calR^4$, which replaces $\Lambda^3$ by a specific
finite  value.   The value  of  $c_{1}$  was  precisely determined  in
\cite{ggv:oneloop,gkv:twoloop} to be
 \be\label{e:c1}    c_{1}=
{2\pi^2\over3}- {4\pi\over3} \,\Lambda^3\, , \ee
which  followed  from  the  fact   that  the  one-loop  terms  in  the
four-graviton  scattering  amplitude in  the  type  IIA  and type  IIB
theories  are  equal   (and  equal  to  the  genus-one   term  in  the
perturbative expansion of $E_{3/2}$).

At two loops ($L=2$) there are new issues.  Firstly, there are new primitive divergences.  The
naive degree of divergence is $\Lambda^{20}$ but
there is now a prefactor of order $S^2\, \calR^4$ \cite{BernDunbar},
which has twelve powers of momentum.  This means that the naive
primitive divergence is reduced from $\Lambda^{20}$ to $\Lambda^8$.
But upon compactification on $\calT^n$ powers of $\Lambda$ may be traded in for inverse powers of the radii of
the compact dimensions.  In addition, there are one-loop sub-divergences behaving as $\Lambda^3$.  We would
like to impose a cutoff on the two-loop Schwinger parameters, $L_k$,  in a manner that is consistent with that
imposed at one loop.  In particular, the two-loop amplitude has massless intermediate two-particle thresholds
arising in the $S$-channel from intermediate states with
$m_I=0$ (or $n_I=0$) in (\ref{iplanar}) and (\ref{inonplanar}).
The discontinuity across this threshold is obtained by
setting $m_I=0$ and is proportional to the one-loop four-point amplitude
multiplied by the tree-level amplitude.  It
is evident from (\ref{iplanar}) and (\ref{inonplanar}) that the Schwinger parameter for the one-loop
sub-amplitude
is $L_1+L_2$.  Therefore, in order to reproduce the same $\Lambda$-dependence as in the one-loop calculation
we must introduce a cutoff $L_1+L_2\ge \Lambda^{-2}$.
By symmetry the individual Schwinger parameters satisfy $L_k\geq \Lambda^{-2}/2$,
which means, using (\ref{paramdefs}), that $V$ is integrated over the range
\be
0\le V\le \Vlambda\equiv {2\over\sqrt 3}\,\Lambda^2\,,
\label{taulimdef}
\ee
Only the fact that the upper cutoff is linear in $\Lambda^2$ will prove to be relevant in this paper, whereas
the precise coefficient of $2/\sqrt 3$  will be irrelevant. The cutoff on $L_1+L_2$ implies
\be
V\tau_2\le \Lambda^2\,,
\label{tautwo}
\ee
which imposes the same ultraviolet cutoff on $\tau_2$ as in \cite{gkv:twoloop},
so that $\tau = \tau_1 + i\tau_2$ is integrated over the cutoff fundamental domain,
\be
\calF_\Lambda: \qquad\{ 1\le |\tau|\,, \qquad \tau_2 \le \tau_2^\Lambda=\Lambda^2/V\,, \qquad -1/2 \le \tau_1 \le 1/2\}\,.
\label{cutoffdom}
\ee

Primitive divergences and sub-divergences
will be subtracted by diagrams containing counterterms, as illustrated in
figure~\ref{fig:TwoBoxDiagram}.
In order to subtract the $\Lambda^3$  sub-divergences we need to include
the one-loop `triangle' diagram  in which there are two supergravity vertices and one $\calR^4$ contact interaction,
which is the counterterm that  cancels
the $\Lambda^3$ one-loop $\calR^4$ divergence and replaces it with a specific finite constant (see
figure~\ref{fig:TwoBoxDiagram}(c)).
The contribution of the
diagram (denoted by  $\delta A$ in section~4.3 of reference~\cite{gkv:twoloop}) is given by
\begin{equation}\label{diag:counter}
A_{sugra\, \triangleright}(S,T,U)=i c_{1}
{\kappa_{11}^6\over (2\pi)^{22}}\, {\pi^3\over 2l_{11}^{12}}\, \left(S^2 \, I_{\triangleright}(S)+ T^2\,
I_{\triangleright}(T)+U^2\, I_{\triangleright}(U)\right)\, \calR^4\ ,
\end{equation}
where
\begin{equation}\label{e:triangle}
 I_{\triangleright}(S)=\pi^{11-n}\int_{\Lambda^{-2}}^\infty dL\, L^{n-7\over2}\int_0^1 du_2\int_0^{u_2}du_1
 \, e^{-L u_1 (1-u_2)\, S}\,\sum_{m\in\ZZ^n} e^{-\pi\, L\, G^{IJ} m_I n_J}
\end{equation}
In addition to the sub-divergences there
are local `primitive' divergences of the form $\Lambda^8 \,S^2\,\calR^4$ and
$\Lambda^6 \,S^3\, \calR^4$, as well
as `overlapping' divergences that lead to an extra $\Lambda^6 \,S^3\, \calR^4$
term.  All of these need to be subtracted by additional local counterterms.
\be\label{e:c2}
\delta_{2} A_{sugra}= - \, {\kappa_{11}^6\over (2\pi)^{22}\, l_{11}^{12}} \,  (a\,
\Lambda^8\, S^2\,\calR^4+ (b \Lambda^6 - c\, \Lambda^3 - d)\, S^3\,\calR^4+\cdots)
\ee
where $a,b,c$ and $d$ are constants.
Obviously, a local counterterm has to be independent of the moduli of the compact space.

\subsubsection{Examples of renormalized interactions}

One can anticipate the kinds of cutoff-dependent terms that arise upon compactification on $\calT^n$ by simple
dimensional considerations.  Such terms translate into particular perturbative terms in type IIA string theory
by use of the dictionary that translates between $M$-theory parameters and string theory parameters.
Such terms will be analyzed in detail later in this paper, but here we will sketch some features
that arise.

The following are examples of terms that will arise after compactification on a circle of radius
$R_{11}$.   In this case the translation of the supergravity results to IIA string theory makes
the identifications $S=R_{11}\, s$, $R_{11}^3 = g_A^2$, and the implicit momentum conservation
delta function, $\delta^{(10)}(\sum_r p_r^\mu)$, transforms in a manner that cancels the transformation of
$\calR^4$.
\begin{itemize}

\item{The lowest-order term with a one-loop sub-divergence has the form
$S^2\, \calR^4\, R_{11}^{-5}\, \Lambda^3$ and corresponds to tree-level IIA.   After renormalizing this by
adding the contribution of the triangle diagram containing
the one-loop counterterm  with coefficient $c_1$ this has a value equal to that of
the tree-level contribution contained in
the modular function $E_{5/2}$ that multiplies $S^2\, \calR^4$ in the IIB theory -- as it should
since the IIA and IIB theories have identical perturbative expansions up to
at least genus four \cite{Berkovits2006}.}
\item{
Expanding the one-loop sub-divergence to next order in $S,T,U$ results in a term of the form
$S^3\, \calR^4\,R_{11}^{-3} \Lambda^3$
that contributes to a genus-one term in IIA string theory and, after renormalization,
 equals the genus-one part of the
the modular function $\calE_{(0,1)}$ (or $\calE_{(3/2,3/2)}$
in the notation of \cite{gv:D6R4}).}
\item{ A term of the form
 $S^4\, \calR^4\,R_{11}^{-1} \Lambda^3$, is associated with  a genus-two IIA string
 contribution that contributes a term of the form $g_A^2\,s^4\, \log(-s)\,\calR^4$.
 [A similar
 finite term of the form $R_{11}^{-4}\, S^4\,\calR^4$ corresponds to a genus-one
 IIA string theory contribution.]}
\item{ A further term arising from one-loop sub-divergences is
 $S^5\, \calR^4\,R_{11}\Lambda^3$, which contributes to a genus-three  term of order
 $g_A^4\,S^5\,\calR^4$  in IIA string theory.}
 \item{
In addition to these contributions from sub-divergences, there is a primitive divergence of the form
 $S^2\, \calR^4\,\Lambda^8 $ which has to be canceled by a new local counterterm.  Since this translates
into a type IIA string-theory term proportional to $g_A^{8/3}$, which is not a consistent power,
 this must have a vanishing  renormalized value.}
\item{Similarly, a possible term $S^4\, \calR^4\, \Lambda^4$ does not have a
  sensible perturbative string theory interpretation and so we will
  set its renormalized value to zero.}
\item{  By contrast,  a term of the form
$S^3\,  \calR^4\, \Lambda^6$ is to be canceled by another new two-loop local counterterm, but the finite
renormalized value must take a specific value
(just as we saw in the cancellation of the one-loop $\calR^4 \, \Lambda^3$ term)
equal to the genus-two term in the IIB modular function, $\calE_{(0,1)}$.}
\item{A new phenomenon that arises at order $S^6\,  \calR^4$ is the
  occurrence of a primitive logarithmic divergence, $\log \Lambda$, which is
the divergence manifested as a pole in $\epsilon$  in dimensional
  regularization of the eleven-dimensional theory.  This should again be subtracted by
  a local counterterm.}
\end{itemize}
An important feature of the  two-loop divergences is that they describe local terms
that are independent of $R_{11}$ and can indeed be subtracted by the addition of new local counterterms.
Whether this continues to be the case at higher loops ($L>2$) is an interesting question that is not addressed here.

Corresponding  terms arise in the $\calT^2$ compactification to
nine dimensional string theory, in which there is also dependence on
the radii $r_B$ or $r_A$.

\subsubsection{Dealing with infrared threshold effects}
  \label{sec:infracut}

Although the four-graviton amplitude has no infrared divergences in nine or ten dimensions, there are subtleties in
extracting the nonanalytic threshold terms, which are infrared consequences of intermediate
massless multi-particle states.  These states are zero Kaluza--Klein modes.

At one loop ($L=1$) there is a complicated $S$, $T$ and $U$-channel
discontinuity  structure  (reviewed  in \cite{grv:oneloop})  with  the
property that the scales of various
logarithmic factors cancel out and the form of the
nonanalytic terms is invariant under rescaling the Mandelstam invariants.

The massless intermediate states originate in the two-loop case ($L=2$)
from zero Kaluza--Klein modes in the factor
$\Gamma_{(n,n)}$ in (\ref{latdef}).  The nonanalytic terms have
discontinuities that arise from the long-time propagation of these
states.  For example, the $S$-channel configuration shown in
figure~\ref{fig:TwoBoxDiagram} has two-particle thresholds associated
with $m_I=0$ or $n_I=0$, or both, arising from
the integration limits $L_1\to \infty$,  $L_3\to \infty$ (or both
simultaneously).  In the low energy expansion this generates
   terms of the form $A_k\, S^k \,\log (-S/C_k)$  (where $A_k$ and
   $C_k$ are constants) that are required by unitarity.  However,
if we were to simply expand the integrand
  in integer powers of $S$ the signature of such thresholds would be
  the occurrence of terms of order $S^k$ with  divergent coefficients,
  $\lim_{\mu \to 0}C_k\, S^k \log \mu$, where $\mu$ is an infrared cutoff.

In the following we will not be interested in the details of the nonanalytic thresholds\footnote{The detailed threshold structure is obtained much more simply
using dimensional regularization than with the cutoff
  procedure we are adopting. }, but will
simply concentrate on the dependence of the nonanalytic terms in the amplitude on the scale $\chi$ ($\chi>0$)
of the Mandelstam invariants, defined by
\be
S= \chi\,S_0\,, \qquad T= \chi\, T_0\,,\qquad U= -\chi (S_0+T_0)\,,
\label{rescale}
\ee
where $S_0$ and $T_0$ are arbitrary constants.
Since the limit $L_k \to \infty$ (for any $k$) translates into the
limit $V\to 0$, the infrared nonanalytic effects
arises when there are sufficient inverse powers of $V$.
The signature of these contributions  is the presence of
divergent  coefficients, $I_{(p,q)}$, in the power series expansion
(\ref{ipqdef}).

A simple model for the parts of our expressions that give rise to nonanalytic thresholds
 is given by considering the convergent integral,
$H = \int_0^{\Lambda^2} dV\,V^a e^{-\chi/V}$,
where $a>-1$.   On the one hand this can be expanded for $\chi/\Lambda^2<1$ as
\be\begin{split}
H&= {\Lambda^{2(a+1)}\over a+1} + {\Lambda^{2a}\, (-\chi) \over a}+ \dots+
{\Lambda^{2a-2r+2}\, (-\chi)^r \over (a-r+1) r!}  + \dots \cr
&  +
{(-\chi)^{a+1}\over (a+1)!}\,(\log(\Lambda^{2}/\chi)- \Gamma'(a+2)/\Gamma(a+2))
+ O(\chi^{a+2}/\Lambda^2)\,.
\label{exactexp}
\end{split}\ee
On the other hand, the analogue of the procedure adopted in this paper is to consider the formal series
obtained by expanding the integrand of $H$,
\be
H = \sum_{r=0}^\infty  \int_0^{\Lambda^2} {1\over r!} V^{a-r} \, (-\chi)^r  dV.
\label{formexp}
\ee
Clearly, terms with $r > a$ are divergent at the small $V$ endpoint,
despite the convergence of the original integral.  However, we are only interested in
the terms with non-negative powers of $\Lambda$ (i.e., terms of order $\chi^r$ with $r\le a+1$) in
the exact expansion given by (\ref{exactexp}).  So the correct result is obtained by simply ignoring all the
divergent terms in the formal expansion, with the
exception of the term with $r=a+1$, which has a logarithmic divergence.
This term is to be interpreted as $\log(\Lambda^2/\chi)$.  An efficient way of describing this is to replace
the formal expression (\ref{formexp}) by the regulated expression
\bea
H_{reg} &=&  \int_0^{\Lambda^2} \sum_{r=0}^{a+1} {1\over r!} V^{a-r} \, (-\chi)^r  \, e^{-\chi f/V}dV\nn\\
&\equiv& \int_{0_\chi}^{\Lambda^2}  \sum_{r=0}^\infty {1\over r!} V^{a-r} \, (-\chi)^r  dV
\label{reguexp}
\eea
with $f\ll 1$.  The second line defines the notation to be used in discussing the analogous integrals that we will meet
later in this paper.
The expression (\ref{reguexp})
 reproduces the exact expansion (\ref{exactexp}), apart from the scale inside the $\log$
factor, which is now proportional to $\chi^{a+1}\, \log(\Lambda^2/\chi f)$.
The above simple example illustrates how the terms in the expansion (\ref{ibdef}) of the integrand in (\ref{funcint})
give the correct series expansion
of the amplitude, including the parts with logarithmic nonanalytic behaviour.  However, the scale of the logarithm
is not determined.

As an example of such a nonanalytic term let us consider the explicit ten-dimensional string loop
calculation \cite{grv:oneloop} that determines a genus-one logarithmic
threshold at order $s^4$.  This can be written as $\calR^4$ multiplied by
\bea
S^4\log(-SR_{11}^2/\mu_2) +T^4\log(-TR_{11}^2/\mu_2)+ U^4\log(-UR_{11}^2/\mu_2) =\sigma_2^2 \, \log(\chi R_{11}^2/C_{(2,0)}),
\label{genonetwo}
\eea
where $\mu_2$ is a specific  constant and
 $C_{(2,0)}(S_0,T_0)$, has a specific dependence on $S_0$,
and $T_0$ (but not on $\chi$).  This scale encodes the precise details of the multiparticle
thresholds.  In this paper we will reproduce the expression on the second line of (\ref{genonetwo}) from
the $\calS^1$ compactification of
$L=2$ eleven-dimensional supergravity, but the scale $C_{(2,0)}$ multiplying $\chi$ inside the $\log$ factor
will not be determined.
More generally, in the compactification to ten dimensions  we will obtain nonanalytic
contributions of the form
$K_{(p,q)}\, \sigma_2^p\, \sigma_3^q\, \log(\chi R_{11}^2 /C_{(p,q)})\,,$
where the constant coefficients, $K_{(p,q)}$, will be evaluated, but
not  $C_{(p,q)}$.  In principle, $C_{(p,q)}$ can be reconstructed, apart from a multiplicative constant, from
two-particle and three-particle unitarity.

In the compactification on $\calT^2$ to nine dimensions most of the nonanalytic
contributions
are characterized by half-integral powers of $S$, $T$ and $U$.  These are
easily separated from the analytic parts and we will not consider them
here.  However, a logarithmic term will also arise at order $l_{11}^8\,\calD^8 \,
\calR^4\sim \sigma_2^2\calR^4$, which is known in detail using dimensional regularization
\cite{BernDunbar} (and is
 reviewed in appendix~\ref{sec:DimReg}) and will also be considered
in the $\Lambda$ cutoff procedure.

\subsection{The general form of the expansion of two loops on $\calT^n$}

After taking care of various normalization constants,
the low-energy expansion of the two-loop supergravity amplitude (\ref{eberndun}) compactified on a $n$-torus
can be written as
\bea
A_{sugra} &=&  \! i{\kappa_{11}^6\over (2\pi)^{22}}\; \calR^4\
 \pi^6
\bigg[\s_2
I_{(1,0)}+{ \sigma_3\over 12} \ I_{(0,1)}+{\s_2^2 \over 2!\cdot 144 }\,
I_{(2,0)}
\non\\
&+& {\sigma_2\sigma_3\over 3!\cdot  15120 }I_{(1,1)}
+{1\over 4! \, 302400 }\big(  \sigma_2^3I_{(3,0)} + {4\over 3}\sigma_3^2
I_{(0,2)} \big)+ \dots \bigg]\ .
\label{eberz}
\eea
where the coefficients are functions of the moduli that are given by the integrals
\be\label{e:DefIn}
I_{(p,q)} = \pi^{N+1}\,\int_{\0chi}^{\Vlambda} dV V^{3-N} \int_{\calF_\Lambda}
     {d^2\tau\over \tau_2^2}  B_{(p,q)}(\tau)\,    \hat \Gamma_{(n,n)}(G_{IJ}V,\tau)\,,
\ee
where $N=2p+3q-2$ and the functions $B_{(p,q)}(\tau)$ are defined via (\ref{bndefs}) and (\ref{ddef}).
These can be evaluated by symbolic computer methods and the resultant expressions are
given up to order $2p+3q=6$
in appendix~\ref{sec:Modular}, together with their decomposition into functions satisfying
Poisson eigenvalue equations.
Recall that the notation for the integration limits in (\ref{e:DefIn}) builds in the $S$, $T$, $U$-dependence
of the nonanalytic thresholds, so that factors of $\log \chi$ are present in certain terms in (\ref{eberz}).
The lattice  factor $\hat \Gamma_{(n,n)}$ defined in (\ref{latticefact}) contains the  information about
the spatial torus.

We will now turn to the evaluation of the
$I_{(p,q)}$'s explicitly for the cases $n=1$ (compactification on
$\calS^1$), and $n=2$ (compactification on $\calT^2$).

%%%%%%%%%%%%%%%%%%%%%
\section{Circle compactification to ten dimensions}
\label{sec:sone}
%%%%%%%%%%%%%%%%%%%%%%%%%

We will now consider the value of the two-loop amplitude after compactification on a circle
of radius $l_{11}R_{11}$ in the special case in which the external momenta are zero in the compact
direction.   The metric of the eleven-dimensional theory is related to the string-frame
ten-dimensional type IIA metric in the usual manner by
\be
ds^2=G^{(11)}_{MN}dx^Mdx^N=
 {l_{11}^2\over l_s^2R_{11}}\; g_{\mu \nu} dx^\mu dx^\nu +
R_{11}^2 l_{11}^2 (dx^{11} - C_\mu dx^\mu )^2,
\label{metdef}
\ee
where $g_{\mu \nu}$ is the string frame metric and $R_{11}\, l_{11}$ is the radius of the eleventh dimension (and we
will not be interested in the one-form $C_\mu$ here).
The dictionary for translating between M-theory and type IIA string theory
relates the string coupling and string-frame Mandelstam invariants to $R_{11}$
and the eleven-dimensional invariants by
\be
l_{11} = R_{11}^{\half}\, l_s\,\qquad g_A^2 = R_{11}^3\,, \qquad S = R_{11} \, s\,.
\label{twoadict}
\ee
In order to evaluate the ten-dimensional amplitude, $A^{(d=10)}$, we will need to evaluate the
integral $I_{(p,q)}$ (\ref{e:DefIn}) in the case $n=1$, which we will call $I_{(p,q)}^{(d=10)}$.

\subsection{Evaluation of $I_{(p,q)}^{(d=10)}$}

In this case the metric of the compact dimension is simply $G_{IJ} =R_{11}^2$ so
from (\ref{edhatdef}) we have
\be
\hat E=V\,v\,{|\hat m+\tau\hat n|^2\over \tau_2}\,,
\label{soneedef}
\ee
which is to be used in (\ref{e:DefIn}), and where we have set
\be
v=R_{11}^2\,.
\label{vdef}
\ee

The integral $I_{(p,q)}^{(d=10)}$ could be evaluated directly by use of the `unfolding trick', but
it is more straightforward to use the method devised in \cite{gv:D6R4} for studying $I_{(0,1)}^{(d=10)}$.
This begins by noting that
\be
\left(    v^2     {\partial^2\over    \partial    v^2}     +    2    v
  {\partial\over \partial v}\right)\, e^{-\pi \hat E}
= \Delta_{\tau}\, e^{-\pi \hat E} \equiv \tau_2^2
\left({\partial^2\over \partial \tau_1^2} +
 {\partial^2\over \partial \tau_2^2}\right)\, e^{-\pi \hat E}\, .
\label{soneintmeth}
\ee
This means that $I_{(p,q)}^{(d=10)}$ satisfies
\bea
\left(  v^2 {\partial^2\over \partial v^2} + 2 v {\partial\over \partial v}\right)&&
I_{(p,q)}^{(d=10)}= \pi^{N+1} \,  \int_\0chi^{\Vlambda} dV V^{3-N} \int_{\calF_\Lambda}
     {d^2\tau\over    \tau_2^2}   B_{(p,q)}(\tau)\,\sum_{(\hat   m,\hat
       n)\in\ZZ^2}\,\Delta_{\tau}\, e^{-\pi \hat E}
\nn\\
&=&\pi^{N+1} \sum_{(\hat m,\hat n)\neq(0,0)}  \int_\0chi^{\Vlambda} dV V^{3-N} \int_{\calF_\Lambda}
     {d^2\tau\over     \tau_2^2}\,\Delta_{\tau}\,    B_{(p,q)}(\tau)\,
     e^{-\pi \hat E}
      -\partial I_{(p,q)}^{(d=10)}\,,\nn\\
\label{ineval}
\eea
where we have integrated by parts and the boundary term is given by
\be
\partial I_{(p,q)}^{(d=10)} = \left. \pi^{N+1} \sum_{\hat m,\hat n}  \int_\0chi^{\Vlambda} dV V^{3-N} \int_{-\half}^{\half}
     d\tau_1\,
     \left(B_{(p,q)}(\tau)\,\partial_{\tau_2}\, e^{-\pi \hat E} - \partial_{\tau_2}\,
     B_{(p,q)}(\tau)\, e^{-\pi \hat E}
     \right)
     \right|_{\tau_2 = \tau_2^\Lambda}\,
\label{ibound}
\ee
(where we recall that $V\,\tau_2^\Lambda = \Lambda^2$).
After substituting $B_{(p,q)} = \sum_{i=1}^{\lceil 3N/2\rceil } b_{(p,q)}^i$ (as in (\ref{modsum})) and writing
\be
I_{(p,q)}^{(d=10)} = \sum_{i=0}^{\lceil 3N/2\rceil} h_{(p,q)}^i(v)\,,
\label{hsum}
\ee
equation~(\ref{ineval}) is replaced by a set of component equations
of the form
\bea
\left( v^2 {\partial^2\over \partial v^2} + 2 v {\partial\over \partial v}\right)\,
h_{(p,q)}^i = j_{(p,q)}^i -\partial b_{(p,q)}^i\equiv J_{(p,q)}^i\,,
\label{inevan}
\eea
where the bulk term is given by
\be
j_{(p,q)}^i=  \pi^{N+1}\int_\0chi^{\Vlambda} dV V^{3-N} \int_{\calF_\Lambda}
     {d^2\tau\over    \tau_2^2}\,\Delta_{\tau}\,   b_{(p,q)}^i(\tau)\,
     \sum_{(\hat m,\hat n)\in\ZZ^2} e^{-\pi \hat E}\,,
 \label{bulkgen}
 \ee
 and $\partial b_{(p,q)}^i$ is the component form of the last term in (\ref{ibound}).
After using the Poisson equation (\ref{poissoeq}) satisfied by $b_{(p,q)}^i$, equation (\ref{inevan}) reduces to a set of
simple second order differential equations.

As we will see later, for terms that are analytic in the Mandelstam invariants
the right-hand side of (\ref{inevan}), $J_{(p,q)}^i$, has a dependence
on $v$ of  the form
\be
J_{(p,q)}^i = v^{N-4} \, J_{(p,q)}^{(4-N)\, i} + \Lambda^3\, v^{N-\fiveh}\,  J_{(p,q)}^{(\fiveh-N)\, i} +
\Lambda^{8-2N} \,  J_{(p,q)}^{(0)\, i}\,,
\label{jsplitv}
\ee
where $J_{(p,q)}^{(\alpha)\, i}$ are constants, the superscript $(\alpha)$
indicates the power of $v$,
and $N=2p+3q-2$.
The first term in (\ref{jsplitv})
is the finite contribution from non-zero winding numbers $\hat m\ne 0$, $\hat n \ne 0$,
and the power $v^{N-4}$ follows by a simple dimensional argument.  The second term, proportional to $\Lambda^3$,
comes from the one-loop sub-divergent contributions in the sectors with $\hat m=0$, $\hat n\ne 0$ and
$\hat m\ne 0$, $\hat n=0$ and the power $v^{N-5/2}$ is again determined by a simple
dimensional argument.  The third term with the power
$\Lambda^{8-2N}$ comes from the sector with zero winding number, $\hat m=0, \hat n=0$.
In the $N=4$ cases,
$(p,q)=(3,0)$ and $(p,q)=(0,2)$, the powers of $v^{4-N}$ and $\Lambda^{8-2N}$ include pieces that should be interpreted
as $\log v$, $\log \Lambda$, as we
will see in the explicit evaluation later in this section.
After substituting the structure (\ref{jsplitv}) into (\ref{bulkgen}) and (\ref{inevan})
each term in (\ref{hsum}) is seen to decompose in the same manner into $v^{N-4} \,h_{(p,q)}^{(4-N)\, i}+
\Lambda^3\, v^{N-\fiveh}\, h_{(p,q)}^{(\fiveh-N)\, i} +
\Lambda^{8-2N} \,  h_{(p,q)}^{(0)\, i}$  and (\ref{ineval}) is solved by substituting
\be
\left( v^2 {\partial^2\over \partial v^2} + 2 v {\partial\over \partial v}\right)\, v^{-\alpha}\,
h_{(p,q)}^{  (\alpha)\, i}
= \alpha(\alpha-1)\,v^{-\alpha}\,h_{(p,q)}^{ (\alpha)\, i} \, ,
\label{alpheq}
\ee
so that
\be
h_{(p,q)}^{ (\alpha)\, i}=  {1\over \alpha(\alpha-1)} J_{(p,q)}^{(\alpha)\, i} \, .
\label{hsols}
\ee
Therefore,  $I_{(p,q)}$ (\ref{e:DefIn}) decomposes into the
sum of three terms with distinct powers of $v$,
\be
I_{(p,q)}^{(d=10)} = v^{N-4} \, I_{(p,q)}^{(4-N)} +
\Lambda^3\, v^{N-\fiveh}\, I_{(p,q)}^{(\fiveh-N)} +
\Lambda^{8-2N} \, I_{(p,q)}^{(0)}\,.
\label{splitv}
\ee
The one-loop sub-divergences proportional to $\Lambda^3$ are canceled by adding the triangle diagram,
$I_{\triangleright\, (p,q)}^{(d=10)}$, with the one-loop
$\calR^4$ counterterm at one vertex (figure~\ref{fig:TwoBoxDiagram} (c)).
The term proportional to $\Lambda^{8-2N}$ is a primitive divergence that has to be subtracted by a
new two-loop local counterterm,
whose contribution to $I_{(p,q)}^{(d=10)}$ will be denoted $\delta_2 I_{(p,q)}^{(d=10)}$.

We will also be concerned, when $N\ge 2$,
 with the situation  in which there are logarithmic terms on the right-hand side of
(\ref{jsplitv}) proportional to $v^\alpha \,\log v$ and $v^\alpha\, (\log v)^2$, with various values of $\alpha$.
In the presence of such source  terms the solutions of (\ref{ineval}) also have logarithms.
To be specific, the general form of the equations in the
examples that follow is
\begin{equation}
(v^2\partial_{v}^2+2v\partial_{v}-\lambda)\, f(v)= a\, \log v + b\, {\log v\over v^\half} + c\, {\log v\over v^2}
+ d\, {(\log v)^2\over v}
\,.
\label{logeq}
\end{equation}
It is easy to verify that the solution of this equation is
\bea
f(v)&=&-a\, {1\over \lambda^2} (1 + \lambda\log v)  - b\, {4\over 1+4\lambda}\,  {\log v\over v^\half}
+ c\, {1\over v^2(2-\lambda)}\, \left(\log v + {3\over 2-\lambda}\right)
\nn\\  &&
- d\,
 {1\over v\lambda^3}(2+ 2\lambda -2\lambda \log v
+ \lambda^2 (\log v)^2) + f_0(v) \,,
\label{gensol}
\eea
where $f_0$ is the solution of the homogeneous equation that will be irrelevant to us.
We will see that these non-analytic terms are logarithmic thresholds expected from unitarity in ten dimensions.

This procedure is implemented in detail in appendix~\ref{sec:s1compact} in the order to evaluate $I_{(p,q)}$
with $N=1,2,3,4$, leading to  terms in the four-graviton amplitude that we will now review.

\subsection{The ten-dimensional type IIA low energy string scattering amplitude}

We will now summarize the expressions deduced in detail in appendix~\ref{sec:s1compact} from the analysis of two-loop
eleven-dimensional supergravity on $\calS^1$.    The complete expressions will also include the contribution
$I_{\triangleright\, (p,q)}^{(d=10)}$  that comes from the triangle diagram where one vertex is the $\calR^4$
one-loop counterterm and the primitive divergence, $\delta_2 I^{(d=10)}_{(p,q)}$.
We will begin by reviewing the $(1,0)$ and $(0,1)$ cases before considering the higher-order terms.

\subsubsection{$(p,q)=(1,0)$}
The contribution to the coefficient of the $\sigma_2\, \calR^4\sim \calD^4 \calR^4$ term
was shown in \cite{gkv:twoloop} to have the form
\bea
I^{(d=10)}_{(1,0)} + I_{\triangleright\, {(1,0)}}^{(d=10)} +
\delta_2 I^{(d=10)}_{(1,0)}&=& {\zeta(5)\over 4\,R_{11}^5}
\,,
\label{i0def}
\eea
which corresponds to the tree-level type~IIA contribution
$\zeta(5)\,\hat\sigma_2\calR^4/g_A^2$.
In this case the finite piece from the $\hat m\ne0$, $\hat n\ne 0$ sector vanishes and
the right-hand side arises entirely from the $\Lambda^3$ sub-divergence of $I^{(d=10)}_{(1,0)}$, together with
$I_{\triangleright\, (1,0)}^{(d=10)}$, which cancels that divergence and replaces it with
a specific finite expression.
The power of $R_{11}$  corresponds to a tree-level
contribution in type IIA string theory, using the identifications in (\ref{twoadict}) (recalling that
$S = R_{11} \,s$).
The $\Lambda^8$ contribution to $I_{(1,0)}^{(d=10)}$ coming  from the $\hat m=0, \hat n= 0$ sector
has been set to zero by subtracting it with a two-loop counterterm, $\delta_2 I_{(1,0)}^{(d=10)}$.
There can be no finite remainder since $S^2\, \calR^4$ has no dependence on $R_{11}$ and
translates into a IIA string contribution
$g_A^{2/3}\,S^2\, \calR^4$, which would not make sense in string perturbation theory.
For completeness, recall that there is also a genus-two contribution
to $R_{11}\, \sigma_2 \calR^4 \sim g_A^2\, \hat\sigma_2\, \calR^4$
that is obtained from the $\calS^1$ compactification of {\it one-loop} ($L=1$)
eleven-dimensional supergravity.  These tree-level and two-loop type IIA string theory contributions are precisely the
same as those contained in the modular function $E_{5/2}$ that arises in the type IIB theory reviewed in the
introduction.  It is also notable that the two-loop supergravity calculation does not generate
a genus-one contribution to  $S^2\, \calR^4$.  Such a term is known to be absent in string perturbation
theory \cite{ggv:oneloop}.

The perfect agreement of the predictions from two-loop eleven-dimensional supergravity with string
theory found in \cite{gkv:twoloop} strongly indicated that higher-loop supergravity does not contribute further
terms at order $\calD^4\calR^4$.
This suggested \cite{gkv:twoloop} that the three-loop amplitude should be of order $\calD^6 \calR^4$
(or higher), as has recently been shown explicitly \cite{Bern:2007hh}.

\subsubsection{$(p,q)=(0,1)$}

The expression $I^{(d=10)}_{(0,1)}$ is the coefficient of the $\sigma_3 \, \calR^4$ (or $\calD^6 \calR^4$) term.
The finite part of $I^{(d=10)}_{(0,1)}$ (the non-zero winding number sector) was considered in \cite{gv:D6R4}.
In addition there is a $\Lambda^3$ sub-divergence (from the sector in which one winding number vanishes)
that needs to be subtracted by $I_{\triangleright\, (0,1)}^{(d=10)}$,
 as well as a $\Lambda^6\, \calD^6 \calR^4$  primitive
 divergence (from the sector in which both winding numbers vanish)
 that is subtracted by $\delta_2 I^{(d=10)}_{(0,1)}$.
 This divergent term translates into a possible IIA string term,
 proportional to $g_A^2\, \calD^6 \calR^4$ term.
 This is not only a possible contribution, but is known to be present
 since it is present in the type IIB theory with a coefficient that is determined by $\calE_{(0,1)}$
 and was reviewed
 in the introduction.   The net result is
\be
I^{(d=10)}_{(0,1)} + I_{\triangleright\, (0,1)}^{(d=10)} + \delta_2 I^{(d=10)}_{(0,1)}=
 {\zeta(3)^2\over 2\,R_{11}^6} + {\zeta(3)\zeta(2)\over R_{11}^3}
+ {6\zeta(2)^2\over 5}\,,
\label{i1def}
\ee
In terms of type IIA string theory, these terms correspond to tree-level, genus-one and genus-two contributions
with coefficients $g_A^{-2}$, $g_A^0$ and $g_A^2$, respectively
Whereas, the coefficients of the first two terms are derived from the $\calS^1$ compactification,
the coefficient of the last term
has been fixed by choosing $\delta_2 I_{(0,1)}^{(d=10)}$ so that
it coincides with the coefficient of the two-loop term in the type IIB theory that came
from the $\calT^2$ compactification.
For completeness, recall that there is a genus-three
contribution to $\calD^6 \calR^4$ that is again obtained as a finite
contribution from the
$\calS^1$ compactification of one-loop ($L=1$)
eleven-dimensional supergravity and which is also contained in the IIB expression
$\calE_{(0,1)}$, with precisely the same coefficient.

\subsubsection{$(p,q)=(2,0)$}

When   $N\geq  4$   the  $L=2$  amplitude   develops  logarithmic
singularities    corresponding     to    string    theory    threshold
contributions.  These contributions  require careful  treatments which
is detailed in the appendices~\ref{sec:s1compact} and in~\ref{sec:Cste}.

The expression $I^{(d=10)}_{(2,0)}$ is the coefficient of
 $\sigma_{2}^2 \, \calR^4\sim \calD^8\, \calR^4$. At this order
 there is a second logarithmic singularity (after the one-loop
 ($L=1$) supergravity threshold of ten-dimensional supergravity, which
 is of order $S\log(-S)$) corresponding to a threshold of
 string perturbation theory.
 The contribution at order $\sigma_{2}^2\, \calR^4$, derived in the appendix~\ref{sec:D8R4circle},
 is
\be\begin{split}
I_{(2,0)}^{(d=10)}+
I_{\triangleright\, (2,0)}^{(d=10)}+\delta_{2}I_{(2,0)}^{(d=10)} =& -{12\over5}\,\zeta(2)\,
\left[ {\zeta(3)\over R_{11}^4}+{2\zeta(2)\over R_{11}} \right] \,
\log(\chi R^2_{11}/C_{(2,0)})\, .
\end{split}\ee
When converting these expressions  to the string frame
$\log(\chi  R^2_{11})$ becomes  $\log(\chi\, g_{A}^2)$, where we have used the relation $S=\chi S_0 = s R_{11}
=\chi s_0 R_{11}$ (in other words, we have rescaled the Mandelstam invariants in the string frame by the same
factor $\chi$ as in the eleven-dimensional frame).
 This non-analytic contribution is seen to correspond to the genus-one and genus-two
normal massless thresholds of the string amplitude with coefficients
$g_{A}^0\, \log(\chi)$ and $g_{A}^2\, \log(\chi)$, respectively
The coefficient of the genus-two threshold contribution arises from a $\Lambda^3$ sub-divergence,
which is regulated by the counter-term $I_{\triangleright\, (2,0)}^{(d=10)}$.
The scale of the logarithms indicated by $C_{(2,0)}$ has not been determined
by this computation and hides the details of the $T$-channel and $U$-channel thresholds, although in this
case the complete calculation is straightforward and leads to (\ref{genonetwo}).

\subsubsection{$(p,q)=(1,1)$}

The coefficient of the  $\sigma_{2}\, \sigma_3 \, \calR^4\sim D^{10}\,  \calR^4$ term is determined by the
integral $I^{(d=10)}_{(1,1)}$, together with the triangle diagram containing the one-loop sub-divergence and the
two-loop counterterm.  These give

\bea
\nonumber I_{(1,1)}^{(d=10)}+
\hat         I_{\triangleright\,        (1,1)}^{(d=10)}+\delta_{2}\hat
I_{(1,1)}^{(d=10)} &=& {448\over  R_{11}^2}\zeta(4)\zeta(3)+
{675\over 2} \zeta(2)\zeta(4)R_{11}+ 182\, \zeta(2)^2\, R_{11}^2\,  \log({\chi R^2_{11}\over C_{(1,1)}})\\
\eea
The first two terms will translate into genus-two and  genus-three
contributions to $S^5\, \calR^4$ in string
theory, while the last term is a genus-two threshold contribution
 In fact, the coefficient of $\log(\chi)$ is precisely the same as in the
ten-dimensional
supergravity calculation in appendix~\ref{sec:tenreg}, which contains the
detailed threshold dependence.

\subsubsection{ $(p,q)=(3,0)$ and $(p,q)=(0,2)$}

In the case of $\sigma_{2}^3\, \calR^4$ and $\sigma_{3}^2\, \calR^4$
the coefficients are determined by the integrals
 $I_{(3,0)}^{(d=10)}$ and  $I_{(0,2)}^{(d=10)}$, together with the contributions from the one-loop and two-loop
 counterterms, that are evaluated in the appendix~\ref{sec:D12R4circle},
\be\begin{split}
I_{(3,0)}^{(d=10)}+
I_{\triangleright\, (3,0)}^{(d=10)}+\delta_{2}I_{(3,0)}^{(d=10)} = - 3465\,\zeta(6)\, & \log(\chi R_{11}^2/C_{(3,0)})
+{100647\over715}\,
 \zeta(3)\,\zeta(6)
 + 210\, \zeta(8) \,
 R_{11}^{3}
\, ,
\end{split}\ee
and
\be\begin{split}
I_{(0,2)}^{(d=10)}+
I_{\triangleright\, (0,2)}^{(d=10)}+\delta_{2}I_{(0,2)}^{(d=10)} = - {6615\over 2}\,\zeta(6)\,
& \log(\chi R_{11}^2/C_{(3,0)})
+{15827\over110}\,
 \zeta(3)\,\zeta(6)
  + 210\, \zeta(8) \,
R_{11}^3
\, ,
\end{split}\ee
corresponding to type-IIA genus-three, genus-four and genus-six contributions, respectively.

\subsection{Connections with string perturbation theory in ten dimensions.}
\label{connectten}

We will now summarize these results and translate them into perturbative terms in IIA string theory.
We will be interested in comparing these terms with direct calculations in string perturbation theory at
genus-one and genus-two.  The fact that the perturbative terms in the type IIA and type IIB theories
are equal  up to at least genus-four will provide additional data for determining the
$SL(2,\ZZ)$-invariant coefficients of the ten-dimensional IIB theory.

\subsubsection{Analytic terms}

First recall the analytic terms in the derivative expansion
of one-loop supergravity ($L=1$) compactified on $\calS^1$
up to the order of interest in this paper are
\cite{ggv:oneloop,Russo:1997mk}
\be\begin{split}
A_{L=1}^{an} =&   i{\kappa_{11}^4\over (2\pi)^{11}\, l_{11}^3}\,
4\pi^4 \; \calR^4
\ \bigg[  {2\zeta(3)\over R_{11}^3}+4\zeta(2)
 +{4\zeta(4)\over 3} R_{11}{\sigma_{2}\over 4^2}
+ {4\zeta(6)\over27}\, R_{11}^3\, {\sigma_{3}\over 4^3}\\
&+{64\zeta(8)\over 2835}\, R_{11}^{5}\, {\sigma_{2}^2\over 4^4}
+{16\zeta(10)\over1125}\,R_{11}^7\,{\sigma_{2}\sigma_{3}\over 4^5}
+{64\zeta(12)\over 31185\cdot 691}\, R_{11}^9\,
{  (675{\sigma_{2}^3}+872{\sigma_{3}^2})\over 4^6}
\cdots\,
\bigg]\, .
\label{aawalone}
\end{split}\ee

\begin{figure}[ht]
\centering
\includegraphics[width=5cm]{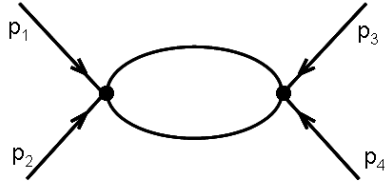}
\caption{The double subdivergence of the three-loop diagrams that contributes at
order $E_{7/2}\, S^4\,\calR^4$ in ten-dimensional type IIB.}
\label{fig:threesub}\end{figure}

The analytic part obtained from two-loop ($L=2$) supergravity on
$\calS^1$, obtained earlier
in this section, together with the known two-loop results of
\cite{gkv:twoloop} and \cite{gv:D6R4}, are given by
\be\begin{split}
A_{L=2}^{an} =&   i{\kappa_{11}^6\over (2\pi)^{22}\,l_{11}^{12}}\; \ 4 \pi^6 \calR^4
\ \bigg[ {\zeta(5) \over R_{11}^5}\   {\s_2\over 4^2}
+ {4\over3}\left({\zeta(3)^2\over 2R_{11}^6}+{\zeta(2)\zeta(3)\over R_{11}^3}
+{6\zeta(2)^2\over 5}\right){\s_3\over 4^3}
\\
&
+{8\over 2835}\,\left({675\over2}\zeta(2)\zeta(4)\,R_{11}
+448\,{\zeta(4)\zeta(3)\over R_{11}^2} \right)
{\s_3\s_2\over 4^5}\\
& + {2\over 14175}\left(70 \, \zeta(8)\, R_{11}^3\,{3\sigma_{2}^3+4\sigma_{3}^2\over 4^6}+
\zeta(6)\zeta(3)\,
\left({100647\over715}  {\sigma_2^3\over 4^6} +{4\over 3}\, {15827\over110}{\sigma_3^2\over 4^6} \right)\right)
+\cdots\bigg]\, .
\label{aawa}
\end{split}
\ee
After conversion to the string frame the $L=1$ and $L=2$ analytic  contributions combine to give
the following terms in type IIA string theory coordinates,
\be\begin{split}
A^{an}_{L=1}+A^{an}_{L=2}& = i\kappa_{10}^2\,\calR^4\,\Big[ {2\zeta(3)\over g_{A}^2 }+4\zeta(2)+
\left(  {\zeta(5)\over g_{A}^2}+{4\zeta(4)\over 3}\,g_{A}^2
 \right)\, \hat\sigma_{2}\\
&+
{2\over3}\left({\zeta(3)^2\over g_{A}^2}
+2\zeta(2)\zeta(3)+{12\zeta(2)^2\over 5}\, g_{A}^2
 + {2\zeta(6)\over 9}\, g_{A}^4  \right)\, \hat\sigma_{3}\\
&+ {64\zeta(8)\over2835}\,g_{A}^6\sigma_{2}^2+
\left({512\over 405} \,\zeta(4)\zeta(3)\, g_{A}^2
+  {20\over 21}\zeta(2)\zeta(4)\,g_{A}^4+{16\zeta(10)\over 1125}\, g_{A}^8\right)
\, \hat\sigma_{2}\hat\sigma_3\\
&+ \left(
{\zeta(3)\zeta(6)}\, \frac{22366}{1126125}\,g_{A}^4+{4\over 135} \zeta(8)g_A^6+
 {320\zeta(12)\over 231\cdot 691}\,g_{A}^{10}\right) \, \hat\sigma_2^3\\
&+
\left(
\zeta(3)\zeta(6)\, \frac{9044}{334125} \,g_{A}^4+ {16\over 405}\zeta(8)g_A^6+
  {64\cdot 872\zeta(12)\over 31185\cdot 691}\, g_{A}^{10}\right)\, \hat\sigma_3^2+\cdots\Big]\, .
\end{split}
\label{lanal}
\ee
These are some of the terms that could, in principle, be obtained from string
perturbation theory. Other perturbative string terms should
emerge from  from higher-loop ($L>2$) supergravity.
For example, tree-level terms beyond order $\calD^6 \calR^4$ are not
obtained from supergravity  Feynman diagrams
of loop number $L\le 2$, but are obtained from higher-loop ($L>2$) contributions.  Thus,
the tree-level $\calD^4 \calR^4$ term can be deduced from a two-loop subdivergent
contribution to the three-loop ($L=3$) Feynman diagrams of eleven-dimensional supergravity
compactified on a circle as shown in figure~\ref{fig:threesub}.

Although the momentum expansion of  string theory at genus greater than one
has not been explicitly considered, several of the low-lying terms in (\ref{lanal}) are known to be in
precise agreement with the expansion of tree-level and genus-one
string theory amplitudes.  So, for completeness,
we now list the analytic terms that have been
extracted from string theory at tree-level ($h=0$) and genus-one
($h=1$)
 \cite{gv:stringloop,grv:oneloop},
\be\begin{split}
A^{an}_{h=0}&+A^{an}_{h=1} = i \kappa_{10}^2\, \calR^4\,\Big[
\left({2\zeta(3)\over g_{A}^2}+4\zeta(2)\right)+
{\zeta(5)\over g_{A}^2}\, \hat\sigma_2
+{2\over 3}\left({\zeta(3)^2\over g_{A}^2}+2\zeta(2)\zeta(3)\right)\, \hat\sigma_3\\
&+{\zeta(7)\over 2g_{A}^2}\, \hat\sigma_2^2
+\left({2\over 3g_{A}^2}\zeta(3)\zeta(5)+{97\over 270}\,
 \zeta(2)\zeta(5)\right)\, \hat\sigma_2\hat\sigma_3\\
&+ \left({\zeta(9)\over 4 g_{A}^2}+{2\zeta(2)\zeta(3)^3\over15} \right)\, \hat\sigma_2^3
 +\left({2\over27g_{A}^2}(2\zeta(3)^2+\zeta(9))
 +{61\over 270}\zeta(2)\zeta(3)^2\right) \,
 \hat\sigma_3^2+\cdots\Big]\, .
\label{stringann}
\end{split}\ee
We see that the terms that overlap with those of
(\ref{lanal}) have precisely the same coefficients.  However, there
are terms that occur in either (\ref{lanal}) or (\ref{stringann})
that do not occur in the other.  For example, the genus-zero term of
order $S^5\, \calR^4$ is not obtained from the
$L=2$ supergravity diagrams described in this paper.  However,  it is
expected to arise from a one-loop sub-divergence (proportional to
$\Lambda^3$) of the three-loop diagrams ($L=3$).  Similarly, its
genus-one partner should arise from a double sub-divergence of the
$L=3$ diagrams.

The above expressions have been obtained in the limit
appropriate for comparison with perturbative ten-dimensional type IIA
string theory.  However, the IIA and IIB theories are known to have
identical four-graviton amplitudes up to at least genus-four
\cite{Berkovits2006} so that,
to the extent that the results match the string theory results, they should also
apply to the ten-dimensional type IIB theory up to genus-four, at least.

\subsubsection{Nonanalytic terms}

The nonanalytic part of the one-loop ($L=1$) supergravity amplitude in
ten dimensions is just the ten-dimensional maximal
supergravity loop amplitude, which is well known
\cite{Green:1982sw} (reviewed in \cite{grv:oneloop}).   In this case
the thresholds obtained by dimensional regularization give rise to
terms of order $S\log(-S)$.  The detailed structure of these terms is
not relevant here, but it is notable that the scales of the
Mandelstam invariants inside the logarithms cancel (the result is proportional to $(S+T+U)\log \chi$).
This means
that the result does not depend on details of the regularization scheme.

The nonanalytic terms obtained from two-loop ($L=2$) supergravity compactified on $\calS^1$
in appendix~\ref{sec:D8R4circle}  are
\be
\begin{split}
A_{L=2}^{nonan} =&    i{\kappa_{11}^6\over (2\pi)^{22}\,l_{11}^{12}}\; \calR^4
\ 4 \pi^6 \bigg[  -{8\over 15\, R_{11}^4}\, \zeta(2)\left({\zeta(3)\over R_{11}^4}+
{2\zeta(2)\over R_{11}}\right)\log(\chi R_{11}^2/C_{(2,0)}){\s_2^2\over 4^4}\\
&- {832\over 1296\, R_{11}^2}\, \zeta(4) \,\log(\chi R_{11}^2/C_{(1,1)}){ \sigma_2\sigma_3\over 4^5}\\
&
-{1\over 45}\zeta(6)\log(\chi  R_{11}^2 /C_{(3,0)})\,
\left(11  {\sigma_2^3\over 4^6}
 + 14\,{\sigma_3^2\over 4^6} \right)+\cdots
\bigg]\,.
\label{astregn}
\end{split}
\ee
As before, the scales of the logarithms, $C_{(p,q)}$, are
complicated functions of   $S_0$ and $T_0$, and contain the information
about the nonanalytic multiple logarithm terms (for example, the term
of order $S^2\, \log(\chi)\, \calR^4$ is given in detail in (\ref{genonetwo})).
This expression involves  factors with logarithms of the Mandelstam invariants, of the form
$S^k\, \log(-S)\, \calR^4$, which are  correlated with discontinuities
of the amplitude that are
determined by unitarity.
 For example, at order $\sigma_2^2\,
\log(-S)\, \calR^4$ there are two terms with coefficients that differ
by a power of $R_{11}^3$.  The first of these arose from a finite
contribution to the two-loop amplitude, while the second arose from
the triangle diagram containing the one-loop counterterm that cancels a
$\Lambda^3$ sub-divergence.

Transforming (\ref{astregn}) to IIA string coordinates leads to
\be\begin{split}
A^{nonan}_{L=2}  =& i \kappa_{10}^2\,\calR^4\, \Big[   -{ \zeta(4)\over (4\pi)^3}\,\left(\zeta(3)+
2\zeta(2)\,g_{A}^2\right)\log(\chi g_{A}^2/C_{(2,0)})\, \hat\sigma_2^2\\
&- {13\over 5184\pi}\, \zeta(4)\,g_{A}^2 \,\log(\chi g_{A}^2/C_{(1,1)})\, \hat\sigma_2\hat \sigma_3\\
&
-{1\over  11520\pi}\zeta(6)\,g_{A}^4\,\log(\chi  g_A^2 /C_{3,0})\big(11\,\hat\sigma_2^3
 + 14\,\hat\sigma_3^2 \big)+\cdots
\bigg]\,,
\label{aawanonstring}
\end{split}\ee
 The first of these thresholds, of order $\hat\sigma_2^2\,\log(\chi)\,\calR^4$
has both a genus-one and genus-two contribution.
The coefficient of the genus-one part matches the value obtained
  in (3.47) of \cite{grv:oneloop} from genus-one string amplitude.
Also present is the expected genus-two  threshold of order
$\hat\sigma_2\hat\sigma_3\,\log(\chi)\,\calR^4$, as well as
the genus-three thresholds of order $\sigma_2^3\,\log(\chi)\,\calR^4$ and
$\sigma_3^2\,\log(\chi)\,\calR^4$.
 The scales $C_{(p,q)}$ are undetermined by our procedure, whereas they are uniquely fixed in string
perturbation theory.  Such scales are also expected to be fixed by the
$SL(2,\ZZ)$ duality of the IIB theory.

Reinterpreting the nonanalytic terms as contributions to
ten-dimensional type IIB and requiring $SL(2,\ZZ)$ duality strongly
suggests how certain perturbative terms combine into nonperturbative modular invariant
coefficients.
Thus, in the type IIB case the two terms in the coefficient of
$\hat\sigma_2^2\, \log(\chi)\, \calR^4$ form the two perturbative terms
in the  expansion of
the modular function $E_{3/2}$, as expected by general
arguments based on
unitarity of string perturbation theory \cite{grv:bigpaper,grv:oneloop}.
In similar manner, unitarity requires that the genus-three coefficients of $\hat\sigma_2^3\,
\log(\chi)\, \calR^4$ and $\hat\sigma_3^2\, \log(\chi)\, \calR^4$ are pair up with genus-one threshold
contributions in the ratio contained in the modular funtion $E_{5/2}$.
Although these genus-one terms do not appear in the $L=2$ supergravity calculation
(dimensional analysis implies that they should arise from a one-loop sub-deivergence of
three-loop $L=3$ supergravity) their value is again known from the direct string theory calculations
in \cite{grv:oneloop}.  Once again the coefficient in (\ref{aawanonstring}) is in accord with
expectations.

%%%%%%%%%%%%%%%%%%%%%
\section{Torus compactification to nine dimensions}
\label{sec:ttwo}
%%%%%%%%%%%%%%%%%%%%%%%%%

Consider now the eleven-dimensional two-loop amplitude  compactified on a
two-torus  of  volume  $\calV_2$  and  complex  structure
$\Omega $, where the external momenta do not have any components in the compact toroidal dimensions.
As before, we want to evaluate
\be
I_{(p,q)}^{(d=9)} = \pi^{N+1}\int_\0chi^{\Vlambda} dV V^{3-N} \int_{\calF_\Lambda}
     {d^2\tau\over \tau_2^2} B_{(p,q)}(\tau)\, \sum_{(\hat m^I,\hat
       n^I)\in \ZZ^4} \, e^{-\pi \hat E}\, .
\label{intwot}
\ee
The metric on the torus that enters into the definition of $\hat E$ in (\ref{edhatdef}) is
\be
G_{IJ} = \frac{\calV_2}{\Omega_2}
\left(
\begin{matrix} 1 & \Omega_1\cr
\Omega_1 & |\Omega|^2
\end{matrix}
\right)\,,
\label{torusmet}
\ee
which leads to
\be
\hat E={\calV_2\, V\over\Omega_2\tau_2}\left| (1\ \Omega)M\left(\begin{matrix}\tau\cr
1\end{matrix}\right)\right|^2-2\calV_2\, V\det M\,,
\ee
where
\be
M=\begin{pmatrix}
\hat m^1 & \hat m^2\\
\hat n^1& \hat n^2
\end{pmatrix}\, .
\label{mnrestrict}
\ee

The dependence on $\calV_2$ can be factored out by rescaling $V$ (with a corresponding rescaling of $V^\Lambda$),
which leads to
\be
I_{(p,q)}^{(d=9)}(\Omega,\calV_2) ={1\over 4} \pi^{3/2} \Gamma(N-3/2) \,\calV_2^{N-4}\, \calE_{(p,q)}^{(N+2)}(\Omega)\,,
\label{calicale}
\ee
with $N=2p+3q-2$.  The normalization has been chosen so that the functions $\calE_{(p,q)}^{(N+2)}(\Omega)$
correspond to those defined in (\ref{analytr}).
The dictionary relating M-theory to type IIB string theory includes the identifications
\be
r_B = \calV_2^{-3/4}\, \Omega_2^{-1/4}\,,\qquad  e^{\phi_B} = g_B = \Omega_2^{-1}\,.
\label{twobm}
\ee

Almost all the massless thresholds in nine dimensions for the terms up to the order we are considering here
involve half-integral powers
of the Mandelstam invariants rather than being logarithmic.  In contrast to the ten-dimensional
case in the previous section, these nonanalytic terms are easily distinguished from the
analytic terms and we will ignore them in the following.  The one exception is the
logarithmic threshold that
arises in the zero Kaluza--Klein ($m_I=n_I=0$) term, which is the genus-two massless supergravity
sector discussed in appendices~\ref{sec:Cste} and \ref{sec:DimReg}.

\subsection{Evaluation of $\Delta_\Omega I_{(p,q)}^{(d=9)}$}
\label{methnine}

Following the method used in \cite{gv:D6R4} we apply the
Laplace operator $\Delta_\Omega \equiv 4\Omega_2^2\, \partial_\Omega\partial_{\bar\Omega}$ to
$I_{(p,q)}^{(d=9)}$ and use
\be
\Delta_\Omega\, e^{-\pi \hat E} = \Delta_\tau\, e^{-\pi \hat E}\,
\label{newdelt}
\ee
to give
\be
\Delta_\Omega I_{(p,q)}^{(d=9)} = \pi^{N+1}\, \int_{\0chi}^{\Vlambda} dV V^{3-N} \int_{\calF_\Lambda}
     {d^2\tau\over  \tau_2^2} \,B_{(p,q)}(\tau)\,  \sum_{(\hat
       m,\hat n)\in\ZZ^4} \Delta_\tau \, e^{-\pi\hat E}\, ,
\label{intwotb}
\ee
where $N=2p+3q-2$.
Integrating by parts gives  the Laplace operator $\Delta_\tau$
acting  on $ B_{(p,q)}(\tau)$, while the boundary term vanishes in the sector with
$(\hat m^1,\hat m^2)\ne(0,0)$
and $(\hat n^1,\hat n^2)\ne(0,0)$.
 For the moment we will restrict our considerations to this sector, which turns out to give terms independent of
the cutoff $\Lambda$, which can therefore be set equal to $\infty$.  We will also (in appendix~\ref{sec:Cste})
need to consider the sector with $(\hat m^1,\hat m^2) = (0,0)$ and  $(\hat n^1,\hat n^2) \ne (0,0)$.

After using (\ref{modsum}) and (\ref{poissoeq}) we see, as in the $\calS^1$ compactification,
that $I_{(p,q)}^{(d=9)}$ is itself a sum of components,
\be
I_{(p,q)}^{(d=9)} = \sum_{i=0}^{\lceil 3N/2\rceil} h_{(p,q)}^{i}(\Omega,\calV_2)\, ,
\label{indecom}
\ee
and $N=2p+3q-2$ as before.
The function $h_{(p,q)}^{i}$  satisfies a Poisson equation
\be
(\Delta_\Omega - \lambda_{(p,q)}^i)\,  h_{(p,q)}^{i} = \left.\pi^{N+1} \int_{\0chi}^{\Vlambda} dV
V^{3-N} \int_1^{\Lambda^2/V}
{d\tau_2\over \tau_2} \, c_{(p,q)}^{i}
(\tau_2)\,\sum_{(\hat m,\hat n)\neq(0,0)} e^{-\pi \hat E}\right|_{\tau_1 = 0}\, ,
\label{innewt}
\ee
where   $c_{(p,q)}^{i}$  is   the   coefficient  of   $\delta(\tau_1)$
in~(\ref{genpoisson}) and~(\ref{poissoeq}) and
\be
\hat E\bigg|_{\tau_1=0} =\calV_2\, V\,
{|\hat m^1+\hat m^2\Omega|^2+\tau_2^2|\hat n^1+\hat n^2\Omega|^2\over\Omega_2\tau_2} \,
\label{tauonee}
\ee
 In order to  determine the right-hand side of  (\ref{innewt}) we will
 use the fact that
     $c_{(p,q)}^i(\tau_2)$    is     a     polynomial    in
 $\tau_2+\tau_2^{-1}$ of degree $N-1=2p+3q-3$
  \be
c_{(p,q)}^{i}(\tau_2)=\sum_{r=0}^{N-1}  \, c_r (\tau_2+
\tau_2^{-1})^r\ .
  \label{biprop}
  \ee
Substituting (\ref{biprop}), using
the symmetry of the integrand under $\tau_2 \to 1/\tau_2$ to extend the
range of integration to $0\le \tau_2 \le \infty$, and changing integration variables to
$x=V/\tau_2$, $y=V\tau_2$
the right-hand side of (\ref{innewt}) becomes
\bea
{\pi^{N+1}\over 4} &\times&\sum_{r=0}^{\lceil 3N/2\rceil}  c_r\,  \int_0^\infty {dx\over x}\, \int_0^\infty
{dy\over y}
x^{2-{N\over 2}+{n_r\over 2}} y^{2-{N\over 2}-{n_r\over 2}} \, \sum_{(\hat m,\hat n)'}
 e^{- \pi \calV_2 \big( y
{|\hat m^1+\hat m^2\Omega|^2\over\Omega_2}+x{|\hat n^1+\hat n^2\Omega|^2\over \Omega_2}\big) }
\non\\
&&\qquad= {\pi^{N+1} \over 4\calV_2^{4-N} }\, \sum_{r=0}^{\lceil 3N/2\rceil} c_r\,E^*_{2-{N\over 2}+{n_r\over 2}}(\Omega ) \
E^*_{2-{N\over 2}-{n_r\over 2}}(\Omega ) \,,
\label{snnn}
\eea
for $-N+1\leq n_r\leq N-1$.
The Eisenstein series $E^*_s$ is defined in terms of $E_s$ by
\begin{equation}
E^*_{s}(\tau)\equiv      {\Gamma(s)\over     \pi^s}\,     E_{s}(\tau)=
2\zeta^*(2s)\, \tau_2^s+2\zeta^*(2-2s)\, \tau_2^{1-s}+O(\exp(-\tau_2)) \,
\end{equation}
and satisfies the symmetry relation $E^*_{s}=E^*_{1-s}$, where  $\zeta^*(2s )=\pi^{-s}\Gamma(s)\zeta(2s)$.
The cutoffs on the integration limits have been removed in (\ref{snnn}) since the result is finite
(if the $E_{s}$ functions for $s\le 1/2$ are defined by analytic continuation from $s> 1/2$).

So we finally obtain the Poisson equations for the components of the $(p,q)$ term,
\be
(\Delta_\Omega - \lambda_{(p,q)}^i)\,  h_{(p,q)}^{i} =
{\pi^{N+1} \over 4\calV_2^{4-N} } \sum_{r=0}^{N-1}  c_r\,E^*_{2-{N\over 2}+{n_r\over 2}}(\Omega ) \
E^*_{2-{N\over 2}-{n_r\over 2}}(\Omega) \, ,
\label{ipoisso}
\ee
where $N=2p+3q-2$ and $1\leq i\leq \lceil 3N/2\rceil$.
The right-hand side of this equation is a sum with a finite number of terms that depends on
the value of $N$.
The solutions of this equation for given values of $(p,q)$
and the corresponding values of the index $i$  determine
$I_{(p,q)}^{(d=9)}$ and hence,
 the coupling constant dependence of the coefficient of the term in $A_{IIB}$ of order  $S^{N+2}\,  \, \calR^4$.
It is notable that the right-hand side of (\ref{ipoisso}) is quadratic in the Eisenstein series, each of which will later
(in section~\ref{sec:modular}) be identified with a coefficient of a lower-order term in the action.

The dependence of $I_{(p,q)}^{(d=9)}$ on the volume, $\calV_2^{N-4}$, in (\ref{snnn}) translates into the IIB string
theory description as
\be
g_B^{N-1}\, r_B^{3-2N}\, ,
\label{stringparams}
\ee
using the correspondence between the supergravity and IIB string parameters given in (\ref{twobm}),
together with the identification $S=R_{11}\, s$.

In the above analysis we only considered terms with $(\hat m^1,\hat m^2) \neq(0,0)$ and
$(\hat n^1,\hat n^2)\neq(0,0)$, which are independent of $\Lambda$.  Certain terms with $\log \Lambda$
dependence also entered into the zero eigenvalue parts of the modular functions $\calE_{(2,0)}^{(2)}$,
$\calE_{(3,0)}^{(6)}$ and $\calE_{(0,2)}^{(6)}$ in appendix~\ref{sec:Cste}.  However, for economy of
space we have not considered the terms that arise from $(\hat m^1,\hat m^2)= (0,0)$ with $(\hat n^1,\hat n^2) \neq(0,0)$,
which correspond to subdivergences and have a power dependence on $\Lambda$ that needs to be subtracted by
counterterms.

In the $N=0$ case ($(p,q)= (1,0)$), which corresponds to the $\calD^4\calR^4$ term, the source on the right-hand
side of (\ref{ipoisso})
vanishes and the equation reduces to the Laplace eigenvalue equation (\ref{lapeig}) for the value $s=5/2$, as
in \cite{gkv:twoloop}.  In the $N=1$ case ($(p,q)=(0,1)$), which corresponds to the $D^6\calR^4$ term,
  the source on the right-hand side of (\ref{ipoisso}) is quadratic in $E_{3/2}$ and there is a single eigenvalue
  $\lambda_{(0,1)}^1=12$, reproducing (\ref{poiss}),
as obtained in \cite{gv:D6R4}.
We will now analyze these solutions for the cases $(2,0)$, $(1,1)$, $(3,0)$ and $(0,2)$,
which raise a number of new issues.

%%%%%%%%%%%%%%%%%%%%%%%%%%%%%%
\subsubsection{$(p,q)=(2,0)$}

The expression for
$B_{(2,0)}(\tau)$  given in (\ref{b22def}) is written as the sum of the $b_{(2,0)}^i$'s in (\ref{eca}).
Applying the method described in the previous subsection, using the explicit Poisson equation (\ref{b2ieq})
satisfied by each  $b_{(2,0)}^i$, we determine that the modular function $\calE_{(2,0)}^{(2)}$
associated with the $\sigma_2^4\, \calR^4$ term has the form
\be
 {4\calV_2^2\over \pi^2}\, I_{(2,0)}(\Omega,\calV_2) \equiv \calE_{(2,0)}^{(2)} (\Omega )=
 \sum_{i=0}^3 \calE_{(2,0)}^{(2)2i}(\Omega )\ ,
\label{caleidef}
\ee
where $ \calE_{(2,0)}^{(2)i}(\Omega )$ are modular functions satisfying
the  Poisson equations
\be
(\Delta_\Omega -r(r+1)) \calE_{(2,0)}^{(2)r}(\Omega )
= - 2 u_r E_{3\over 2}
E_{1\over 2}\ ,\qquad \textrm{for}\ r=2i = 2,4,6
\label{primas}
\ee
and $u_r$ are constants given in  appendix~\ref{sec:ModD8R4}.

The function  $E_{1/2}$ is defined as the limit
\be
\lim_{s\to 1/2} E_s= 2\Omega_2^{\half}\, \log{\Omega_2 \over 4\pi c_e}
+4 \Omega_{2}^{\half}\sum_{n\neq0} d_{|n|}\, K_{0}(2\pi |n|\Omega_{2})\, e^{2i\pi\, n\,\Omega_{1}} \ ,
\qquad c_e =e^{-\gamma}\, ,
\ee
where $\gamma $ is Euler's constant and $d_{|n|}$ is the number of divisors of $n$.
This is a very special Eisenstein series which has a large-$\Omega_2$ expansion that has no purely power-behaved terms,
but starts with $\Omega_2^{1/2}\, \log \Omega_2$.
The interpretation of the $\log\Omega_{2}$ factor will be given in the section~\ref{sec:string}
where it will shown to be associated with the presence of massless thresholds.

The  $i=0$  term, $\calE_{(2,0)}^{(2)0}$  associated with the  constant
$b_{(2,0)}^0=-13/21$, is shown in
appendix~\ref{sec:Cste} to be equal to
\be
\calE_{(2,0)}^{(2)0}= -{104\over 21} \zeta(2)
\log{(-S\, \calV_2/\Omega_2\, C_{(2,0)}})\,,
\label{zeroconst}
\ee
where $C_{(2,0)}$ is, as before, an undetermined
function of $S_0$ and $T_0$, but is independent of $\calV_2$ and $\Omega_2$.  The
logarithm comes from
the contribution of the zero Kaluza--Klein modes, $m_I=n_I=0$ in the
$\calT^2$ reduction from eleven dimensions and should coincide with the supergravity
calculation in nine dimensions discussed in section~\ref{sec:DimReg}.
It is notable that the coefficient of $\log(\chi)$ in (\ref{zeroconst}) does indeed coincide with the coefficient of
the $\epsilon$ pole in dimensional regularization of two-loop maximal supergravity around nine
dimensions.  However, in our case the scale
 depends on the compactification  moduli rather
than an arbitrary cutoff.

When translated to IIB coordinates the $\calE_{(2,0)}$
 contribution has the form $r_B^{-1}\,\calE_{(2,0)}^{(2)}(\Omega)\,
S^4\,\calR^4$.

\subsubsection{$(p,q)=(1,1)$}
With some effort one can use (\ref{b3detail}) to write $B_{(1,1)}= \sum_{j=0}^4 b_{(1,1)}^j$ where the $b_{(1,1)}^j$'s satisfy
the Poisson equations (\ref{ecb}).   It is straightforward to extend
the general method described in sub-section~\ref{methnine}
to determine the coefficient, $\calE_{(1,1)}^{(4)}$, of the
$\sigma_2\sigma_3\, \calR^4$ term in the amplitude.  This is given by
\be
 { 8\calV_2\over \pi^2 }\,I_{(1,1)}\equiv
\calE_{(1,1)}^{(4)}(\Omega )= \sum_{j=0}^4
\calE_{(1,1)}^{(4)2j+1}(\Omega )\ ,
\ee
where $\calE_{(1,1)}^{(4)j}(\Omega )$ are modular functions
satisfying
\be
(\Delta_\Omega -r(r+1))
\calE_{(1,1)}^{(4)r}(\Omega )=- 2v_r E_{3\over 2} E_{3\over 2}- 4\pi^2 w_r E_{1\over 2} E_{1\over 2}\,,
\qquad\quad r=2j+1 = 1,3,5,7,9\,,
\label{fffm}
\ee
The coefficients  $v_j,\ w_j$ are given in  appendix~\ref{subsec:modten}.

When translated to IIB coordinates this contribution has the form
$r_B^{-3}\,\calE_{(1,1)}^{(4)}(\Omega)\, S^5\,\calR^4$.

\subsubsection{The cases $(p,q) = (3,0)$ and $(p,q)=(0,2)$}
In this case there are two modular functions,
$\calE_{(3,0)}^{(6)}$ and $\calE_{(0,2)}^{(6)}$, multiplying the independent
kinematical structures $\sigma_2^3$ and $\sigma_3^2$.
Equations in (\ref{b4ydef}) determine that $B_{(3,0)}^k= \sum_{k=0}^6
b_{(3,0)}^{2k}$
and $B_{(0,2)}^k= \sum_{k=0}^6 b_{(0,2)}^{2k}$ where $ b_{(3,0)}^k$ and  $b_{(0,2)}^k$ satisfy the
Poisson equations  (\ref{xecb}).  This leads to the expressions for the coefficients
of the two kinematic structures at order $S^6\,\calR^4$,
\bea
{16 \over 3\pi^2}I_{(3,0)}\equiv
\calE_{(3,0)}^{(6)}(\Omega )
&=&
\sum_{k=0}^6 \calE_{(3,0)}^{(6)2k}(\Omega )\,,
\nn\\
{16 \over 3\pi^2}I_{(0,2)}\equiv
\calE_{(0,2)}^{(6)}(\Omega )
&=&
\sum_{k=0}^6 \calE_{(0,2)}^{(6)2k}(\Omega )\,,
\label{ggiinn}
\eea
where $\calE_{(3,0)}^{(6)k} (\Omega )$ and  $\calE_{(0,2)}^{(6)k}(\Omega)$
are modular functions satisfying,
\bea
(\Delta_\Omega -r(r+1))
\calE_{(p,q)}^{(6)r}(\Omega )
&=&   -  2f_{(p,q)}^r  E_{3\over   2}  E_{5\over   2}- 16\zeta(2)\,
(f_{(p,q)}^r+g_{(p,q)}^r)\, E_{1\over 2} E_{3\over 2}\,,
\label{calgdefs}
\eea
where $r=2k= 2,4,6,8,10,12$ and $(p,q)=(3,0)$ or $(p,q)=(0,2)$ and the
coefficients $f_{(p,q)}^r$ and $g_{(p,q)}^r$ are given in appendix~\ref{subsec:modtwelve}.
The expressions for the functions $\calE_{3,0}^{(6)\,0}$ and
$\calE_{0,2}^{(6)\,   0}$  associated   with  the   constant  function
$b_{(3,0)}^0=12264/715$ and $b_{(0,2)}^0=2716/165$ can be obtained by direct evaluation of the
integrals as in (\ref{kzerodef}) and (\ref{lzerodef}).

When   translated  to IIB coordinates  these   $\calD^{12}\calR^4$
contributions  have the form $r_B^{-5}\,
(\calE_{(3,0)}^{(6)}\,\hat\sigma_2^3   +\,\calE_{(0,2)}^{(6)}\,   \hat
\sigma_3^2)\, \calR^4$.

\subsection{The nine-dimensional type IIB low energy string scattering amplitude}

To summarize, we have determined a number of terms in
 the expansion of the the $\calT^2$ compactification of two-loop ($L=2$)
eleven-dimensional supergravity
up to  order $S^6\, \calR^4$. Adding these to the terms found previously, gives the following
expression in terms of the  type IIB string
theory parametrization:
 \bea
A_{L=1}^{(d=9)}+A_{L=2}^{(d=9)} &=& r_B (g_B^{-\half}\,  \calE_{(0,0)} (\Omega )\, \calR^4
+ g_B^{\half}\,  \calE_{(1,0)} (\Omega )\, {\hat\sg_2 }\calR^4
+ g_B\,  \calE_{(0,1)}(\Omega )\, {\hat\sg_3 }\calR^4)\nn\\
&+&
(4\pi)^2{ g_B^2\over r_B}\,  \calE_{(2,0)}^{(2)} (\Omega )\, {\hat\sg_2^2\over 288} \calR^4
+(4\pi)^2\, {g_B^3\over r_B^3}\, \calE_{(1,1)}^{(4)}(\Omega )\, {\hat\sg_2\hat\sg_3\over 3!\, 15120}\calR^4\nn\\
&+& {(4\pi)^2\over 4!\, 302400}\, {12g_B^4\over r_B^5}\, \Big( \calE_{(3,0)}^{(6)} (\Omega )\, \hat\s_2^3 \calR^4
+{4\over 3} \calE_{(0,2)}^{(6)}(\Omega ) \, \hat\s_3^2 \calR^4\Big)
\label{unxxx}
\eea
Here we have included the coefficients given in (\ref{eberz}) and the powers of $r_B$ and $g_B$
in (\ref{stringparams}).   The terms in the first line are the ones found in previous work,
namely, the function $\calE_{(0,0)}$, which was derived from the $L=1$ amplitude \cite{ggv:oneloop}, and the
functions ${\cal E}_{(1,0)}$ and ${\cal E}_{(0,1)}$ that were derived from the $L=2$ amplitude in
\cite{gkv:twoloop,gv:D6R4}.  We have not included terms that arise from renormalised subdivergent contributions
(apart from the $\hat\sigma_2 \calR^4$ term), although these are easy to evaluate.

We emphasize again that the Feynman diagrams of supergravity are only expected to be an
approximation to low energy string theory in a limited range of moduli space, although some some very
special processes are presumably protected by supersymmetry.  This requires, in particular,
that $r_B\ll 1$ or $r_A \gg 1$ with $\alpha's\, r^2_A \ll 1$.
The type IIA expression follows by use of the usual T-duality relations
\be
r_A = r_B^{-1}\,,\qquad g_A = r_B^{-1} \, g_B\,.
\label{tdual}
\ee

The functions $\calE_{(2,0)}^{(2)}$, $\calE_{(1,1)}^{(4)}$, $\calE_{(3,0)}^{(6)}$ and $\calE_{(0,2)}^{(6)}$
are the unique $SL(2,\ZZ)$-invariant solutions of the Poisson
equations obtained earlier, subject to the condition that they are no worse than power-behaved in $g_B$ as
$g_B\to 0$.
Our interest here is in obtaining the terms in these functions that are
power-behaved in the string coupling $g_B= \Omega_2^{-1}$, which is the subject of the following
sub-section.

%%%%%%%%%%%%%%%%%%%%%%%%%%%%%%%%%%%%%%%%%%%%%%
\subsubsection{The perturbative expansion of $A_{IIB}^{(d=9)}$}
\label{sec:pertmod}
%%%%%%%%%%%%%%%%%%%%%%%%%%%%%%%%%%%%%%%%%%%%%%

In analyzing the perturbative parts of the solutions to the preceding Poisson equations
we may replace $\Delta_\Omega$ by $\Omega_2^2\, \partial_{\Omega_2}^2$ since the perturbative terms
are independent of $\Omega_1$.
The cases $(1,0)$ and $(0,1)$ were discussed in \cite{gkv:twoloop, gv:D6R4}
and reviewed in the introduction, so we will begin with the next term in the expansion.

\medskip
\noindent{ \it $(p,q)=(2,0)$}
\smallskip

We start with the coefficient, $\calE_{(2,0)}^{(2)}=
\sum_{i=0}^3 \calE_{(2,0)}^{(2)\, 2i}$ , of the $\sigma_2\calR^4\sim\calD^4 \calR^4$ terms.
In this case, alone among the nine-dimensional terms  that we are considering, there is a
logarithmic singularity of order $S^4\, \log(\chi)\, \calR^4$, which arises from the sector with
zero Kaluza--Klein modes,
$m_I=n_I=0$, which enters into the function $\calE_{(2,0)}^{(2)\, 0}$.  In addition to the analytic part,
proportional to $\calE_{(2,0)}^{(2)}\, \sigma_2^2\, \calR^4$, the amplitude therefore
contains a nonanalytic part,
\be
A^{nonan}_{(2,0)} =\calE_{(2,0)}^{(2)\,0}\, S^4\, \log(\chi/C_{(2,0)})\, \calR^4\,,
\label{sfourterm}
\ee
For $r=2,4,6$, we see from (\ref{primas}) that the perturbative parts of $\calE_{(2,0)}^{(2)r}$  satisfy
\be
(\Omega_2^2\partial_{\Omega_2}^2 -r(r+1))\calE_{(2,0)}^{(2)r\, \rm pert}=-2 u_{r}  \big(
 4\zeta(3) \Omega_2^2 +
8\zeta(2) \big) \log {\Omega_2\over 4\pi c_e} \,,
\label{poissei}
\ee
Hence
\bea
\Omega_2^{-2}\calE_{(2,0)}^{(2)r\, \rm pert} &=&
\a_{(2,0)}^{(r)}  \Omega_2^{r-1} +\beta_{(2,0)}^{(r)}
 \Omega_2^{-r-2} + {8 u_r\zeta(3)\over (r(r+1)-2)^2}
\left( 3+ (r(r+1) -2)\log {\Omega_2\over 4\pi c_e} \right)
\non\\
&-& {16\zeta(2)\, u_r\over (r(r+1))^2}\Omega_2^{-2}\left( 1 - r(r+1)\, \log {\Omega_2\over 4\pi c_e} \right)\, .
\label{epertu}
\eea
where $u_2=20/21$, $u_4=90/ 77$, $u_6=640/165$
and $\alpha_{(2,0)}^{(r)}$ and $\beta_{(2,0)}^{(r)} $ are integration constants and must be fixed by boundary conditions.
Since the term proportional to $\alpha_{(2,0)}^{(r)}$ is an odd power of the string coupling, which does not appear
in string perturbation theory, we deduce that $\alpha_{(2,0)}^{(r)}$ must be zero.
As shown in \cite{gv:D6R4} and in
appendix~\ref{sec:Diff},
 this uniquely determines the value of $\beta_{(2,0)}^{(r)}$.

Summing all contributions and using the values of $u_1$, $u_2$ and
$u_3$ given above leads to the complete
contribution (including the $i=0$ term in (\ref{zeroconst}))
\bea\label{e:calE}
\Omega_2^{-2}\calE_{(2,0)}^{(2)\rm pert} &=&
{16\over 5} \zeta(3) \log {\Omega_2 \over 4\pi c_e}
+ {4\over 9}\zeta(4)   \Omega_2^{-4}  +{4\over 945} \zeta(6)   \Omega_2^{-6}
+ {512\over 496125} \zeta(8)   \Omega_2^{-8}
\,.
\eea
Notice that the sum of the $\zeta(2)\,\Omega_2^{-2}\,  \log\Omega_2$
terms  appearing in each $\calE_{(2,0)}^{(2)i\, \rm pert}$ in (\ref{epertu}) have canceled with
the $\log \Omega_2$ factor in $\calE_{(2,0)}^{(2)0}$
(\ref{zeroconst}) and the only remaining genus-two term is the non-analytic term (\ref{sfourterm}),
which is proportional to $g_B^2\,s^4\,\log(\chi/r_B^2 C_{(2,0)})$
in the IIB string parametrization).

Furthermore, we see that in the language of type IIB string theory, where $\Omega_2^{-1} = g_B$,
the $S^4\, \calR^4$
coefficient contains perturbative string contributions from genus-one to genus-five.
The genus-one and genus-two terms (proportional to $\Omega_2^0$ and $\Omega_2^{-2}$, respectively),
are simply obtained by equating the corresponding terms on the left-hand and right-hand sides of
(\ref{poissei}). The
power-behaved terms at order $\Omega_2^{-4}$,  $\Omega_2^{-6}$ and $\Omega_2^{-8}$
 have been evaluated using the method described in appendix~\ref{sec:Diff}.
We recognize part of the genus-one contribution to $\hat\sigma_2^2\,\calR^4$ in nine dimensions
derived in \cite{grv:oneloop}.
Indeed, the $\log\Omega_{2}$ term originates from the stringy
corrections to the massless threshold
and is associated with the $\log r$ term found in \cite{grv:oneloop}.
It is notable that the scale of the logarithm is absolutely determined
in this expression.

\medskip
\noindent{\it $(p,q)=(1,1)$}
\smallskip

Now we turn to the power-behaved terms in $\calE_{(1,1)}^{(4)\rm pert}$, the coefficient of the
$\sigma_2\sigma_3\, \calR^4\sim \calD^{10}\calR^4$ contribution,
which are determined by (\ref{fffm}).
In this case the source term (the right-hand side of (\ref{fffm})) contains the powers $\Omega_2^0,\dots,\Omega_2^{-4}$
which lead to terms with the same powers in $\calE_{(1,1)}^{(2)\rm pert}$.  In addition there are $\beta$ terms that are
again  deduced from  the expressions  in  appendix~\ref{sec:Diff}. The
result is
\bea\label{e:calF}
\Omega_2^{-3}\calE_{(1,1)}^{(4)\rm pert} &=& {180}\zeta(3)^2 +
{7168\over 15}\,  \zeta(2) \zeta(3) \Omega_2^{-2}  -
 {1456\over 3} \, \zeta(2) \Omega_2^{-2} \log {\Omega_2\over 4\pi c_e}
+  {3248\over 3}\zeta(4) \Omega_2^{-4}\nn\\
 &+&{98\over 9}\zeta(6)  \Omega_2^{-6}
+{896\over 405}\zeta(8)  \Omega_2^{-8}+{304\over 1875}\zeta(10)  \Omega_2^{-10}
+{185600\over 15802479}\zeta(12) \Omega_2^{-12}\,.
\eea
Note that the scale of the $\log \Omega_2$ term is determined in this expression.
Thus  $\Omega_2^{-3}\calE_{(1,1)}^{(4)\rm pert}$ contains terms that are interpreted as
perturbative string theory contributions from genus-one up to genus-seven.
Note, in particular, the presence of  the genus-two $\log\Omega_2 $ term.  This is directly related to the
presence of a  $S^5\, \log (-S)$ term at  genus-two in ten-dimensional
supergravity, as we saw in (\ref{aawanonstring}) and which is required by unitarity.  This
will be discussed in the analysis of ten-dimensional $L=2$ supergravity in
appendix~\ref{sec:tenreg}.
Importantly, the $\log$ square terms -- present in each $\calE_{(1,1)}^{(4)j}$ contribution -- have canceled out in the sum.
This corresponds to the cancelation of the leading $1/\epsilon^2$ pole also described in
appendix~\ref{sec:tenreg}.

\medskip
\noindent{\it $(p,q) =(3,0)$ and $(p,q)= (0,2)$}
\smallskip

Finally, we turn to the coefficients of the two order
$S^6\,\calR^4$ terms,  $\calE_{(3,0)}^{(6){\rm pert}}$
and  $\calE_{(0,2)}^{(6){\rm pert}}$.   These are determined by (\ref{calgdefs}).  In this case the source term on the
right-hand side of each of these equations contains the powers $\Omega_2^0\,\dots, \Omega_2^{-6}$, which determine the
corresponding powers of $\Omega_2$ in the solutions.  In addition, there are $\beta$ terms with powers $\Omega_2^{-8}\,
\dots, \Omega_2^{-16}$ that are determined by the expressions in appendix~\ref{sec:Diff}.  The resulting
perturbative terms in the solutions are
\bea
\Omega_2^{-4}\calE_{(3,0)}^{{(6)\rm pert}} &=& 96\zeta(3)\zeta(5) +{8828\over 77}\zeta(2)\zeta(5)\Omega_2^{-2}-
{28096\over 77} \zeta(3) \zeta(2)\Omega_2^{-2}
\non\\
&+& 1760\, \zeta(4)\,\Omega_2^{-4}
 \log {\Omega_2\over C_{(3,0)}}
+ {280\over 3} \zeta(6) \Omega_2^{-6} +{1792\over 135}\zeta(8) \Omega_2^{-8}
+{32\over 45}\zeta(10) \Omega_2^{-10}
\non\\
&+& {30720\over 53207} \zeta(12) \Omega_2^{-12} -{707584\over 7432425}\zeta(14) \Omega_2^{-14}
+{973635584\over 41937606711}\zeta(16) \Omega_2^{-16}
\label{e:calGx}\eea
and
\bea
\Omega_2^{-4}\calE_{(0,2)}^{(6){\rm pert}} &=& 96\zeta(3)\zeta(5) +
{81148\over 693}\zeta(2)\zeta(5)\Omega_2^{-2}-{233512\over 693} \zeta(3)
\zeta(2)\Omega_2^{-2}
\non\\
&+&1680 \zeta(4)\, \Omega_2^{-4}\, \log {\Omega_2\over C_{(0,2)}}
+ {1120\over 9}  \zeta(6)  \Omega_2^{-6}
+{3584\over 405}\zeta(8) \Omega_2^{-8}
+{32\over 15}\zeta(10) \Omega_2^{-10}
\non\\
&+& {2432\over 159621} \zeta(12) \Omega_2^{-12} + {356268544\over 3277699425}\zeta(14) \Omega_2^{-14}
+{215105536\over 9677909241} \zeta(16) \Omega_2^{-16}\,.
\label{e:calGy}
\eea
Thus,  these   modular  functions  contain  perturbative  string
contributions from  genus-one up to genus-nine.   Note, in particular,
that the  $\Omega_2^{-2}\, \log \Omega_2$ terms which  are present for
each individual eigenvalue have canceled in the sum.  There remain two
genus-two  terms in  (\ref{e:calGx}) and  (\ref{e:calGy}),  which have
coefficients proportional to $\zeta(2)\zeta(3)$ and $\zeta(2)\zeta(5)$. Here we see another 
example of the  lack of transcendentality.  The example described in the introduction arose in comparing
contributions of different genera whereas here it arises purely at  genus-two.
The only $\log$ terms  in (\ref{e:calGx}) and
(\ref{e:calGy}) are the ones associated with the power $\Omega_2^{-4}$,
which correspond  to genus-three terms in string theory.  As we  will see,
these  have  the  numerical  values  expected  on  the  basis  of  string
unitarity.  The undetermined constants $C_{(3,0)}$ and $C_{(0,2)}$ are
once again associated with the scale of these $\log$ terms.

In addition to the terms in (\ref{e:calGx}) and (\ref{e:calGy}) there are $\Omega$-independent terms arising from
$\calE_{(3,0)}^{(6)\, 0}$ and $\calE_{(0,2)}^{(6)\, 0}$ of the form $\zeta(2)^2 \,\log(\calV_2\Lambda)$, as given in
(\ref{laplaceonee}).  This is the same $\log \Lambda$ divergence that we found in the case of the $\calS^1$ compactification
to ten dimensions in (\ref{h4zerox}) and (\ref{h4zeroy}).  This new $\Lambda$-dependent term
should be canceled by a local counterterm.  The values of the
earlier local counterterms, such as the one that cancels the $\Lambda^3$
behaviour of the one-loop amplitude, were determined by enforcing T-duality and the equality of perturbative
type IIA and IIB four-graviton scattering at low genus.  Whether this argument
 can be extended to the case of the $\log \Lambda$
terms is not clear.

%%%%%%%%%%%%%%%%%%%%%%%%%%%%%%%%%%%%%%%%%%%%%%%%%%%%%
\subsection{Connections with string perturbation theory in nine dimensions}\label{sec:string}
\label{sec:stringres}
%%%%%%%%%%%%%%%%%%%%%%%%%%%%%%%%%%%%%%%%%%%%%%%%%%%%%
We can now compare the perturbative terms in the modular functions with known features of string
perturbation theory.  Contributions to terms that contribute in the ten-dimensional limit $r_B\to \infty$
up to order $\calD^6\calR^4$
arise from
one-loop and two-loop eleven-dimensional supergravity  compactified on $\calT^2$, as discussed in
\cite{ggv:oneloop,Russo:1997mk,gkv:twoloop,gv:D6R4}.
No further ten-dimensional  terms arise from the expansion
of $L=2$ supergravity to higher orders in momenta as considered
in this paper.  A dimensional argument shows that in order to generate higher-order
ten-dimensional string theory terms one needs to consider
higher-loop supergravity amplitudes with $L>2$, together with corresponding counterterm diagrams that cancel
divergences. One example that is easy to extract explicitly is a contribution
$E_{7/2}\, \calD^4 \calR^4$ that emerges from a diagram involving two $\calR^4$ counterterms
that cancels the contribution of a pair of one-loop sub-divergences in three-loop supergravity diagrams,
as shown in figure~\ref{fig:threesub}.  The emergence of this
term follows from a simple dimensional argument that takes into account the fact that the
double-divergence behaves as  $\Lambda^6$ (i.e., $\Lambda^3$ for each loop).
However, there is no reason to expect this to be the complete $(\alpha's)^4\, \log(\chi)\,\calR^4$ contribution.

As we saw in the last subsection,
the perturbative expansions of the modular functions considered in this paper all begin with
genus-one terms followed by a finite series of higher-genus
corrections.  Some of these terms may be compared with the known
string theory results, which mostly come  from the low-energy expansion of the
genus-one amplitude \cite{grv:oneloop}.

\subsubsection{Comparison with genus-one string theory}

The terms in (\ref{unxxx}), apart from $\hat\sigma_2\, \calR^4$ and $\hat\sigma_3\, \calR^4$,
disappear in the ten-dimensional IIB limit,  $r_B\to \infty $.
However, as discussed earlier, if the supergravity approximation does make contact with string theory
this would happen for large values of $r_A$, which is described by T-duality
from the IIB expression in the small-$r_B$ limit.
In this limit there are terms with both negative and  positive powers of $r_{A}$.
Those proportional to $r_A$ give rise to perturbative contributions
of type IIA theory in ten dimensions.  These comprise  a genus-one contribution to
$\hat \sigma_2^2\calR^4\sim\calD^8\calR^4$, a genus-two contribution to $\hat\sigma_2\hat\sigma_3\, \calR^4\sim\calD^{10}\calR^4 $
and a genus-three contribution to the $\calD^{12}\calR^4$ terms $\hat\sigma_2^3\, \calR^4$ and $\hat\sigma_3^2\, \calR^4$, which
will be discussed in the following subsection. There are also terms
which behave as $r_A^{1+k}$, which diverge in the decompactification limit $r_{A}\to \infty$ and
must be resummed in order to reconstruct
the string thresholds in ten dimensions, as explained in \cite{grv:bigpaper}.
In addition to terms that are power-behaved in $r_A$,  in type IIA string perturbation theory
 there are exponentially suppressed terms of the form
 $e^{-r_A}$. Such highly-suppressed terms do not appear in the compactification of the
 perturbative supergravity amplitude, which is not sensitive to terms
 of the form $e^{-\calV_n}$ that decrease exponentially with the
 compactification volume.

\begin{itemize}
\item
Consider the type IIA interpretation of the analytic contributions obtained in the previous subsection.
The genus-one terms of the modular functions in (\ref{unxxx}) have  the following form,
\be
\begin{split}
A^{an}_{h=1}
=  2\zeta(2)\,\bigg(&{1\over r_A}+
{1\over 3 r_A}\,\zeta(3)\hat\sigma_3+
 {r_A^3\over 21}\zeta(3)^2 \hat\sigma_2\hat\sigma_3
+ {2 r_A^5 \over 525}
\zeta(3)\zeta(5)
\big( \hat\sigma_2^3 + {4\over 3}\hat\sigma_3^2\big)\\
&- {4\over 15}\, r_A \zeta(3)\log(g_A/r_A)\, \hat\sigma_2^2\bigg)\,.
\label{gensug}
\end{split}
\ee
 The analytic terms in this expression are exactly the same as those obtained from the genus-one
string theory calculation in \cite{grv:oneloop}, so the $L=2$ eleven-dimensional
 supergravity on $\calT^2$ precisely reproduces these genus-one terms in string theory.

The $\log r_A$ contribution in the last line of (\ref{gensug}) comes from the
function $\calE_{(2,0)}^{(2)}$ multiplying the $\hat\sigma_2^2\,\calR^4$ contribution, which contains
 $\log\Omega_{2}=-\log(g_{s}^B)$ factors of the form
\be
{1\over r_{B}}\, {16\over5}\, \zeta(3)\, \log(g_{s}^B)= r_{A}\,{16\over5}\, \zeta(3)\,
\log(g_{s}^{A}/ r_{A})\, .
\ee
The coefficient of the $\log r_A$ piece agrees with that calculated in genus-one
string perturbation theory on a circle of finite radius $r_{A}=r^{-1}_{B}$
in \cite{grv:oneloop}.  This leaves a term proportional to $\log g_A$.

\item
The genus-two term
at order  $\hat\sigma_2\hat\sigma_3\, \calR^4\sim\calD^{10}\calR^4$ contained in
$\calE_{(1,1)}^{(4)}$ in equation~(\ref{e:calF})
has a $\log\Omega_{2}$ factor, whereas
the terms proportional to $\log^2\Omega_{2}$ in each $\calE_{(1,1)}^{(4)\,j}$
cancelled after summing up all contributions.
This is consistent with the absence of the
$1/\epsilon^2$ pole  in the total two-loop supergravity calculation detailed in appendix~\ref{sec:DimReg}.
After T-duality, this $h=2$ term transforms into a term in type IIA that is proportional to $r_A$ and therefore
survives the $r_A\to \infty$ limit.  It is therefore gratifying that its coefficient agrees with that
evaluated by dimensional regularization in ten dimensions, as
described earlier.

\item
The functions $\calE_{(3,0)}^{(6)}$ and $\calE_{(0,2)}^{(6)}$
of equations~(\ref{e:calGx}) and~(\ref{e:calGy})
exhibit a genus-three logarithmic term of order $S^6\,\calR^4$ of the form
\be
r_B^{-5} \zeta(2)^2 g_B^4\log g_B =
 r_A \zeta(2)^2 g_A^4 \log {g_A\over r_A} \,,
\ee
which shows that these terms are again proportional to terms nonanalytic  in $r_A$ in the IIA theory that are
proportional to $r_A$.  Therefore these terms survive the ten-dimensional IIA limit, which was
obtained in the $\calS^1$ compactification in (\ref{astregn}).  Since the IIA and IIB amplitudes are equal
up to at least genus four, it follows that
these terms also arise in ten-dimensional IIB with the same coefficients.
This, in turn,  is consistent with two-particle unitarity \cite{grv:bigpaper}, which relates the order
$S^6\, \calR^4$ threshold contributions at genus-one and genus-three in
string perturbation theory.  The precise coefficients of the
genus-one ($h=1$) massless thresholds at order $S^6\, \calR^4$ have been
evaluated in string theory \cite{grv:oneloop}.  The coefficients of the genus-three terms deduced above
imply that the $h=1$ and $h=3$
terms combine into the nonanalytic term proportional to
\be
g_B^{5/2}\, E_{5/2}\, ({11\over 210}\hat\sigma_2^3 + {1\over 15}\hat\sigma_3^2)\log(\chi)\, \calR^4\,,
\label{genthree}
\ee
 which is precisely
the anticipated non-perturbative threshold term \cite{grv:bigpaper}.
\end{itemize}
%%%%%%%%%%%%%%%%%%%%%%%%%%%%%%%%%%%%%%%%%%%%%%%%%%%%%%%%%%

%%%%%%%%%%%%%%%%%%%%
\section{ Supersymmetry and  higher-derivative couplings -- a schematic discussion.}
\label{sec:modular}
%%%%%%%%%%%%%%%%%%%%%%%%%%%%%%%%%%%%%%%%%%%%%%%%%%%%

In this paper we have analyzed the momentum expansion of the two-loop
four-graviton amplitude in eleven-dimensional supergravity
up to order $S^6\, \calR^4$.  We considered the compactification on
$\calS^1$ to make contact with the ten-dimensional IIA theory, and on $\calT^2$
to make contact with the nine-dimensional IIB theory.  In the $\calS^1$ case
we obtained a number of higher-momentum terms that correspond to terms of
particular genus in string perturbation theory.  In the $\calT^2$ case we
obtained a number of higher-momentum terms with coefficients that
are specific $SL(2,\ZZ)$-invariant functions of the complex scalar coupling
multiplying particular powers of $r_B$.
We have found some impressive matches with perturbative string-theory
results at different genera that are obtained from  direct calculations in string perturbation
theory \cite{grv:oneloop}
combined with unitarity constraints \cite{grv:bigpaper}.  However, it is clear that there would be
immense problems in going further in this manner.  To begin with, the pattern of
ultraviolet divergences of Feynman diagrams becomes much more complicated at higher values of $L$,
which raises questions about how to implement the cutoff on the Schwinger parameters at higher
loops.  Furthermore, it is unclear whether this procedure of computing supergravity
amplitudes with an ultraviolet cutoff and determining the finite part by using string dualities,
can account for the details of intrinsically M-theory quantum effects,
such as quantum effects of membranes,
to all orders in the low energy expansion of string theory.
Interestingly, according to the argument in \cite{Berkovits2006} that uses the pure spinor
formalism, the terms of order $S^6R^4$ are the first terms for which one does not
expect a non-renormalization theorem to hold just on the basis of
supersymmetry.  It is therefore of interest that the genus-one pieces and the threshold pieces of the genus-three
terms in the $S^6\calR^4$ coefficient functions  match the string theory results.

More generally, it is of interest to consider to what extent the structure of
the coefficients in the momentum expansion
might be determined by symmetry constraints that might generalize to higher orders.
In particular, it would be of interest to determine the extent to which maximal
supersymmetry controls the form of the inhomogeneous Laplace equations satisfied by the
coefficients.

%%%%%%%%%%%%%%%%%%%%%%%%%%%%%%%%%
\subsection{Supersymmetry}
%%%%%%%%%%%%%%%%%%%%%%%%%%%%%%%%%

The structure of the Poisson equations satisfied by the coefficient functions should
be highly constrained by maximal supersymmetry, although this has not been explored in
detail beyond the lowest order term in the momentum expansion.
In the case of the $\calR^4$ term the  supersymmetry constraints are indeed
known to determine  that the coefficient function is the modular function $E_{3/2}$ \cite{Green:1998by}.
At general order in the momentum expansion the requirement is that the full effective action
be invariant under the modified supersymmetry transformation with spinor parameter $\epsilon$
acting on any field $\Phi$ is
\be
\label{susymod}
\delta\, \Phi = \left(\delta^{(0)} + {\a'}^3 \, \delta^{(3)} + {\a'}^5 \, \delta^{(5)}
+ \dots\right)\, \Phi\, ,
\ee
where $\delta^{(0)}$ is the classical supersymmetry transformation
and $\delta^{(n)}\, \Phi$
denotes the modified transformation at $O({\a'}^n)$.  Invariance of the modified action, $
 (\alpha')^4\, S = S^{(0)} + \alpha'\,  S^{(1)} + \ldots + (\alpha')^n \, S^{(n)} +
\ldots$ (where $S^{(n)}$ is the action at order ${\alpha'}^n$) requires
\be
 \left( \sum_{m=0}^r {\alpha'}^m \delta^{(m)} \right) \sum_{n=o}^r
{\alpha'}^n S^{(n)} = 0 ,
\label{invaract}
\ee
Furthermore, the modified supersymmetry transformations must form a closed algebra when acting
on $\Phi$, modulo terms proportional to the modified $\Phi$ equation of motion and local symmetry transformations.
This means that the commutator
of two supersymmetry transformations with spinorial parameters $\epsilon_1$ and $\epsilon_2$ is given by
\be
\left[\delta_1\,,\delta_2\right]\, \Phi = -2{\rm Im} (\bar\epsilon_2\gamma^\mu\epsilon_1)\,
\partial_\mu\Phi + \Phi\ {\rm eqn. \ of\ motion}+ \delta_{local}\, \Phi\,,
\label{modeq}
\ee
where $\gamma_\mu$ is a Dirac Gamma matrix for the ten-dimensional theory, the second term is proportional to an
equation of motion and the third term represents local symmetry transformations.

In \cite{Green:1998by} these equations at order  ${\alpha'}^3$
were used to determine that the ten-dimensional type IIB $\calR^4$ coefficient  satisfies a Laplace
eigenvalue equation of the form (\ref{lapeig}) that has as solution
the modular function $\calE_{(0,0)}=E_{3/2}$.
   A similar argument at $O({\a'}^5)$ involving $\delta^{(5)}$
determines the modular function $\calE_{(1,0)} =E_{5/2}/2$
\cite{sinha}.  Similarly,
the form of the Poisson equation with a quadratic source term (\ref{poiss}) that
determines $\calE_{(0,1)}=\calE_{(3/2,3/2)}/6$ is at least in qualitative accord with supersymmetry
at $O({\a'}^6)$ \cite{gv:D6R4}. However, in  this case, not only do the classical supersymmetries
mix the ${\alpha'}^6\, S^{(6)}$ with the $O({\a'}^6)$
supersymmetry transformations, $\delta^{(6)}$, but there is also mixing with the $O({\a'}^3)$
variations, $\delta^{(3)}$, of the terms in $S^{(3)}$,
\be
\label{alphap6}
\delta^{(6)}\, S^{(0)} + \delta^{(3)}\, S^{(3)}+ \delta^{(0)}\, S^{(6)}  =0\,,
\ee
as well as in the closure of the algebra, where we require (ignoring detailed coefficients)
\be
[\delta_1^{(0)},\, \delta_2^{(6)}] + [\delta_1^{(6)},\, \delta_2^{(0)}] + [\delta_1^{(3)},\, \delta_2^{(3)}]
= 0 + \frac{\delta S^{(6)}}{\delta \Phi^*} + \delta_{local}\, \Phi\,.
\label{sixalg}
\ee
We may refer to terms such as $\delta^{(3)}\, S^{(3)}$ and their generalizations at higher order as
`intermediate mixing terms'. These are terms of intermediate order in $\alpha'$ that mix
with the $\delta^{(0)}$ (i.e., classical) variation of a higher-order term.
The detailed analysis of these constraints is very
cumbersome and has not been carried out.  However, the structure of
 (\ref{alphap6}) and (\ref{sixalg}) is
just what is needed for the coefficient function $\calE_{(0,1)}$ to satisfy a Poisson equation with a source
term that is proportional to $E_{3/2}E_{3/2}$ arising from the presence of the contributions from  intermediate mixing,
$\delta^{(3)}\, S^{(3)}$ and $[\delta_1^{(3)},\, \delta_2^{(3)}]$.

More generally, at order ${\alpha'}^p$ the modified supersymmetry conditions,
\be \sum_{k=0}^p \delta^{(p-k)}\, S^{(k)}=0\,,
\label{gensusy}
\ee
mix all terms at orders $k\le p$.  The Poisson equations can, in general, have a number of distinct source terms that are
quadratic in different lower order terms, as we have seen.  There may also be degeneracies in which several terms
of the same order mix under supersymmetry.

%%%%%%%%%%%%%%%%%%%%%%%%%%%%%%%%%%%%%%%%%%%%%%%%%%%%%%%%%%5
\subsection{Systematics of the nine-dimensional amplitude}

These arguments suggest how the pattern might continue to higher derivatives.
The general structure should involve Poisson equations with quadratic source terms that
are determined by commuting two supersymmetries.  Each factor that appears in the
source is itself a modular function associated with a lower-order interaction or modified supersymmetry
transformation.
The fact that the source terms in the Poisson equations found in section~\ref{sec:ttwo} should
be consistent with supersymmetry should therefore provide information concerning classes of terms
in the nine-dimensional amplitude.

We can illustrate this in a very schematic manner by listing the subset of terms required to
reproduce the Poisson equations that we earlier obtained by analyzing $L=2$ diagrams of eleven-dimensional
supergravity.  In the language of the effective action, and
ignoring coefficients, the effective action contains the following terms,
\be
S^{(9)} = S^{(9)}_{subset} + S^{(9)}_{rest}\,,
\label{twosets}
\ee
where $S^{(9)}_{subset}$ is a subset of terms of the form $D^{2k}\, \calR^4$
that will mix with each other under the intermediate supersymmetries, such as $\delta^{(3)} S^{(3)}$
in (\ref{alphap6}) and its higher order generalizations.  The following set of terms is needed
\bea
S_{subset}^{(9)} &=& \int d^9x \sqrt{-G^{(9)}}\, r_B\, \bigg( R+ {\a'}^3 E_{3\over 2}\,\calR^4+
{\a'}^4\, r_B^{-2}\, E_{1\over 2}\, D^2\calR^4\non\\
&+&{\a'}^5\, (E_{5\over 2} + r_B^{-4}\, E_{3\over 2})\, D^4\calR^4+{\a'}^6 (\calE_{(0,1)}^{(0)}
+ r_B^{-6}\,E_{5\over 2})\, D^6\calR^4\non\\
&+& {\a'}^7\, (r_B^{-2}\,\calE_{(2,0)}^{(2)} + r_B^{-8}\,E_{7\over 2})\,D^8\calR^4+{\a'}^8
(r_B^{-4}\, \calE_{(1,1)}^{(4)}+ r_B^{-10}\, E_{9\over 2})\, D^{10}\calR^4
\non\\
&+& {\a'}^9(r_B^{-6}\calE_{(3,0)}^{(6)} + r_B^{-12}\, E_{11\over 2})\, D^{12}\calR^4\bigg)\, ,
\label{nineffact}
\eea
where $D^{12}\calR^4$ stands  for both kinematic structures $\sigma_2^3\calR^4$ and $\sigma_3^2 \calR^4$
(and $G^{(9)}$ is the metric in the nine-dimensional space transverse to the torus).
The coefficient functions  are various modular functions, including some
that have been discussed in this and earlier papers. We have included
the interaction $r_B^{-1}\, E_{1/2}\, D^2\calR^4 \sim r_B^{-1}\, E_{1/2}\, \sigma_1\,\calR^4$, where
 $\sigma_1 = S+T+U$,  even though it vanishes on shell when the dilaton is constant, because
it is important for the structure of $\a'$--corrected supersymmetry transformations.  In considering the
supersymmetry variations of the fields in the action we need to consider general infinitesimal transformations
(that are not on-shell).
This is the $k=1$ term in  the series of terms,
$r_B^{1-2k}\, E_{k-1/2}\, D^{2k}\calR^4$, that
arises from $L=1$ supergravity on $\calT^2$ \cite{Russo:1997mk,gkv:twoloop}.

The remaining terms, which are contained in $S^{(9)}_{rest}$, include a host of further contributions that mix with $S^{(9)}_{subset}$
under both the classical and higher-order supersymmetry transformations.  Such terms, which are not of the form
$D^{2k}\, \calR^4$ but involve the other fields in
the supergravity multiplet, generally carry nonzero $U(1)$ charge, $u$ (where $U(1)$ is the $R$-symmetry
of the IIB theory).  The moduli-dependent coefficients of terms of this type are modular forms that transform with
a phase under $SL(2,\ZZ)$ that compensates for the non-zero phase associated with the charge $u$.  An example of such
a term is $\calE_{(0,0)}^{-2}\, G^2\, R^3$, where $G$ is the complex type IIB three-form that carries unit $U(1)$ charge
\cite{Schwarz:1993} and the modular form $\calE_{(0,0)}^u$ is given by acting with a
$U(1)$-covariant derivative $u$ times
on the Eisenstein series $E_{3/2}$ \cite{Green:1998by}\footnote{The superscript $u$ was suppressed for the
coefficients $\calE_{(p,q)}$ of the  $U(1)$-conserving terms considered explicitly earlier in
this paper, which all have $u=0$.}
Such $U(1)$-violating interactions are not present in classical IIB supergravity
and are believed to arise in string theory only in $n$-point functions with $n>4$.

The double expansion in powers of $\a'$ and powers of $r_B^{-2}$ in (\ref{nineffact})
fits in with the general structure expected from supersymmetry.
Demanding supersymmetry at a given order ${\a'}^{6+p} r_B^{-2p}$ gives conditions that can
schematically be argued to associate modular functions with source terms as shown in the table.
\begin{table}[h]
\begin{center}
\begin{tabular}{|c|c||c|}
\hline
ORDER & COEFFICIENT & SOURCE\\
\hline
${\a'}^7 r_B^{-2}$&$\calE_{(2,0)}^{(2)}$ & $E_{\half}E_{\threeh}$\\ \hline
${\a'}^8 r_B^{-4}$ & $\calE_{(1,1)}^{(4)}$&$ E_{\threeh}E_{\threeh} + E_{\half}E_{\half}$\\ \hline
${\a'}^9 r_B^{-6}$& $\calE_{(3,0)}^{(6)}$ & $E_{\threeh}E_{\fiveh}+ E_{\half}E_{\threeh}$\\ \hline
\end{tabular}
\caption{Summary of source terms associated with the inhomogeneous Laplace equations for various
coefficient functions.}
\end{center}
\end{table}
In the first line the source arises from the presence of
${\alpha'}^3\,E_{3/2}\,\calR^4$ and ${\alpha'}^4\, r_B^{-2}\, E_{1/2}\,D^2\calR^4$ in (\ref{nineffact}),
together with their supersymmetric partners, which we have not determined.  The powers of
both $\alpha'$ and $r_B$ are such that these terms can
mix with the $\delta^{(0)}$ transformation of the $O({\alpha'}^7\, r_B^{-2})$ terms.
In the second line, the first source term comes from  ${\alpha'}^3\, E_{3/2}\, \calR^4$ with
${\alpha'}^5\,r_B^{-4}\, E_{3/2}\,D^4\calR^4$,
while the second source term comes from the  ${\alpha'}^4\,r_B^{-4}\, E_{1/2}\,D^2\calR^4$ (more precisely,
from the  term $\delta^{(4)}S^{(4)}$ in the supersymmetry
transformation at order $r_B^{-4}$).
In the third line the first source term comes from  ${\alpha'}^3\, E_{3/2}\,\calR^4$ and
${\alpha'}^6\, r_B^{-6}\, E_{5/2}\,D^6\calR^4$
while the second source term comes from ${\alpha'}^4\, r_B^{-2}\, E_{1/2}\, D^2\calR^4$ and
${\alpha'}^5\,r_B^{-4}\, E_{3/2}\, D^4\calR^4$.
In this manner we can see how the structure of the source terms in
the Poisson equations of section~\ref{sec:ttwo} arise.

These very sketchy arguments do not explain why the modular invariant coefficients in
(\ref{nineffact}) are generally {\it sums} of modular functions satisfying Poisson equations, as we have seen
in the examples derived from $L=2$ supergravity in this paper. This could well arise from the possible
degeneracies in terms that mix with each other under supersymmetry mentioned earlier, which obviously merits further study.

Finally, even the set of $D^{2k}\, \calR^4$ terms shown explicitly in $S^{(9)}_{subset}$ in (\ref{nineffact})
 is not complete. In the case of the lowest derivative terms,
 $\calR^4$, $\hat \sigma_2 \calR^4$ and  $\hat\sigma_3 \calR^4$
the complete coefficients can be deduced
by imposing T-duality on the expressions obtained by compactifying $L=1$ and $L=2$-loop
supergravity on a circle\footnote{Furthermore, the exact form of the coefficients of these terms
is known in eight dimensions, where they are $SL(3,\ZZ)\otimes SL(2,\ZZ)$-invariant functions
\cite{Kiritsis:1997em,Basu:2007ru,Basu:2007ck}. The exact nine-dimensional expression can therefore be
deduced by decompatifying these expressions.}.   The terms of higher order in $\alpha'$
have not been completed,  although T-duality, together with the tree-level and one-loop perturbative
string theory `data',
do lead to some very suggestive constraints on the missing terms.
However, we expect significant generalizations in the structure of the Poisson equations satisfied by the
coefficients of the higher order terms, and a complete determination
will almost certainly need an extension of the considerations of this
paper.

%%%%%%%%%%%%%%%%%%%%%%%%%%%%%%%%%
\subsection{Concluding remarks}
%%%%%%%%%%%%%%%%%%%%%%%%%%%%%%%%%

We have determined terms in the derivative expansion of type II superstring theory that
arise via duality
from compactification of two-loop ($L=2$) eleven-dimensional supergravity on a circle and on a two-torus up
to order $S^6\, \calR^4$.
In the case of the two-torus compactification these coefficients are sums of modular functions of the scalar fields,
satisfying   an intriguing set of Poisson equations
on moduli space with source terms that are bilinear in lower-order coefficients.  This is the
principle message of this paper.  The structure of these equations has a form that is in line with the expectations
based on implementing maximal supersymmetry.   Although the terms that we have determined in this manner are incomplete,
there are many intriguing correspondences with results directly obtained from string perturbation theory at tree-level and
genus one in nine and ten dimensions.  This structure should generalize to the larger moduli spaces that become
relevant upon compactification to lower dimensions.  Examples of this are the
 $SL(3,\ZZ)\otimes SL(2,\ZZ)$-invariant functions relevant
to the compactification on $\calT^3$ to eight dimensions that were mentioned in the previous footnote.

As emphasized in the introduction, supersymmetry guarantees that this structure should also apply to the low-energy
expansion of the four-particle amplitudes in which the external states are any of the 256 states in the supermultiplet.
These amplitudes conserve the $U(1)$ charge, $u$.  However, 
as we have discussed, the full nonlinear supersymmetry relates such processes 
to amplitudes with total $u\ne 0$, and should therefore provide interesting constraints on these $U(1)$ non-conserving
processes.  However, the analysis of the 
complete set of conditions implied by supersymmetry is far from complete.

All this suggests that the exact expressions for the moduli-dependent coefficients at higher orders in the 
low-energy expansion are given by duality-invariant functions that are
solutions of  generalizations of  the Poisson equations  obtained from
two-loop ($L=2$) supergravity (\ref{ipoisso}). 

%%%%%%%%%%%%%%%%%%%%%%%%%%%%%%%%%%%%%%%%%%%%%%%%%%%%%%%%%%%%%%% 
\section*{Acknowledgements}
We would like to thank Don Zagier, Nathan Berkovits and Sav Sethi for useful discussions.
P.V.  would  also like  to  thank the  Niels  Bohr  Institute for  the
hospitality during the completion of this work.
This work was partially supported by the RTN contracts MRTN-CT-
2004-503369, MRTN-CT-2004-512194 and MRTN-CT-2004-005104, the ANR grant
BLAN06-3-137168, MCYT FPA 2007-66665 and by  NORDITA.

%%%%%%%%%%%%%%%%%%%%%%%%%%%%%%%%%%%%%%%%%%%%%%%%%%%%%%%%%%
\appendix
%%%%%%%%%%%%%%%%%%%%%%%%%%%%%%%%%%%%%%%%%%%%%%%%%%%%%%%%%%

%%%%%%%%%%%%%%%%%%%%%%%%%%%%%%%%%%%%%%%%%%%%%%%%%
\section{Properties of the integrands $B_{(p,q)}$}\label{sec:Modular}
%%%%%%%%%%%%%%%%%%%%%%%%%%%%%%%%%%%%%%%%%%%%%%%%%

In this appendix we will describe properties of the functions $B_{(p,q)}$ that enter in the integrands
of the coefficients $I_{(p,q)}$ in (\ref{e:DefIn}).
The coefficients $I_{(1,0)}$ and $I_{(0,1)}$ were computed in
\cite{gkv:twoloop,gv:D6R4}, respectively. The higher order coefficients
of interest here are $I_{(2,0)}$, $I_{(1,1)}$, $I_{(3,0)}$, $I_{(0,2)}$.
Recall that the functions $B_{(p,q)}$ are proportional to the functions $\tilde B_{(p,q)}$
 that enter into the expansion of the integrand in
(\ref{funcint}),
\be
B_{(p,q)}=d_{(p,q)}\, \tilde B_{(p,q)}\,,
\label{ddefnew}
\ee
where the coefficients $d_{(p,q)}$ are arbitrarily chosen integers that avoid the
occurrence of unwieldy coefficients in the main equations.  The values of $d_{(p,q)}$ of relevance to
the examples in this paper were given in (\ref{bbtildef}).

After mapping the integrand from the domain in figure~\ref{fig:ModularRegions}(a) to
figure~\ref{fig:ModularRegions}(b)  the functions $B_{(p,q)}(\tau_1,\tau_2)$ are manifestly invariant under the
transformation $\tau_1 \to -\tau_1$, which is equivalent to the symmetry $\tau\to 1-\tau^*$ in the original region.
This means that the dependence on $\tau_1$
enters via the combination
\be
T_1 = -\tau_1^2+|\tau_1|\, ,
\label{betadeff}\ee
and there is a discontinuity in $\partial_{\tau_1}$ at $\tau_1=0$.
The coefficient of the $\sigma_2\,\calR^4\sim \calD^4 \calR^4$ term is simply $B_{(1,0)}(\tau)=1$ \cite{gkv:twoloop}.
In this notation the coefficient of the $\sigma_3\,\calR^4\sim \calD^6 \calR^4$ term \cite{gv:D6R4} in (
\ref{bonedef}) and (\ref{bonedefb}) is given by
\be
B_{(0,1)}(\tau) = \tau_2  + {1-6T_1\over \tau_2} + {5T_1^2\over \tau_2^3}
\,.
\label{bonedefnew}
\ee
We will here show that the higher order functions $B_{(p,q)}(\tau)$ are given by sums of the form\footnote{We  would  like to  thank  Don  Zagier  for
  explaining us the mathematical significance of this decomposition}
\be
B_{(p,q)}(\tau) = \sum_{ i=0}^{\lceil \threeh N\rceil}\, b_{(p,q)}^{3N-2i}(\tau)\,,
\label{modsum}
\ee
where $N=2p+3q-2$ and $b_{(p,q)}^{i}$
satisfies a Poisson equation with delta function source of general structure
\be
\Delta b_{(p,q)}^{i}(\tau) =
i(i+1)
\, b_{(p,q)}^{i}(\tau)  -\tau_2\,
c_{(p,q)}^{i}(\tau_2)\, \,\delta(\tau_{1})\,,
\label{poissoeq}
\ee
where $c_{(p,q)}^i(\tau_2)$ is a polynomial of order $N-1$ in $\tau_2+\tau_2^{-1}$.
The index $i$ takes values $\lceil 3N/2\rceil$.
The range of the summation index in (\ref{modsum}) is
determined by the powers of $1/\tau_{2}$ in the expansion of
$B_{(p,q)}$ which has the general form
\be
B_{(p,q)} (\tau)      =\sum_{i=0}^{2N}       \,      q_{2i}(|\tau_1|)\,
\tau_2^{N-2i}
\label{bngen}
\ee
where $q_{2i}(|\tau_1|)$ are polynomials of degree $i$
 in $T_1$.
 The highest inverse power of $\tau_2$   in    this   sum   is   given   by    a
 constant   times $T_1^{2N}\,\tau_2^{-3N}$.

An important feature for later considerations is that
$q_{2}(|\tau_1|)=q_2^{(0)}\,  (1-6T_1)/6$      where
$q_2^{(0)}$ is  a constant.  Since
\be
\int_{-\half}^\half d \tau_1 (1-6T_1)=0\, ,
\label{importlat}
\ee
it follows that the zero mode with respect to $\tau_1$ satisfies
\be
\int_{-\half}^\half d\tau_1 \, (B_{(p,q)} - q_0\tau_2^{2p+3q-2}) = O(\tau_2^{2p+3q-5})\,.
\label{zermod}
\ee

In the following subsections
we will present the rather unwieldy complete expressions for the $B_{(p,q)}$ functions up to order $N=2p+3q-2
=4$ of interest in this paper, which result from computer evaluations.  However, it is worth noting two general
features of these functions that are straightforward to derive to all orders.

Firstly, for the special value $\tau_1=0$ (or $L_2=0$ in terms of the original Schwinger parameters),
only the planar diagrams contribute to the amplitude
and the integrals over the vertex positions $t_r$  can be  computed explicitly.   The result is
 \be
\tilde     B_{(p,q)}(L_2/L_1=0,,L_3/L_1)=    \alpha_{p,q}    \sum_{k=0}^N
c(k)c(N-k) \ \tau_2^{N-2k}\,,
\label{tauonezero}
\ee
where
\be
\alpha_{p,q}= {N!\,(N+2)(p+q-1)!\over  p! q! 2^p 3^q}\ ,\qquad    c(k)=\frac{\sqrt{\pi } }{2^{2 k+1}(k+1)
\Gamma \left(k+\frac{3}{2}\right)}\, .
\label{tauonecoeff}
\ee
The coefficient $\alpha_{p,q}$ arises from the conversion of $\sigma_{N+2}$ to $\sigma_2^p\sigma_3^q$
using the identity \cite{gv:stringloop}
\be
\sigma_n = n\, \sum_{2p+3q=n}{(p+q-1)!\over  p! q! 2^p 3^q}\, \sigma_2^p\sigma_3^q\,,
\label{sigident}
\ee
while the coefficients $c(k)$ come from further combinatorics in the expansion of the integrand (\ref{funcint}).

Secondly, for arbitrary values of the Schwinger parameters, $L_k$, the leading terms in the expansion of
$\tilde B_{(p,q)}$ for large $\tau_2=\Delta^{\half}/(L_1+L_2)$ are
\be
\tilde B_{(p,q)}(L_2/L_1,L_3/L_1)=
\alpha_{p,q} \left( a_{N} \ \tau_2^{N}+ b_{N}\big( 1-6T_1\big)
\tau_2^{N-2}+O\big(\tau_2^{N-4}\big) \right)\,,
\label{leadingtau}
\ee
with
\be
a_N = {\sqrt\pi\over 2^{2 N+1} (N+1)\,\Gamma \left(N+ {3\over
      2}\right)}\ ,\qquad b_N = {\sqrt\pi\,
      (N+1)\Gamma(N-1)\over 3\cdot 2^{2 N+1}  \ \Gamma \left(N+ {1\over 2}\right)}\,.
\label{leadingcoeff}
\ee
%%%%%%%%%%%%%%%%%%%%%%%%%%%%%%%%%%%%%%%%%%
\subsection{Properties of $B_{(2,0)}$}\label{sec:ModD8R4}
%%%%%%%%%%%%%%%%%%%%%%%%%%%%%%%%%%%%%%%%%%%

The modular function  $B_{(2,0)}$ that enters the $\sigma_2^2\,\calR^4\sim\calD^4\calR^4$ interaction has the form
\begin{equation}
B_{(2,0)}     =   {4\over5}\,     \tau_2^2+(1  -
  6\,T_1)    +   {2\over5}\, \frac{2    -   15\,T_1    +
    40\,T_1^2}{\tau_2^2}\
+  {2\over5}\, \frac{T_1^2\,\left(
      11     -    43\,T_1     \right)     }{\tau_2^4}    +
  {32\over5}\,\frac{T_1^4}{\tau_2^6}\ .
   \label{b22def}
   \end{equation}

We will now describe the iterative process for writing $B_{(2,0)} = \sum_{i=0}^6
b_{(2,0)}^{2i}$, where each of the functions
$b_{(2,0)}^{2i}$ satisfies a Poisson equation with delta function
source of the form given in~(\ref{poissoeq}).  The procedure will be the same in the
cases with $N>2$.
First    consider  the    action    of   the    laplacian
$\Delta_{\tau}=\tau_{2}^2\,(\partial_{\tau_{1}}^2+\partial_{\tau_{2}}^2)$
on   a   function    of   the   form   $q_n(|\tau_1|)/\tau_2^r$   with
$q_n(|\tau_1|)$ polynomials of degree $n$ in  the decomposition
of the $B_{(2,0)}$ in (\ref{bngen}).
The action of the laplacian gives two types of
contributions
\be\label{ediffone}
\Delta_{\tau}{q_n(|\tau_1|)   \over\tau_2^r}   =r(r+1)  {q_n(|\tau_1|)
\over\tau_2^r}+ {q_n^{\prime\prime}(|\tau_1|)\over\tau_2^{r-2}}\ .
\ee
The first  contribution is proportional to the  original function times
an `eigenvalue' determined by the power of $\tau_2$. The
second contribution is of the same type as the original function but
with the power  of   $\tau_2$  increased   by  $2$   and  the numerator is a  polynomial
$q^{\prime\prime}_n(|\tau_1|)$ of degree $n-2$.  The linear term
 $|\tau_1|$  in $q_n$  contributes to the $\delta(\tau_1)$ source in the Poisson
equation using $\partial_{\tau_1}^2 |\tau_1| = 2\delta(\tau_1)$. Splitting off
 this contribution by writing
\be
 q_n^{\prime\prime}(|\tau_1|) = \hat q_n(|\tau_1|) +q_n^{(1)}\, \delta(\tau_1)
\ee
one finds that (\ref{ediffone}) can be rewritten as
\be\begin{split}
\Delta_\tau    \left({q_n(|\tau_1|)   \over\tau_2^r}    +  {\hat
q_n(|\tau_1|) \over(4r-1)\,\tau_2^{r-2}} \right)=& r(r+1)\left({q_n(|\tau_1|)   \over\tau_2^r}    +  {\hat
q_n(|\tau_1|) \over (4r-1)\,\tau_2^{r-2}} \right)\cr
&+{\hat q^{\prime\prime}_n\over(4r-1)\, \tau_2^{r-4}}+
{q_n^{(1)}\over\tau_2^{r-2}}\delta(\tau_1)\, .
\end{split}\ee
By iterating this procedure until the degree of the polynomial in
$|\tau_1|$  is $1$  or  $0$, one
can construct an eigenfunction of the laplacian $\Delta_\tau$
together with a delta function source term.  This defines the function $b_{(p,q)}^{3N}$
that contains the most negative power, $\tau_2^{-3N}$.   After subtracting this function
from $B_{(2,0)}$, the most negative remaining power is $\tau_2^{-3N+2}$ and the
above procedure may be repeated to determine the function $b_{(p,q)}^{3N-2}$, and so on until
the complete set of functions has been determined.

Applying this procedure to  $B_{(2,0)}$ leads to a sum of the following $b_{(2,0)}^i$ functions,
\begin{eqnarray}
b_{(2,0)}^{0}(\tau)&=&-{13\over21}\nn\\
  b_{(2,0)}^{2}(\tau)&=&{10\over21}\left(\tau_2^2+ 1-2T_1
     +   \frac{(1-T_1)^2}{ \tau_{2}^2}\right)\nn\\
b_{(2,0)}^{4}(\tau)&=& {10\over 77}\Big(\tau_2^2+\frac{3}{5}\, (4-15T_1)
     +   \frac{1-9T_1+15T_1^2}{ \tau_{2}^2}
+
 7T_1^2\, \frac{1-T_1}{{\tau_2}^4}\Big)  \nn \\
 b_{(2,0)}^{6}(\tau)&=&{32\over165}\Big( {\tau_2}^2+\frac{10}{7}(3-14T_1)+ \frac{1-20T_1+70T_1^2}{{\tau_2}^2} \nn \\
&+& 6T_1^2\, \frac{3-14T_1}{ {\tau_2}^4} +
  \frac{33\, T_1^4}{ {\tau_2}^6} \Big) \, .
\label{eca}
\end{eqnarray}
These functions satisfy the inhomogeneous Laplace equations for $r=0,2,4,6$
\begin{equation}
\Delta b_{(2,0)}^{r}(\tau) =r (r+1)\, b_{(2,0)}^r(\tau) -2u_r\,\tau_2\,
(\tau_{2}+\tau_2^{-1}) \,\delta(\tau_{1})\, ,
\label{b2ieq}
\end{equation}
where $u_0=0$ and, for $r=2,4,6$,
\be
u_r={1\over 4} q_r\, (r(r+1) -2) =({10\over 21},{45\over 77},{64\over 33})\ , \qquad q_r=({10\over 21},
{10\over 77},{32\over 165})
\label{qdefs}
\ee
The value $u_r$ can be computed by
\be
\partial_{\tau_1} b_{(2,0)}^{r}\bigg|_{\tau_1=0} =-2u_n (1+\tau_2^{-2})
\ee

%%%%%%%%%%%%%%%%%%%%%%%%%%%%%%%%%%%%%%%%%%
\subsection{Properties of $B_{(1,1)}$}\label{subsec:modten}
%%%%%%%%%%%%%%%%%%%%%%%%%%%%%%%%%%%%%%%%%%%
The modular function  associated the coefficient of $\sigma_2\sigma_3\,\calR^4\sim\calD^{10}\calR^4$
is given by
\begin{equation}\begin{split}
B_{(1,1)}= & \frac{45\,\tau_2^3}{2} + 35\,\left( 1 - 6\,T_1 \right)  \,\tau_2+
\frac{7\,\left( 10 - 75\,T_1 + 191\,T_1^2 \right) }{2\,\tau_2}+
  \frac{45   -  420\,T_1   +  1372\,T_1^2   -
    2086\,T_1^3}{2\,\tau_2^3} \cr
&+ \frac{T_1^2\,\left( 285 - 1264\,T_1 + 1761\,T_1^2 \right) }{2\,\tau_2^5}
+ \frac{T_1^4\,\left( 347 - 782\,T_1 \right) }{2\,\tau_2^7}+\frac{145\,T_1^6}{2\,\tau_2^9}
\end{split}\label{b3detail}\end{equation}
Following the previous iterative procedure
this function can straightforwardly be shown to be a sum of five modular functions
$B_{(1,1)}=\sum_{j=0}^4\,b_{(1,1)}^{2j+1}$
that again satisfy Poisson equations with delta-function source terms,
\bea
b_{(1,1)}^1&=&-\frac{245}{66}\, \left( \tau_2+ {1 - T_1 \over \tau_2} \right)\\
b_{(1,1)}^3&=&-\frac{7}{429}\, \Big( - 679\, \tau_2^3 + 3\, \left(
   176 + 679\, T_1 \right) \, \tau_2+ {-679 + 2037\, T_1 +
          4677\, T_1^2 +
            679\, T_1^3 \over \tau_2^3}\\
\nn& -& \frac{3\, \left( -176 + 1735\,
              T_1 + 679\, T_1^2 \right) }{\tau_2}\Big)\\
\nn b_{(1,1)}^5&=&{49\over39}\,\Big( 7\,\tau_2^3- 12\,\left( -2 + 7\,T_1 \right) \,\tau_2 +
  \frac{6\,\left( 4 - 25\,T_1 + 35\,T_1^2 \right) }{\tau_2} - \frac{7\,\left( -1 + 12\,T_1 - 36\,T_1^2 + 28\,T_1^3 \right) }{\tau_2^3} \\
 &+&
\frac{63\,{\left( -1 + T_1 \right) }^2\,T_1^2}{\tau_2^5}\Big)\\
b_{(1,1)}^7&=&-{1862\over7293}\,\Big(- 9\,\tau_2^3 +
  5\,\left( -11 + 45\,T_1 \right) \,\tau_2
 -  \frac{5\,\left(  11   -  98\,T_1  +  210\,T_1^2  \right)
}{\tau_2}
 \\
\nn&+& \frac{9\,\left( -1 + 25\,T_1 - 140\,T_1^2 + 210\,T_1^3 \right) }{\tau_2^3}
-  \frac{33\,T_1^2\,\left( 6 - 38\,T_1 + 45\,T_1^2 \right) }{\tau_2^5}
+\frac{429\,\left( -1 + T_1 \right) \,T_1^4}{\tau_2^7} \Big)\\
\nn b_{(1,1)}^9&=&{1\over4862}\Big( 11172\,\tau_2^3 -
  \frac{18620\,\left( -11 + 45\,T_1 \right) \,\tau_2}{3} +   \frac{18620\,\left(  11  -   98\,T_1  +   210\,T_1^2  \right)
 }{3\,\tau_2}\\
\nn& -&
  \frac{11172\,\left(  -1  +  25\,T_1  -  140\,T_1^2  +  210\,T_1^3  \right)
  }{\tau_2^3}
 +  \frac{40964\,T_1^2\,\left( 6 - 38\,T_1  + 45\,T_1^2 \right)
}{\tau_2^5}
\\
 &-&\frac{532532\,\left(  -1  + T_1  \right)
  \,T_1^4}{\tau_2^7} +\frac{352495\,T_1^6}{\tau_2^9}\Big)  \,.
\eea
The inhomogeneous  Laplace equations satisfied by  these functions are
given by, for $r=1,3,5,7,9$
\be
\Delta_\tau b_{(1,1)}^r =r(r+1)\,  b_{(1,1)}^r-  2\tau_2\big(v^{r} (\tau_2^2+
\tau_2^{-2})
+w^r\big)\ \delta(\tau_1 )\,,
\label{ecb}
\ee
where the constants $v^r$ and $w^r$ are given by
\bea
\nn v^1&=&0\ ,\ \ \ v^3={14\cdot 679\over 143}\ ,\ \ \ v^5={196\cdot
14\over 13}\ ,\ \ \
v^7={18620\cdot 45\over 7293}
\ ,\ \ \  v^9={6090\cdot 11\over 2431}\, ,\\
w^1&=&-{245\over 33} \ ,\qquad w^3=-{14\cdot 1735\over 143}\ ,\qquad
w^5=
{196\cdot 25\over 13} \,,\\
\nn w^7&=&{18620\cdot 98\over 7293} \ ,\qquad w^9={6090\cdot 30\over 2431}\,.
\eea

%%%%%%%%%%%%%%%%%%%%%%%%%%%%%%%%%%%%%%%%%%
\subsection{Properties of $B_{(3,0)}$ and $B_{(0,2)}$}\label{subsec:modtwelve}
%%%%%%%%%%%%%%%%%%%%%%%%%%%%%%%%%%%%%%%%%%%
The integrands  that define the  coefficients $I_{(3,0)}$, $I_{(0,2)}$
of the $S^6\,\calR^4$ contributions
$\sigma_2^3\,\calR^4$ and $\sigma_3^2\, \calR^4$, respectively, are
\begin{equation}\begin{split}
    B_{(3,0)}=&   24\,\tau_2^4+  45\,\left(  1   -  6\,T_1
    \right) \,\tau_2^2 + 4\,\left(  14 - 105\,T_1 + 270\,T_1^2
    \right)  + \frac{3}{\tau_2^2} \,\left( 15 - 140\,T_1 +
      462\,T_1^2 - 756\,T_1^3 \right)\cr
 &  +  \frac{6
    }{\tau_2^4}\,\left(   4   -   45\,T_1   +   190\,T_1^2   -
      390\,T_1^3   +    558\,T_1^4   \right)   +
    \frac{3\,T_1^2}{\tau_2^6}\,\left(   58  -  334\,T_1  +
      715\,T_1^2 - 1402\,T_1^3 \right) \cr
  &   +
    \frac{24\,T_1^4   }{\tau_2^8}\,\left(  10   -   41\,T_1  +
      167\,T_1^2             \right)         +
    \frac{24\,T_1^6}{\tau_2^{10}}\,\left( 7  - 93\,T_1 \right)
    + \frac{516\,T_1^8}{\tau_2^{12}}
\end{split}\label{b4xdef}\end{equation}
and
\begin{equation}\begin{split}
B_{(0,2)}&=24\,\tau_2^4 + 45\,\left( 1 - 6\,T_1 \right) \,\tau_2^2+14\,\left( 4 - 30\,T_1 + 75\,T_1^2 \right)
 + \frac{  45  - 420\,T_1 +  1358\,T_1^2  - 1904\,T_1^3 }{\tau_2^2}\,\cr
&+
  \frac{6}{\tau_2^4}\,\left( 4 -  45\,T_1 + 185\,T_1^2 - 330\,T_1^3  + 197\,T_1^4 \right)
+     \frac{T_1^2    \,\left(    174     -
  942\,T_1        +        1265\,T_1^2      +
  1418\,T_1^3 \right) }{\tau_2^6}\cr
&+
  \frac{2T_1^4}{\tau_2^8}\,\left( 105 - 96\,T_1 - 1529\,T_1^2 \right)
 - \frac{76\,T_1^6 }{\tau_2^{10}}\,\left( 1 - 27\,T_1 \right)
 - \frac{494\,T_1^8}{\tau_2^{12}}
\end{split}\label{b4ydef}\end{equation}

The functions $B_{(3,0)}$ and $B_{(0,2)}$ are each given by a sum of seven  functions
 $b_{(3,0)}^{2k}$ and $b_{(0,2)}^{2k}$ with $k=0,\dots,6$,
 which satisfy Poisson equations, The detailed form of these functions
 is
straightforward to determine using the iterative process described earlier, giving
\bea
b_{(3,0)}^0&=&{12264\over715}\,,
\eea
\bea
b_{(3,0)}^2&=&-{2408\over143}\,\big( \tau_2^2+1  - 2\,T_1 +  \frac{\left( 1  - T_1
        \right) ^2}{\tau_2^2}\big)\,,
\eea
\bea
\nn b_{(3,0)}^4&=&{42\over12155}\,\Big(3915\,\tau_2^4 - 20\,\left( -181 + 783\,T_1 \right) \,\tau_2^2+3\,\left( 547 - 3030\,T_1 + 7830\,T_1^2
\right) \\
\nn& -&
  \frac{10\,\left( -362 + 909\,T_1  - 732\,T_1^2 + 1566\,T_1^3 \right)
  }{\tau_2^2}\\
&+&  \frac{5\,\left( -1  + T_1 \right)  \,\left( -783 +  2349\,T_1 +
    413\,T_1^2 + 783\,T_1^3 \right) }{\tau_2^4}  \Big)\,,
    \eea
    \bea
 b_{(3,0)}^6&=&-{1\over 3553}\Big( - 20322\,\tau_2^4 + 5\,\left( -10889 + 60966\,T_1 \right) \,\tau_2^2-10\,\left( -1827 - 7280\,T_1 +
  101610\,T_1^2  \right) \nn
\\
\nn& +&
  \frac{5\,\left(   -10889   +   14560\,T_1   +   91294\,T_1^2   +
      284508\,T_1^3 \right) }{\tau_2^2}\\
\nn & -&
  \frac{6\,\left(  3387  -  50805\,T_1  + 112530\,T_1^2  +  152260\,T_1^3  +
      152415\,T_1^4 \right) }{\tau_2^4}\\
 &+&  \frac{33\,T_1^2\,\left(  -6774 +  20322\,T_1  +
    13295\,T_1^2 + 6774\,T_1^3 \right) }{\tau_2^6}\Big)\,,
    \eea
\bea
\nn b_{(3,0)}^8&=&{2\over2717}\Big(
7920\,\tau_2^4- 3600\,\left( -17 + 66\,T_1 \right) \,\tau_2^2+
25200\,\left(  4   -  30\,T_1  +  55\,T_1^2
\right) \\
\nn& -&   \frac{3600\,\left(  -17  +  210\,T_1  - 798\,T_1^2  +  924\,T_1^3  \right)
  }{\tau_2^2}
+
  \frac{7920\,\left(  1   -  30\,T_1  +  225\,T_1^2   -  600\,T_1^3  +
      495\,T_1^4 \right) }{\tau_2^4}\\
 &- &\frac{102960\,T_1^2\,\left(  -1 + 2\,T_1  \right) \,\left(  2 -  12\,T_1 +
      11\,T_1^2 \right) }{\tau_2^6}
+ \frac{514800\,{\left( -1 + T_1 \right) }^2\,T_1^4}{\tau_2^8}
  \Big)\,,
  \eea
\bea
\nn b_{(3,0)}^{10}&=&{2\over1062347}\Big(- 756756\,\tau_2^4
 + 407484\,\left( -22 + 91\,T_1 \right) \,\tau_2^2-333396\,\left(  50  -  451\,T_1  +
  1001\,T_1^2   \right) \\
\nn&  +&
  \frac{407484\,\left( -22 + 369\,T_1 - 1881\,T_1^2 + 3003\,T_1^3 \right) }{\tau_2^2}\\
\nn& -&
  \frac{756756\,\left( 1  - 49\,T_1 +  567\,T_1^2 - 2310\,T_1^3  + 3003\,T_1^4
    \right) }{\tau_2^4}\\
\nn& +&  \frac{2270268\,T_1^2\,\left(  -15  +  215\,T_1  - 880\,T_1^2  +  1001\,T_1^3
    \right) }{\tau_2^6}\\
&-&
  \frac{12864852\,T_1^4\,\left( 15 - 87\,T_1 + 91\,T_1^2 \right) }{\tau_2^8}
  +  \frac{244432188\,\left(   -1  +   T_1  \right)
  \,T_1^6}{\tau_2^{10}}\Big)\,,
  \eea
\bea
\nn b_{(3,0)}^{12}&=&{2\over96577}\Big(16770\,\tau_2^4+
 - 40248\,\left( -7 + 30\,T_1 \right) \,\tau_2^2
+119196\,\left( 5 - 52\,T_1 + 130\,T_1^2
\right)\\
\nn&  -&
  \frac{281736\,\left(  -1  +  22\,T_1  - 143\,T_1^2  +  286\,T_1^3  \right)
  }{\tau_2^2}
+
  \frac{16770\,\left( 1 - 72\,T_1 + 1188\,T_1^2 - 6864\,T_1^3 + 12870\,T_1^4 \right) }{\tau_2^4}\\
 \nn&-&
  \frac{1140360\,T_1^2\,\left( -1 + 22\,T_1  - 143\,T_1^2 + 286\,T_1^3 \right)
  }{\tau_2^6}+
  \frac{2166684\,T_1^4\,\left(  5 -  52\,T_1 +  130\,T_1^2 \right)  }{\tau_2^8}\\
&-&  \frac{4333368\,T_1^6\,\left( -7
    + 30\,T_1 \right) }{\tau_2^{10}}+ \frac{24916866\,T_1^8}{\tau_2^{12}}\Big)\,,
\eea
and
\bea
b_{(0,2)}^0&=&{12264\over715}\,,
\eea
\bea
b_{(0,2)}^2&=&-{2128\over143}\,\left( \tau_2^2+1  - 2\,T_1 +  \frac{\left( 1  - T_1
        \right) ^2}{\tau_2^2} \right)\,,
        \eea
\bea
\nn b_{(0,2)}^4&=&{42\over12155}\,\Big( 3385\,\tau_2^4
- 20\,\left( -69 + 677\,T_1 \right) \,\tau_2^2
+3\,\left( -927 + 2630\,T_1 + 6770\,T_1^2 \right) \\
\nn&- &  \frac{10\,\left( -138 - 789\,T_1 +  1992\,T_1^2 + 1354\,T_1^3
  \right) }{\tau_2^2}\\
&+& \frac{5\,\left( -1 + T_1 \right) \,\left( -677 + 2031\,T_1 + 2807\,T_1^2 + 677\,T_1^3 \right) }{\tau_2^4}
 \Big)\,,
 \eea
\bea
\nn b_{(0,2)}^6&=&-{1\over 561}\Big( - 6090\,\tau_2^4+ 525\,\left( -43 +
  174\,T_1   \right)  \,\tau_2^2-350\,\left(   61  -   420\,T_1  +
  870\,T_1^2 \right)
\\\nn &+&
  \frac{525\,\left( -43  + 280\,T_1  - 574\,T_1^2  + 812\,T_1^3
  \right) }{\tau_2^2}
  -  \frac{210\,\left(  29  - 435\,T_1  +  1500\,T_1^2  -
    1200\,T_1^3 + 1305\,T_1^4 \right) }{\tau_2^4}
\\
&+&
   \frac{1155\,T_1^2\,\left( -58 + 174\,T_1    - 65\,T_1^2 + 58\,T_1^3 \right) }{\tau_2^6}
\Big)\,,
\eea
\bea
\nn b_{(0,2)}^8&=&{1\over143}\Big( 22\,\tau_2^4 - 10\,\left( -17 +
  66\,T_1 \right) \,\tau_2^2+
70\,\left( 4  - 30\,T_1 + 55\,T_1^2  \right)
\\\nn &  -&
  \frac{10\,\left( -17 + 210\,T_1 - 798\,T_1^2 + 924\,T_1^3 \right) }{\tau_2^2}+
  \frac{22\,\left( 1 - 30\,T_1  + 225\,T_1^2 - 600\,T_1^3 + 495\,T_1^4
    \right) }{\tau_2^4}\\
&-& \frac{286\,T_1^2\,\left(  -1  + 2\,T_1  \right)  \,\left(  2  - 12\,T_1  +      11\,T_1^2 \right) }{\tau_2^6} + \frac{1430\,
{\left(      -1 + T_1 \right) }^2\,T_1^4}{\tau_2^8}  \Big)\,,
\eea
\bea
\nn b_{(0,2)}^{10}&=&-{1\over55913}\Big( - 90948\,\tau_2^4+ 48972\,\left( -22 + 91\,T_1 \right) \,\tau_2^2
-40068\,\left( 50 - 451\,T_1 + 1001\,T_1^2 \right)
\\ \nn&+&
  \frac{48972\,\left( -22 + 369\,T_1 - 1881\,T_1^2 + 3003\,T_1^3
    \right) }{\tau_2^2}\\
\nn&-&
  \frac{90948\,\left( 1  - 49\,T_1 + 567\,T_1^2  - 2310\,T_1^3 +
      3003\,T_1^4 \right) }{\tau_2^4}
\\
&+&\frac{272844\,T_1^2\,\left( -15 + 215\,T_1 - 880\,T_1^2
    + 1001\,T_1^3 \right) }{\tau_2^6}\\ \nn&-&
  \frac{1546116\,T_1^4\,\left( 15 - 87\,T_1 + 91\,T_1^2 \right) }{\tau_2^8}   + \frac{29376204\,\left( -1 + T_1 \right) \,T_1^6}{\tau_2^{10}}
\Big)\,,
\eea
\bea
b_{(0,2)}^{12}&=&-{1\over391}\Big(  130\,\tau_2^4 -  312\,\left(  -7 +
  30\,T_1   \right)  \,\tau_2^2   +924\,\left(  5   -   52\,T_1  +
  130\,T_1^2 \right)\\
\nn &-&
  \frac{2184\,\left( -1 + 22\,T_1 - 143\,T_1^2 + 286\,T_1^3 \right) }{\tau_2^2} +
  \frac{130\,\left(  1 - 72\,T_1  + 1188\,T_1^2  - 6864\,T_1^3  + 12870\,T_1^4
    \right) }{\tau_2^4}\\ \nn&-&
  -\frac{8840\,T_1^2\,\left( -1 + 22\,T_1 - 143\,T_1^2 + 286\,T_1^3 \right) }{\tau_2^6}+
  \frac{16796\,T_1^4\,\left( 5 - 52\,T_1 + 130\,T_1^2 \right) }{\tau_2^8}
\\
\nn&- & \frac{33592\,T_1^6\,\left( -7 + 30\,T_1 \right) }{\tau_2^{10}}
+ \frac{193154\,T_1^8}{\tau_2^{12}}\Big)\,.
\eea
The inhomogeneous Laplace equations satisfied by these functions have the form
\be
\Delta_\tau b_{(p,q)}^r = r(r+1)\,  b_{(p,q)}^r - 2\tau_2 \,\big(\tau_2+\tau_2^{-1}\big) \big[
f_{(p,q)}^r (\tau_2^2+\tau_2^{-2})+g^r_{(p,q)} \big]\ \delta(\tau_1 )\,,
\label{xecb}
\ee
where
\bea
&&f_{(3,0)}^{12}={15\cdot 12384 \over 7429}\ ,\ \ f_{(3,0)}^{10}=-{91\cdot 74088  \over 96577}\ ,\ \
f_{(3,0)}^{8}={11\cdot 43200  \over 2717}\ ,\ \ f_{(3,0)}^{6}= {10\cdot 30483  \over 3553 }\ ,\ \non\\
&&f_{(3,0)}^{4}={174\cdot 756 \over 2431}\ ,\ \ \ \ f_{(3,0)}^{2}=f_{(3,0)}^{0}=0\ ,\non\\
&&g_{(3,0)}^{12}={62\cdot 12384 \over 7429}\ ,\ \ g_{(3,0)}^{10}=-{278\cdot 74088  \over 96577}\ ,\ \
g_{(3,0)}^{8}={24\cdot 43200  \over 2717}\ ,\ \ g_{(3,0)}^{6}=- {10\cdot 23203  \over
  3553 }\ ,\ \non\\
&&g_{(3,0)}^{4}=-{73\cdot 756 \over 2431}\ ,\ \ g_{(3,0)}^{2}=-{4816\over 143}\ ,\ \ \
g_{(3,0)}^{0}=0\ ,
\eea and
\bea
&&f_{(0,2)}^{12}=-{15\cdot 624 \over 391}\ ,\ \ f_{(0,2)}^{10}={91\cdot 4452  \over 5083}\ ,\ \
f_{(0,2)}^{8}={11\cdot 60  \over 143}\ ,\ \ f_{(0,2)}^{6}= {87\cdot 350  \over 187 }\ ,\
\non\\
&&f_{(0,2)}^{4}={84\cdot 1354 \over 2431}\ ,\ \ \ \ f_{(0,2)}^{2}=f_{(0,2)}^{0}=0\ , \non\\
&&g_{(0,2)}^{12}=-{62\cdot 624 \over 391}\ ,\ \ g_{(0,2)}^{10}={278\cdot   4452 \over 5083 }\ ,\ \
g_{(0,2)}^{8}={24\cdot 60 \over 143}\ ,\ \ g_{(0,2)}^{6}= {53\cdot 350  \over 187 }\ ,\
\non\\
&&g_{(0,2)}^{4}=-{84\cdot 2143 \over 2431}\ ,\ \ g_{(0,2)}^{2}=-{4256\over 143}\ ,\ \ \
g_{(0,2)}^{0}=0\ .
\eea

%%%%%%%%%%%%%%%%%%%%%%%%%%%%%%%%%%%%%%%%%%%%%%%%%
\section{Interactions from circle compactification to ten dimensions}
\label{sec:s1compact}
%%%%%%%%%%%%%%%%%%%%%%%%%%%%%%%%%%%%%%%%%%%%%%%%%

We will here evaluate the integrals $I_{(p,q)}^{(d=10)}$ for ${(p,q)}=(2,0), (1,1), (3,0), (0,2)$ for the
circle compactification that relates eleven-dimensional supergravity
to ten-dimensional type IIA string theory.  The method used to evaluate these
integrals is an extension of that  used for the $(0,1)$ case in \cite{gv:D6R4}, which
we will review in the appendix~\ref{sec:D6R4circle} ($I_{(1,0)}^{(d=10)}$ was evaluated in \cite{gkv:twoloop}).
First we will discuss some expressions that need to be evaluated in the course of the
calculations.

%%%%%%%%%%%%%%%%%%%%%%%%%%%%%%%%%%%%%%%%%%%%%%%%%
\subsection{Some basic sums}\label{sec:integrals}
%%%%%%%%%%%%%%%%%%%%%%%%%%%%%%%%%%%%%%%%%%%%%%%%%
In the course of these calculations we will encounter both analytic and nonanalytic terms, as discussed
in section~\ref{sec:infracut}.  The calculation will reduce to the evaluation of expressions of the form
\be
 \Sigma_\alpha (v,\Lambda,\chi)= \sum_{\hat m\in \mathbb{Z}} \int_{0}^{\Lambda^2}
 dx\,  x^{-\alpha}\, e^{-\pi \hat m^2 v x}\,e^{-{\chi f\over x}}\,.
\label{sumalphval}
\ee
in the limit $\chi\,v \ll 1$ and $\Lambda^2 v\to \infty$ with $v$ fixed,
where $f$ may be a function of $S_0$ and $T_0$, but its exact form will be irrelevant in the following
(since, in general,  we will not keep track of the scale of logarithmic thresholds). The regulating factor
$e^{-\chi f/x}$ is inserted to regulate the infrared logarithmic factor as in (\ref{reguexp}).

First consider the case $\alpha<1/2$.  In this case we can safely set $S=0$ in (\ref{sumalphval}),
giving
\bea
 \Sigma_{\alpha<\half}  =  \sum_{\hat m\in \mathbb{Z}} \int_{\0chi}^{\Lambda^2}
 dx\,x^{-\alpha}\, e^{-\pi \hat m^2 v x} = {\Lambda^{2-2\alpha}\over 1-\alpha} +
  2\Gamma(1-\alpha)\, \zeta(2-2\alpha)\, (\pi v)^{\alpha-1}
 \,,
\label{sumalphv}
\eea
where the $\Lambda$-dependence comes from the $\hat m=0$ term.  For later reference we
note
\bea
\Sigma_{-\half} = {2\over 3} \, \Lambda^3 + {1 \over \pi}\,\zeta(3)\, v^{-\threeh}\,,&&\qquad
\Sigma_{-\threeh} = {2\over 5} \, \Lambda^5 + {3\over 2\pi^2 }\,\zeta(5) \, v^{-\fiveh}\nn\\
\Sigma_0 &=& \Lambda^2 + {2 \over \pi}\,\zeta(2)\, v^{-1}
\,.
\label{exsum}
\eea
Now we consider the case $\alpha =1/2$,
\be
\Sigma_\half= \sum_{\hat m\in \mathbb{Z}} \int_{0}^{\Lambda^2 }
 {dx\over\sqrt x }\, e^{-\pi \hat m^2 v x}\,e^{-{\chi f\over x}}\,.
\label{sumfirst}
\ee
Here we cannot simply set $\chi =0$ since this leads to a  singular sum,
even though each term in the sum is finite,
\be
\Sigma_\half (\chi =0) = \sum_{\hat m\in \mathbb{Z}} \int_{0}^{\Lambda^2 }
 {dx\over\sqrt x }\, e^{-\pi \hat m^2 v x} = \sum_{\hat m\ne 0} {1\over \hat m}\,
  {\Gamma(\half)\over  \sqrt {\pi v}}+ 2\Lambda \sim
{2\over v^\half}\, \zeta(1)+ 2\Lambda \,.
 \label{sigmadeff}
 \ee
The presence of a divergence in the form of $\zeta(1)$ shows the importance
of keeping the factor of $e^{-\chi f/ x}$ in (\ref{sumfirst}).

Separating the $\hat m=0$ contribution and computing the integral we find
(for $Sfv\ll 1$, or $\alpha' s g_A^2 \ll 1$)
\bea
\Sigma_{\half} &=& 2\Lambda +\sum_{\hat m \neq 0}  \int_0^\infty {dx\over \sqrt{x}} e^{-\pi \hat m^2 v x -\chi f/x}
= 2\Lambda+  \sum_{\hat m\neq 0}
 {1\over \sqrt{v \hat m}}e^{-2\sqrt{\pi \chi f v}|\hat m|}
\non\\
&=& 2\Lambda -{2\over \sqrt{v}} \log \big( 1-e^{-2 \sqrt{\pi \chi f v}}\big)
\non
\\
&\cong & 2\Lambda  - {1\over \sqrt{v}} \log (4\pi \chi f v)
\label{sigres}
\eea
Notably, the scale of the logarithmic depends on $v=R_{11}^2$ but is independent of the
cutoff $\Lambda$.

Another special case that will be needed later is
\bea
\Sigma_1 &=&  \sum_{\hat m\in \mathbb{Z}} \int_{0}^{\Lambda^2}
 {dx\over x}\, e^{-\pi \hat m^2 v x}\, e^{-{\chi f\over x}}
\nn\\
 &=& \log(v\, \Lambda^2/C') - {2\pi^{-\half}}\, (\chi f)^{-\half}\,  v^{-\half}\,
\label{sigone}
\eea
(where $C'$ is independent of $v$ and $\chi$),
as can be checked by differentiating the first expression with respect to $\Lambda$ and with
respect to $S$.  The inverse power of $\chi$ will be ignored in the following as described in
subsection (\ref{sec:infracut}).

We now turn to consider $\alpha >1$.  The integrand of $\Sigma_\alpha$ is more singular at small $x$ so
\bea
 \Sigma_{\alpha>1} &=& \sum_{\hat m\in \mathbb{Z}} \int_{0}^{\Lambda^2}
 dx\,x^{-\alpha}\, e^{-\pi \hat m^2 v x}\,e^{-{\chi f\over x}}
 \nn\\
 &=&{1\over v^\half}\, \sum_{ m\in \mathbb{Z}} \int_{0}^{\Lambda^2}
 {dx\over x^{\half} }\,x^{-\alpha}\, e^{-\pi  m^2/ v x}\, e^{-{\chi f\over x}}\nn\\
 &=& {1\over v^\half}\,  \sum_{ m\in \mathbb{Z}} \int_{1/\Lambda^2}^{\infty}
 dw\, w^{\alpha-\threeh }\, e^{-\pi  m^2 w/v}\,e^{-\chi f w}
 \nn\\
 &=& {1\over v^\half}\, (\chi f)^{\half -\alpha}\, \Gamma(\alpha-\half)
 + {2\over v^\half}\,  \,\left({v\over \pi}\right)^{\alpha-\half}\, \Gamma(\alpha -\half)\, \zeta(2\alpha-1) \,,
\label{sumalphmore}
\eea
where we have performed a Poisson resummation to express the sum in terms of Kaluza--Klein integers $m$,
and separated the $m=0$ term, which is proportional to $(-S)^{\half -\alpha}$ and set $S=0$ in the terms with $m\ne 0$.
We note, in particular, that after again dropping negative powers of $\chi$,
\be
\Sigma_{\threeh} = {\pi v^\half\over 3}\,,
\quad\qquad \Sigma_{\fiveh} =  {\pi^2 v^{\threeh}\over 45}\, .
\label{examplesii}
\ee

%%%%%%%%%%%%%%%%%%%%%%%%%%%%%%%%%%%%%%%%%%%%%%%%
\subsection{Evaluation of $I_{(0,1)}^{(d=10)}$}\label{sec:D6R4circle}
%%%%%%%%%%%%%%%%%%%%%%%%%%%%%%%%%%%%%%%%%%%%%%%%

The integral $I_{(0,1)}^{(d=10)}$ of relevance to the $\sigma_3\,\calR^4$ interaction
decomposes into three distinct pieces of the form
\be
I_{(0,1)}^{(d=10)}={I_{(0,1)}^{(3)}\over v^3}
+ I_{(0,1)}^{(3/2)}\,{\Lambda^3\over v^{3/2}} +I_{(1,0)}^{(0)}\, \Lambda^6\, ,
\ee
where $v=R_{11}^2$.
The contribution $I_{(0,1)}^{(3)}$ is the finite part of the amplitude, which comes from
non-zero winding numbers, and  which was evaluated in
in section~4 of \cite{gv:D6R4}.   This corresponds
to the tree-level string contribution to the $\calS^{(3)}\calR^4$ term in the amplitude
(or the $\calD^6\calR^4$ interaction).
The $I_{(0,1)}^{(3/2)}$ term proportional to $\Lambda^3$ comes from a one-loop sub-divergence
that needs to be subtracted by the addition of the
 triangle diagram where one vertex is the $\calR^4$ one-loop
counterterm.  The $I_{(0,1)}^{(0)}$ term proportional to $\Lambda^6$ comes from a new two-loop
divergence that also needs to be subtracted by the addition of a local counterterm.

Each of these contributions satisfies a second order differential equation of the form,
\be
\big(v^2{\p^2\over \p v^2}+2v{\p\over \p v}\big){ I_{(0,1)}^{(\alpha)}\over v^{\alpha}}=
\alpha(\alpha-1)\, { I_{(0,1)}^{(\alpha)}\over v^{\alpha}}\, .
\label{gensoll}
\ee
Applying the operator on the left-hand side of this equation to the explicit integral
$I_{(0,1)}$ and using the explicit form of $\hat E$,
\be
\hat E(\tau,V) = v\, V {|\hat m+\hat n\tau|^2\over\tau_{2}}\,,
\label{eexplic}
\ee
 leads to
\be
\big(v^2{\p^2\over \p v^2}+2v{\p\over \p v}\big) I_{(0,1)}^{(d=10)}=
{\pi^2} \sum_{(\hat m,\hat n)\in\ZZ²} \int _\0chi^{\Vlambda} dV\, V^2\int _{\calF_\Lambda}
{d^2\tau \over \tau_2^2} B_{(0,1)}(\tau) \Delta_\tau e^{-\pi \hat E(\tau,V)}\,.
\label{laplaceone}
\ee
 After integration by parts,
and using the Laplace equation (\ref{laplaceA}) satisfied by $B_{(0,1)}$
this equation can be reexpressed as
\be
\begin{split}
\big(v^2{\p^2\over \p v^2}+2v{\p\over \p v}-12\big) I_{(0,1)}^{(d=10)}=& j_{(0,1)}- \partial I_{(0,1)}^{(d=10)}\, ,
\end{split}
\label{laplacetwo}
\ee
where $j_{(0,1)}$ is the bulk term
\be
j_{(0,1)} = - 12 {\pi^2}
\sum_{(\hat m,\hat n)\in\ZZ^2} \int _0^{\Lambda^2} dV\, V^2\int _1^{\Lambda^2\over V}
{d\tau_2 \over \tau_2}\,
 e^{-\pi \hat E}\, ,
\label{bulk1}
\ee
and $ \partial I_{(0,1)}$ is the boundary term
\be
 \partial I_{(0,1)} = {\pi^2} \left. \sum_{(\hat m,\hat n)\in\ZZ^2} \int_{0}^{\Lambda^2}dV\, V^2
\left(\partial_{\tau_{2}}B_{(0,1)}(\tau) \, e^{-\pi\hat E}-
B_{(0,1)}(\tau)\partial_{\tau_{2}} e^{-\pi\hat E}\right) \right|_{\tau=\tau_2^\Lambda}\, ,
\label{bounterm}
\ee
which receives contributions from
$\tau_{2}= \tau_2^\Lambda=\Lambda^2/V$.   Note that the upper limit
on $V$ is equal to $\Lambda^2$ (whereas $V_\Lambda= 2\Lambda^2/\sqrt 3$) since
$\tau_2^\Lambda = \Lambda^2/V \ge 1$.

The boundary contributions with $\hat n\neq0$ are
exponentially suppressed as $\Lambda^2\to\infty$ because
they are proportional to
\be
e^{-v\, \hat n^2 \,\Lambda^2}\, .
\ee
Therefore only terms with $\hat n=0$ contribute to $\partial I_{(0,1)}$.
These zero winding number terms contribute to the sub-leading divergence proportional to
$\Lambda^3$, which is canceled
by the diagram with the one-loop counterterm of
(\ref{diag:counter}).  Only the leading positive power of $\tau_2$ in $B_{(p,q)}$ contributes in an essential way to the
boundary term
 (\ref{bounterm}).
More explicitly, we may write
$B_{(0,1)}(\tau)=\tau_{2} + \alpha_1(\tau_1) \tau_{2}^{-1} + o(\tau_2^{-3})$
where  $\int_{-1/2}^\half d\tau_1 \alpha_1(\tau_1) =\tilde \alpha_1=0$.  In that case, after
some manipulations (\ref{bounterm}) becomes
\bea
\partial I_{(0,1)} &=& \pi^2 \sum_{\hat m\in\ZZ}\int_{0}^{\Lambda^2}dV\, V^2
\big(   e^{-\pi \hat E}-
  (\tau_2^\Lambda)^{-1} \,(\pi\hat m^2 v V)  \,  e^{-\pi\hat E}\big)\nn\\
  &=& \pi^2 \sum_{\hat m\in\ZZ}\int_{0}^{\Lambda^2}dV\, V^2
\big(1-   {1\over \Lambda^2}\pi v V^2\hat m^2 \big)e^{-\pi {vV^2 \hat m^2\over \Lambda^2}}
\nn\\
  &=&
{\pi^2\over 3}\, \Lambda^6 - {3\over 2\pi } \, \zeta(2)\zeta(3)\,v^{-\threeh}\, \Lambda^3\, ,
 \label{boundthree}
\eea
using (\ref{exsum}) in the last step.

The contribution from the bulk term in (\ref{laplacetwo}) is
\be
j_{(0,1)} = -12 \pi^2\,\sum_{(\hat m,\hat n)\in\ZZ^2} \int _0^{\Lambda^2}
dV\, V^2\int _1^{V_\Lambda\over V}
{d\tau_2 \over \tau_2} \,e^{-\pi  v\, V\, ( {m^2\over \tau_{2}}+ n^2 \tau_{2})}\, .
\label{bulkss}
\ee
 We now change variables to
\be
x=V/\tau_{2}\, , \qquad y=V\tau_{2}\, ,
\label{xyvar}
\ee
which are integrated over the domain
\be
0 < y < \Lambda^2 \, , \qquad 0 < x < y\, ,
\label{limsxy}
\ee
with measure
\be
dV\, d\tau_2 = {1\over 2y}\, dx dy\, .
\label{xymeas}
\ee
Noting that since the integrand is symmetric we can double the region of integration and
integrate over $x$ and $y$ independently.
In these variables we have
\bea
j_{(0,1)} &=& -3 \pi^2\, \sum_{(\hat m,\hat n)\in\ZZ} \int _0^{\Lambda^2} dx \int _0^{\Lambda^2} dy
\sqrt{x\, y}\
  e^{-\pi v\,(\hat m^2 y +\hat n^2 x) }\nn\\
  &=& -3\pi^2 \, (\Sigma_{-\half})^2 = -{3\over v^3}\, \zeta(3)^2 - {24 \over \pi v^{\threeh}}
  \, \zeta(2)\zeta(3)\, \Lambda^3 -{4\over 3}\, \pi^2 \, \Lambda^6\, .
\label{fullj}
\eea
Substituting the  contributions to $j_{(0,1)}$  and $\partial I_{(0,1)}^{(d=10)}$ into  (\ref{laplacetwo}) gives
the Poisson equation
\begin{equation}
\begin{split}
(v^2\partial^2_{v}+2v\partial_{v}-12)I_{(0,1)}^{(d=10)}=& - {3\over v^3}\,\zeta(3)^2
 - {45\over 2\pi v^{\threeh}}\,\zeta(2)\zeta(3)\, \Lambda^3  - 10\, \zeta(2) \,\Lambda^6\,.
  \end{split}
\end{equation}
This equation is simple to solve using the general formula (\ref{alpheq}), (\ref{hsols}), (\ref{gensol}), giving
\begin{equation}
I_{(0,1)}^{(d=10)} =
{5\over 6}\zeta(2)\, \Lambda^6 +\Lambda^3 {2\zeta(2) \zeta(3)\over\pi v^{3/2}} + {\zeta(3)^2\over 2\, v^3}\, .
   \end{equation}
%%%%%%
The $\Lambda^3$ divergence in $I_{(0,1)}$  is canceled by the counter
term  $\delta A_{\triangleright}$ of equation~(\ref{diag:counter}) which, at order
$\sigma_3\,\calR^4$, contributes
\begin{equation}\label{e:cc}
I_{\triangleright\,(0,1)}^{(d=10)}= {\pi\over4}\,c_{1}\,
\left({2\Lambda^3\over3}+ {\zeta(3)\over \pi \, v^{3/2}}\right) =
{\zeta(2) \zeta(3) \over  v^{\threeh}}
 - \Lambda^{3}{2\zeta(2)\zeta(3)\over \pi v^{\threeh}}
- \Lambda^6  {4\zeta(2) \over 3}\, ,
\end{equation}
where we have used the
value of $c_{1}$ given in (\ref{e:c1}).
The relative normalisation of the counter-term triangle diagram with respect to the
double box diagram,
which is fixed by unitarity,
is such that the $\Lambda^3$ divergence cancels.  We also need to subtract
the superficial $\Lambda^6$ divergence with a new counterterm
\be
\delta_{2}    I_{(0,1)}^{(d=10)}   ={4\zeta(2)\over3}    \Lambda^6   +
{6\zeta(2)^2 \over 5}\, ,
\label{countertwob}
\ee
where the value of the constant last term is determined from the value of the genus-two coefficient of the
$\sigma_3\, \calR^4$ interaction in
type IIB string theory, which is contained in the modular function $\calE_{(0,1)}$ \cite{gv:D6R4}
(using the fact that the four-graviton amplitudes in the IIA and IIB theories are identical up to four loops).
The total contribution
\be
I_{(0,1)}^{(d=10)}+I_{\triangleright\,(0,1)}^{(d=10)}+\delta_{2}I_{(0,1)}^{(d=10)} =
 {\zeta(3)^2\over   2\,  v^3}+   {   \zeta(2)\zeta(3)\over
   v^{3/2}}+ {6\zeta(2)^2 \over 5}\, ,
\label{ionetot}
\ee
Using the dictionary between M-theory and string variables
the first two terms coincide with the perturbative string tree-level and genus-one results.  These
are also reproduced by the first
 two terms of the perturbative expansion of $\calE_{(0,1)}$ (while the last term in (\ref{ionetot})
 is the genus-two term).

%%%%%%%%%%%%%%%%%%%%%%%%%%%%%%%%%%%%%%%%%%%%%%%%
\subsection{Evaluation of $I_{(2,0)}^{(d=10)}$}\label{sec:D8R4circle}
%%%%%%%%%%%%%%%%%%%%%%%%%%%%%%%%%%%%%%%%%%%%%%%%

In a similar fashion to the treatment of $I_{(0,1)}^{(d=10)}$, we may write $I_{(2,0)}^{(d=10)}$ as the sum of three
terms with different powers of $v$ (recalling that $v=R_{11}^2$)
\be
I_{(2,0)}^{(d=10)}={I^{(2)}_{(2,0)}\over v^2}
+ I^{(1/2)}_{(2,0)}\,{\Lambda^3\over v^{1/2}} +I^{(0)}_{(2,0)}\, \Lambda^4\ .
\ee
The contribution $I^{(2)}_{(2,0)}$ is the finite part of the amplitude that comes from non-zero windings.
The piece that diverges as $\Lambda^3$ comes from the sub-divergences in which there is zero winding in one loop
and non-zero in the other. The leading $\Lambda^4$
divergence does not make sense in string perturbation and is subtracted
(just as the $\Lambda^8\, \calD^4\calR^4$ term was subtracted
 in \cite{gkv:twoloop}).

Each of the contributions satisfies
$$
\big(v^2{\p^2\over \p v^2}+2v{\p\over \p v}\big){ I^{(\alpha)}_2\over v^{\alpha}}=
\alpha(\alpha-1)\, { I^{(\alpha)}_2\over v^{\alpha}}\ .
$$
Now we write $B_{(2,0)}$ as a sum of the four functions $b_{(2,0)}^0, b_{(2,0)}^2, b_{(2,0)}^4, b_{(2,0)}^6$
satisfying Poisson equations with delta function sources.  $I_{(2,0)}^{(d=10)}$ is then naturally
written as
\be
I_{(2,0)}^{(d=10)}=\sum_{i=0}^3\, h_{(2,0)}^{2i}\ .
\label{kuno}
\ee
The integral $h_{(2,0)}^0$ needs separate treatment, because
$b_{(2,0)}^0=-13/21$ is a constant and will be considered later.
 The integrals $h_{(2,0)}^2,
h_{(2,0)}^4, h_{(2,0)}^6$ can be computed by following the analogous computation to that given in the last
sub-section (and
section 4 of \cite{gv:D6R4}). By definition they satisfy the equations
\be
\big(v^2{\p^2\over \p v^2}+2v{\p\over \p v}\big) \i_2^i=
{\pi^3} \sum_{(\hat m,\hat n)\in\ZZ^2} \int _\0chi^{\Vlambda} dV\, V\int _{\calF_\Lambda}
{d^2\tau \over \tau_2^2} b_{(2,0)}^i(\tau)  \Delta_\tau e^{-\pi\hat E}\ .
\ee
where $\calF_\Lambda$ is once again the cutoff fundamental domain
 $\tau_{2}\leq \tau_2^\Lambda=\Lambda^2/V$ and $V_\Lambda = 2\Lambda^2/\sqrt 3$.
Integrating by parts gives for $i=2,4,6$
\be
\begin{split}
\big(v^2{\p^2\over \p v^2}+2v{\p\over \p v}-i(i+1)\big) h_{(2,0)}^i= j_{(2,0)}^i - \partial h^i_{(2,0)}\,,
\label{i3tot}
\end{split}\ee
where the bulk term is
\be
j_{(2,0)}^i = - 2 u_i {\pi^3}
\sum_{(\hat m,\hat n)\in\ZZ^2} \int _\0chi^{\Vlambda} dV\, V\int _1^{\Lambda^2\over V}
{d\tau_2 \over \tau_2^2} \big(\tau_2^{2}
+1\big)  e^{-\pi E(0,\tau_{2})}
\label{bulk2}
\ee
and  the boundary term is
\be
\partial h^i_{(2,0)}={\pi^3} \left. \sum_{(\hat m,\hat n)\in\ZZ^2}\int_\0chi^{\Vlambda}dV\, V
\left(\partial_{\tau_{2}}b_{(2,0)}^i(\tau) \, e^{-\pi\hat E}- b_{(2,0)}^i\,(\tau)
\partial_{\tau_{2}} e^{-\pi\hat E}\right) \right|_{\tau=\tau_2^\Lambda}\,
\label{i2bound}
\ee
(where $\lambda_{(2,0)}^i$ and $u_i$ are defined in the appendix~\ref{sec:ModD8R4}).

This boundary term again receives contributions from the  region $\tau_{2}\sim \tau_2^\Lambda=\Lambda^2/V$
and in the parametrization where $\hat E = v\, V |\hat m+\hat n\tau|^2/\tau_{2}$
the contribution with $\hat n\neq0$ is again exponentially suppressed as $\Lambda^2\to\infty$.
The $\hat n=0$ terms contribute to the sub-leading divergence which is regularised
by the diagram with the one-loop counter-term of
equation~(\ref{diag:counter}). As before, the only boundary contributions that matter are
the leading ones, which in this case are given by using the expansion
\be
b_{(2,0)}^i(\tau)=q_i \tau_{2}^2 + \alpha_2^i(\tau_{1})+ o(\tau_{2}^{-1})\,, \qquad i=2,4,6\ ,
\label{btwodef}
\ee
to give
\be\begin{split}
\partial h^i_{(2,0)} = u_i {3\pi^3\over4}\,\Lambda^3\, \Sigma_{\half}  - {3\pi\over 4}\, \zeta(3)\,
 \tilde \alpha_2^{i}\, {\Lambda\over v^{3\over2}}\,,
\end{split}\ee
where $\tilde \alpha_2^{i}=\int_{-1/2}^{1/2} d\tau_{1} \alpha_2^{i}(\tau_{1})$,
and $\Sigma_{\half}$ was defined in (\ref{sumfirst}).
The contribution from the bulk term (\ref{bulk2}) is
\bea
j_{(2,0)}^i&=& - 2u_i\, \pi^3\,\sum_{(\hat m,\hat n)\in\ZZ^2} \int _\0chi^{\Vlambda} dV\, V\int _1^{\Lambda^2\over V}
{d\tau_2  \over  \tau_2^2} \big(\tau_2^2  +1\big)  e^{-\pi  v\, V\,  (
  {\hat m^2\over \tau_{2}}+ \hat n^2 \tau_{2})}\\
\nn&=& -u_i\pi^3\,\sum_{(\hat m,\hat n)\in\ZZ^2} \int _0^{\Lambda^2} dx\int _0^{\Lambda^2} dy
 \sqrt{x\over y}\,  e^{-\pi  v\,  ( \hat m^2\, x+ \hat n^2 \,y)}\nn\\
 &=& - u_i\, \pi^3 \, \Sigma_{\half}\,\Sigma_{-\half}\nn\\
&=&- u_i\, \pi^3\, \left(2\Lambda - {1\over v^\half}\log(\chi v/2\pi^2c_{e}) \right)
\,\left({2\over 3}\Lambda^3 + {\zeta(3)\over \pi \, v^{3\over2}}\right) \, .
\label{qwa}
\eea
Therefore (\ref{i3tot})  becomes for $i=2,4,6$
\begin{equation}
\begin{split}
(v^2\partial^2_{v}+2v\partial_{v}&-i(i+1))h_{(2,0)}^i=u_i\,\big(
   {3 \pi^3 \over 2}\, \Lambda^4
  + \frac{17\pi \,\Lambda^3\, \,\log(\chi  v/2\pi^2c_{e})\,\zeta(2)}{2 \,\sqrt{v}}
 \, \\
 &+ {1\over v^{\threeh}}  \zeta(3)\zeta(2) \,\Lambda \left(\frac{9\,\hat\alpha_2^{i}}{2\pi} - 12\right)
      + \frac{6\,\log(\chi  v/2\pi^2c_{e})\,\zeta(2)\,\zeta(3)}{v^2}\big)\,.
  \end{split}
\end{equation}
The terms proportional to $\Lambda$ will eventually cancel due to the relation
(\ref{importlat}). Furthermore, the $\Lambda^4$ terms are primitive divergences that we
will cancel with a counterterm, so their precise coefficients are
not of relevance (there can be no finite
remainder since this term does not correspond to a sensible term in string perturbation theory).
These equations are of the form (\ref{logeq}) (with $a=d=0$), which
have solutions (\ref{gensol}).  The explicit expressions will not be
given here but their sum enters the complete expression for $
I_{(2,0)}^{(d=10)}$.

Now consider the case of  $h_{(2,0)}^0$
for which  the integrand is a total derivative. Integration by parts shows that the integral only gets
contributions from $\tau_{2}$ boundary $\tau_2^\Lambda=\Lambda^2/V$, so that
\bea
\nn (v^2\partial^2_{v}+2v\partial_{v} )h_{(2,0)}^0&=&-{13\over 21}\,  \pi^3\
\sum_{(\hat m,\hat n)\in\ZZ^2}\int_0^{V_\Lambda} dV\, V\int_{\calF_{\Lambda}} {d^2\tau\over\tau_{2}^2}\Delta_{\tau}e^{-\pi \hat E}\\
\nn &=&-{13\over 21}\,  \pi^4\, {v\over \Lambda^4}\, \, \sum_{\hat m\neq0}\,\hat m^2\,
\int_{0}^{\Lambda^2} dV \, V^4\,  e^{-\pi V^2\, v\hat m^2/\Lambda^2}\\
&=&-{39\over 14}\, \zeta(2)\zeta(3)\, {\Lambda\over v^{3\over2}}\, .
\label{hzerodef}
\eea

Summing all the contributions to $I_{(2,0)}^{(d=10)}$ gives
\begin{equation}
I_{(2,0)}^{(d=10)} = \frac{8}{5\,{\sqrt{v}}}\,\pi \,\Lambda^3\,\log(\chi v/ \tilde C_{(2,0)})
\,\zeta(2)
  - \frac{12}{50\,v^2}\,\log(\chi  v/\tilde C_{(2,0)} ))\,\zeta(2)\,\zeta(3)\,,
\label{fullsoltwo}
\end{equation}
where $\tilde C_{(2,0)}$ is an unknown function of $z$.
Note that the term with coefficient $1/v^2=1/R_{11}^4$ corresponds to a finite
genus-one contribution in IIA string theory,
while the term with coefficient $1/v^\half = 1/R_{11}$ corresponds to a genus-two string theory term,
that comes from the sub-divergences (as indicated by the factor of  $\Lambda^3$).

The $\Lambda^3$ divergence in $I_{(2,0)}^{(d=10)}$  is canceled by the counter-term
$I_{\triangleright\,(2,0)}^{(d=10)}$  of equation~(\ref{diag:counter}) which contributes
\begin{equation}\label{e:ccb}
I_{\triangleright\,(2,0)}^{(d=10)}=  {\pi^3\zeta(2)\over 15}
\, c_{1}\,  v^{-\half}\,\left(-\log(\chi v/\tilde{\tilde C}_{(2,0)})\right)\,
\end{equation}
to the coefficient of $\sigma_2^2\,\calR^4$, where $c_{1}$ given in (\ref{e:c1})
(and $\tilde{\tilde C}_{(2,0)}$ is another unknown function of $z$).
The relative normalisation of the counter-diagram with respect to the double box diagram
is such that the $\Lambda^3$ sub-divergence cancels. Furthermore, we need to introduce a counterterm that subtracts
the primitive $\Lambda^4$ divergence,
\bea
I_{(2,0)}^{(d=10)}+
I_{\triangleright\, (2,0)}^{(d=10)}+\delta_{2}I_{(2,0)}^{(d=10)} &=&
-{12\over5}\,\zeta(2)\,
\left[ {\zeta(3)\over v^2}+{2\zeta(2)\over \sqrt{v}} \right] \, \log(-S v/C_{(2,0)})\, .
\label{primdiv}
\eea
The $\log(-S v)$ terms are threshold contributions that correspond to the genus-one and genus-two string theory
thresholds expected from unitarity, as described in the body of this paper.

%%%%%%%%%%%%%%%%%%%%%%%%%%%%%%%%%%%%%%%%%%%%%%%%
\subsection{Evaluation of $I_{(1,1)}^{(d=10)}$}\label{sec:D10R4circle}
%%%%%%%%%%%%%%%%%%%%%%%%%%%%%%%%%%%%%%%%%%%%%%%%

We now consider $I_{(1,1)}^{(d=10)}$ for the term of order $\sigma_2\sigma_3\,\calR^4$
in the expansion of the amplitude.
We saw earlier that $B_{(1,1)}= \sum_{j=0}^4 b_{(1,1)}^{2j+1}$, where $b_{(1,1)}^j$ satisfies the Poisson equation (\ref{ecb}).
Extending the earlier cases, this leads to the decomposition
\be
I_{(1,1)}^{(d=10)}=h_{(1,1)}^1+h_{(1,1)}^3+\cdots+h_{(1,1)}^9\ .
\label{ktres}
\ee
In this case we have
\be
\big(v^2{\p^2\over \p v^2}+2v{\p\over \p v}\big) h_{(1,1)}^j=
 \pi^{4}
\sum_{(\hat m,\hat n)\in\ZZ^2} \int_0^{\Vlambda} dV \int _{\calF_\Lambda}
{d^2\tau \over \tau_2^2} b_{(1,1)}^j(\tau) \Delta_\tau e^{-\pi\hat E}\ .
\ee
Proceeding as in the previous section we obtain, for $j=1,3,5,7,9$
\be
\begin{split}
(v^2\partial^2_{v}+2v\,\partial_{v}-j(j+1)) h_{(1,1)}^j=& j_{(1,1)}^j - \partial h_{(1,1)}^j\, ,
\end{split}
\label{hthree}
\ee
where
\be
j_{(1,1)}^j = -  2\pi^{4}\sum_{(\hat m,\hat n)\in\ZZ^2} \int_0^{\Vlambda} dV
\int _1^{\Lambda^2/V}
{d\tau_2 \over \tau_2^2}  \big(v_j (\tau_2^3+{1\over\tau_{2}})
+w_j \tau_2\big)\ e^{-\pi \hat E(i\tau_{2})}
\label{jthree}
\ee
and
\be
\partial h_{(1,1)}^j =
-\pi^4\,\sum_{(\hat m,\hat n)\in\ZZ^2}\int_{0}^{\Lambda^2} dV  \, (\partial_{\tau_{2}}
b_{(1,1)}^j\, e^{-\pi  \hat E}- b_{(1,1)}^j\partial_{\tau_{2}} e^{-\pi
  \hat E})\big|_{\tau=\tau_2^\Lambda}\, ,
\label{parth3}
\ee
which again only gets contributions from the $\hat n=0$ sector.  Furthermore,
only the leading terms of
\be
b_{(1,1)}^j\sim e_3^j\tau_{2}^3+ \tau_{2}\, \alpha_3^j(\tau_{1})+o(\tau_{2}^{-1})\,,
\label{btwoone}
\ee
 contribute significantly to $\partial h_{(1,1)}^j$.  Setting $x=V^2/\Lambda^2$ we get
\bea
\partial h_{(1,1)}^j &=&\pi^4\, {\Lambda\over 2}\,\sum_{(\hat m,\hat n)\in\ZZ^2}
 \int_{0_\chi}^{\Lambda^2} {dx\over x^{\half}}\,
\left({3e_3^j\Lambda^2\over x} + \tilde\alpha_3^j
-\pi\,v\,\hat m^2\, e^j\,\Lambda^2 -\pi\,v\,\hat m^2\,x\, \tilde\alpha_3^j\right)\, e^{-\pi \hat m^2 x v}
\nn\\
&=&{7\pi^4 e_3^j\over 4}\Lambda^3\,\Sigma_{3\over2} -  {\pi^4\over 2}\,\tilde\alpha_3^j\,  \Sigma_{\half}\, \Lambda\nn\\
&=&{7\pi ^4 e_3^j\over 4}\Lambda^3\,\left({1\over v^\half}\, (\chi f)^{-1}+{\pi v^\half\over 3}\right)
-  {\pi^4\over 2}\,\tilde\alpha_3^j\,  \left(2\Lambda^2 - {1\over 2v^\half}\, \Lambda\, \log(\chi v/C)
\right)\,,
\label{boundthreeb}
\eea
where $\tilde\alpha_3^j\equiv \int_{-1/2}^{1/2} d\tau_{1}\, \alpha_3^j(\tau_{1})$.
As before the $\log(\chi)$ term arises from the massless threshold associated with $m=0$ Kaluza-Klein charge in the
intermediate states.
Turning to the bulk term  (\ref{jthree}) we write $j^j_{(1,1)} = -2\pi^4\, (v_j\, K_1+ w_j\, K_2)$, where
\be
v_j K_1+   w_j K_2 \equiv 2\pi^{4}\sum_{(\hat m,\hat n)\in\ZZ^2} \int_0^{\Lambda^2} dV
\int _1^{\Lambda^2\over V}
{d\tau_2 \over \tau_2}  \big(v_j (\tau_2^2+{1\over\tau^2_{2}})
+w_j\big)\ e^{-\pi \hat E(i\tau_{2})}\ ,
\ee
where (after introducing $x=V/\tau_2,\ y=V\tau_2$)
\bea
\nn K_1 &=& {\pi^{4}\over 2}\sum_{(\hat m,\hat n)\in\ZZ^2} \int_0^{\Lambda^2}{dx dy\over \sqrt{xy}}
\big( {x\over y}+{y\over x}
\big)\   e^{-\pi v\,(\hat m^2 y +\hat n^2 x)}\nn\\
&=& \pi^4 \Sigma_{\threeh}\, \Sigma_{-\half}
\nn\\
&=&\pi^4\,\left({2\over 3} \, \Lambda^3 + {1 \over \pi}\,\zeta(3)\, v^{-\threeh}\right)\,
 \left({1\over v^\half}\, (\chi f)^{-1} + {\pi v^\half\over 3}\right)
\nn\\
&=& {2\over3}{\pi^4\over v^\half}\, \Lambda^3\,(\chi f)^{-1}
+
{4\over3} \pi^3\, \zeta(2) \, \Lambda^3\, v^{\half}+ {\pi^3\zeta(3)\over v^2}
 (\chi f)^{-1}
+{ \pi^{4}\over  3\, v}\zeta(3)\, ,
\label{Konedef}
\eea
and
\bea
\nn K_2&=& {\pi^{4}\over 2}\sum_{(\hat m,\hat n)\in\ZZ^2} \int
_0^\infty {dx dy\over \sqrt{xy}}
\   e^{-\pi v\,(\hat m^2 y +\hat n^2 x)}\\
&=& {\pi^4\over 2}\, (\Sigma_{\half})^2
={\pi^4\over 2}\, \left(2\Lambda -{1\over v^\half}\log(\chi v/C)\right)^2
\nn\\
&=& 2\pi^4\, \Lambda^2 - 2\pi^4\,\Lambda\,{1\over v^{\half}}\, \log(\chi v/C) + {\pi^4\over 2v}\, \log^2(\chi v/C)\,.
\label{e:zetaOne2}\eea
The inverse powers of $S$ in (\ref{boundthreeb})  and (\ref{Konedef}), originate, as anticipated, from the attempt to expand
the nonanalytic amplitude in powers of $S$.  We can drop these terms, which are singular in the limit $f\to 0$,
following the argument in section~\ref{sec:infracut}.
Summing all the other contributions of order $S^5$ and $S^5 \log(-S)$ gives
\be\begin{split}
(v^2\partial_{v}^2+2v\partial_{v}-j(j+1))h_{(1,1)}^j&=-
\frac{\pi ^5}{36}\,\left(21\,e^{j} + 8\,v_j \right)   \,\sqrt{v}\,\Lambda^3  -
{ \pi^4\over2}\,(\hat e_j +8w_{j}) \,\Lambda^2\\
  & +
  {\pi^4\over 4 v^{\half}} \,(\hat \alpha^j_3 +8w_{j}) \,\log(\chi v/2\pi^2 c_{e})\,\Lambda \\
  &-
  \frac{{\pi }^4}{3v}\,\left(3\,w_j\,\log(\chi  v/2\pi^2 c_{e})^2 + v_j\,\zeta(3) \right) \, .
  \end{split}\ee
These equations involve $\log v$ and $(\log v)^2$ factors are again of the form (\ref{logeq})
(this time with $a=c=0$) with solutions (\ref{gensol}).
Exploiting these solutions together with the following facts that follow from the explicit coefficients in
appendix~\ref{sec:Modular}
\be
\sum_{j=0}^4 \ { w_{2j+1}\over (2j+1)(2j+2)}=0\ , \qquad \sum_{j=0}^4 {\hat f_{j}+8 w_{j}\over {1\over 4}+(2j+1)(2j+2)}=0\,,
\label{sumss5}
\ee
and
\bea
&& \sum_{j=0}^4 {v_{2j+1}\over (2j+1)(2j+2)} = {224\over 15}\,, \qquad
\sum_{j=0}^4  {\hat  \alpha^{2j+1}_3\over  (2j+1)(2j+2)}  =  -{91\over
  18}\, ,\nn\\
&&
\sum_{j=0}^4{ (21q_{2j+1}+8v_{2j+1})\over ({3 \over 4} - (2j+1)(2j+2))}  =180\,,
\label{moresums}
\eea
and recalling that $I_{(1,1)} = \sum_{j=0}^4h_{(1,1)}^{2j+1}$, the total result is
\begin{equation}
\begin{split}
I_{(1,1)}^{(d=10)}=& - {91\over36} \, \pi^4\, \Lambda^2
+  \frac{675}{2}\,\zeta(4)\,\sqrt{v} \,{4\pi\Lambda^3\over3}+
  \frac{448\,\zeta(4)\,\zeta(3)}{v}\\
&
\qquad +\frac{273\pi^4}{54\,v}\,\log(\chi  v/\tilde C_{(1,1)}) \,.
\end{split}
\end{equation}
  The cancelation of the  $\log^2(-S v)$ contributions corresponds to the fact that  $1/\epsilon^2$ terms
  cancel in the two-loop diagrams of ten-dimensional type II supergravity, as we will see in detail in
  appendix~\ref{sec:tenreg}.
The $\Lambda^2$ contribution is the leading superficial divergence at two string loop and must be subtracted with no
finite  residue   since  it   is  not  accompanied   by  a   power  of
$v=R_{11}^2 = g_A^{2/3}$ that is an integer power of
$g_A^2$ and therefore cannot contribute in string theory.   The $\Lambda^3$
contribution is a subleading divergence regulated by the counter term~(\ref{diag:counter})
leaving a finite genus-three string contribution.  The total contribution at order $\sigma_2\sigma_3\,\calR^4$ is
\be
\begin{split}
I_{(1,1)}^{(d=10)}+I_{\triangleright\,(1,1)}^{(d=10)}+\delta_{2}I_{(1,1)}^{(d=10)}
=& {4725\over8}\,\zeta(6)\,\sqrt{v}+
  \frac{448\,\zeta(4)\,\zeta(3)}{v}\\
  &\qquad +\frac{455\zeta(4)}{v}\, \log(\chi  v/C_{(1,1)})\, .
\end{split}\ee
The first term in this expression  corresponds to a  genus-three IIA string
contribution while the remaining terms (with the $1/v$ factor) are genus-two IIA string contributions.
These contributions are distinguished by their distinct zeta function coefficients, so it would look very unnatural
to associate the analytic $1/v$ term with the unknown scale of the logarithm in the nonanalytic $1/v$ term.

Substituting this result  into the  expansion for the amplitude
gives the terms of order $\sigma_2\sigma_3\, \calR^4$ as summarized in the text.  It is worth noting, in particular, the presence of
the logarithmic term
\begin{equation}
i {\kappa_{(11)}^6\over (4\pi)^{10}}\, {13\over 466560}\, \sigma_2\sigma_3\,\log(\chi)\, \calR^4\,,
\end{equation}
which reproduces  the   result  obtained for  the coefficient of the  $1/\epsilon$   pole
obtained by dimensional regularization around nine dimensions in (\ref{sec:ninereg}).

%%%%%%%%%%%%%%%%%%%%%%%%%%%%%%%%%%%%%%%%%%%%%%%%
\subsection{Evaluation of $I_{(3,0)}^{(d=10)}$
and $I_{(0,2)}^{(d=10)}$}
\label{sec:D12R4circle}
%%%%%%%%%%%%%%%%%%%%%%%%%%%%%%%%%%%%%%%%%%%%%%%%

In order to analyze the integrals $I_{(3,0)}$ and $I_{(0,2)}$
we  write $B_{(p,q)}=\sum_{k=0}^6  b_{(p,q)}^{2k}$  with $(p,q)=(3,0)$
and $(p,q)=(0,2)$, where
$b_{(p,q)}^k$ satisfy Poisson equations (\ref{xecb}).
This leads to a decomposition
\be
I_{(p,q)}= h_{(p,q)}^{0}+h_{(p,q)}^2+\cdots+h_{(p,q)}^{12}\, ,
\label{i4decom}
\ee
where, for $k=2,4,\dots,12$, the components satisfy the equations
\be
\big(v^2{\p^2\over \p v^2}+2v{\p\over \p v}\big)\, h_{(p,q)}^k=
 {\pi^{5}}
\sum_{(\hat m,\hat n)\in \ZZ^2} \int_\0chi^{\lambda}{ dV\over V} \int _{\calF_\Lambda}
{d^2\tau \over \tau_2^2}\, b_{(p,q)}^k(\tau) \,\Delta_\tau e^{-\pi \hat E}\,.
\label{compeqs}
\ee
The components $h_{(3,0)}^0$ and  $h_{(0,2)}^0$
need separate treatment because they are associated with the constant contributions
 $b_{(3,0)}^0=12264/715$ and $b_{(0,2)}^0=2716/165$.
 In this case both the left-hand and right-hand sides of (\ref{compeqs}) vanish.  We
 will return to these cases later.

 \subsubsection{$h_{(3,0)}^k$, $h_{(0,2)}^k$ with $k>0$}
 \label{sec:hfoureval}
Integrating (\ref{compeqs}) by parts and using (\ref{xecb})  gives
\be
\big(v^2{\p^2\over \p v^2}+2v{\p\over \p v}-k(k+1)\big) h_{(p,q)}^k= j_{(p,q)}^k - \partial h_{(p,q)}^k
\,,
\label{hfourieq}
\ee
for $(p,q)=(3,0)$ and $(p,q)=(0,2)$ and $k=2,4,\dots,12$
where
\be
j_{(p,q)}^i = - 2{\pi^{5}}\sum_{(\hat m,\hat n)\in \ZZ^2} \int
_\0chi^{\Vlambda} { dV\over V}
\int _1^\infty
{d\tau_2 \over \tau_2^2}  \big(1+\tau_2^2\big) \big[
f_{(p,q)}^k (\tau_2^2+
\tau_2^{-2}) + g_{(p,q)}^k \big]\
\delta(\tau_1 )
 e^{-\pi E}
 \label{jfourdef}
 \ee
 and
 \be
\partial h_{(p,q)}^k =-\pi^5\, \int_\0chi^{\Vlambda} {dV\over V} \big(\partial_{\tau_{2}}b_{(p,q)}^k
e^{-\pi \hat E}-b_{(p,q)}^k\partial_{\tau_{2}} e^{-\pi\hat E}\big)\big|_{\tau_{2}=\tau_2^\Lambda}\, .
\label{boundfourdef}
\ee
with $(p,q)=(3,0)$ and $(p,q)=(0,2)$.
We recall that the eigenvalues are the same for the two tensorial structures, so
\bea
(v^2\partial_{v}^2+2v\partial_{v}-k)(k+1)h_{(p,q)}^{2k}&=&        -2\,(
f_{(p,q)}^{2k} H_1+ g_{(p,q)}^{2k} H_2) -\partial h_{(p,q)}^{2k}\, ,
\label{poissxy}
\eea
where (after introducing $x=V/\tau_2,\ y=V\tau_2$)
\bea
\nn H_1&=&  {\pi^{5}\over 2}\,\sum_{(\hat m,\hat n)\in\ZZ^2} \int
_0^\infty {dx dy\over xy} \big( \sqrt{x\over y}+\sqrt{y\over x}\big)
\big( {x\over y}+{y\over x}\big)\   e^{-\pi v\, (\hat m^2 y +\hat n^2 x)}\\
\nn &=& {\pi^5\over 2}\, (\Sigma_{-\half}\, \Sigma_{\fiveh} + \Sigma_\half \, \Sigma_\threeh)
\\
\nn&=&  {\pi^{5}\over 2}\, \left({2\over 3}\Lambda^3 + {1\over \pi}\zeta(3)\,v^{-\threeh}\right)
\,\left(v^{-\half}(\chi f)^{-2}+{\pi^2\over 45}v^\threeh\right)\nn\\
&&  {\pi^{5}\over 2}\, \left(-v^{-\half}\log(\chi v/C) + 2\Lambda\right)
\,\left(v^{-\half}(\chi f)^{-1}+{\pi\over 3}v^\half\right)\,,
       \eea
and
\bea
\nn H_2 &=& {\pi^{5}\over 2}\sum_{(\hat m,\hat n)\in\ZZ^2} \int
_0^\infty {dx dy\over {xy}} \big( \sqrt{x\over y}+\sqrt{y\over x}\big)
\   e^{-\pi v\,(\hat m^2 y +\hat n^2 x)}\\
&=&  {\pi^5\over 2}\, \Sigma_\half\, \Sigma_\threeh\nn
\\
&=&  {\pi^{5}\over 2}\, \left(-v^{-\half}\log(\chi v/C) + 2\Lambda\right)
\,\left(v^{-\half}(\chi f)^{-1}+{\pi\over 3}v^\half\right)\, ,
  \eea
  where the terms with inverse powers of $\chi$ will once more be dropped.

The relevant contributions to the boundary term come from the positive
powers of $\tau_{2}$ in the expansions
\be
b_{(p,q)}^{k}=e_{(p,q)}^k\, \tau_{2}^4+
\alpha_{(p,q)}^{k}(\tau_{1})\,\tau_2^2+O(1)\,.
\label{edefs}
\ee
Substituting in (\ref{boundfourdef}) gives
\bea
 \partial h_{(p,q)}^k &=& -{\pi^5\over 2}\, \left({11\over 2}\Lambda^3 e_{(p,q)}^k\, \Sigma_{\fiveh}
       +{5 \over 2}\,\Lambda \, \tilde \alpha_{(p,q)}^k \, \Sigma_\threeh \right) \, ,
\label{boundfour}
\eea
where $\tilde \alpha_{(p,q)}^k=\int_{-1/2}^{1/2} \, \alpha_{(p,q)}^k(\tau_{1}) \, d\tau_{1}$.
Putting the various contributions together (\ref{hfourieq}) gives, for $k\ne 0$,
\bea
\nn (v^2\partial_{v}^2+2v\partial_{v}-k(k+1))\, h_{(p,q)}^{2k}&=&
-\frac{315}{4}\,{\sqrt{v}}\,\Lambda \,\left( 8\,f_{(p,q)}^{2k} + 8\,g_{(p,q)}^{2k} +       5\,\tilde \alpha_{(p,q)}^{2k} \right) \,\zeta(6)\\
       &-&
  \frac{7}{4}\,\pi \,v^{\frac{3}{2}}\,\Lambda^3\,
     \left( 8\,f_{(p,q)}^{2k} + 33\,e_{(p,q)}^{2k} \right) \,\zeta(6)
\\
  &-&\left( -
       315\,(g_{(p,q)}^{2k}+f_{(p,q)}^{2k})\,\log(\chi  v/2\pi^2c_{e}) + 21 \,f_{(p,q)}^{2k}\,\zeta(3)
       \right) \,\zeta(6)\,.\nn
\eea
for $(p,q)=(3,0)$ and $(p,q)=(0,2)$.

Once again these equations have $\log v$'s on the right-hand side and the solutions were obtained in (\ref{gensol}).
We also note the values of the sums,
\begin{eqnarray}
&&\sum_{k=1}^6 { 8\,f_{(3,0)}^{2k} + 8\,g_{(3,0)}^{2k} +  5\,\tilde\alpha_{(3,0)}^{2k}\over 3/4-2k(2k+1)}=0\,, \qquad
\sum_{k=1}^6 { 8\,f_{(3,0)}^{2k} + 33\,e_{(3,0)}^{2k} \over 15/4-2k(2k+1)}=-96\,,\\
&&\sum_{k=1}^6 { 8\,f_{(0,2)}^{2k} + 8\,g_{(0,2)}^{2k} +  5\,\tilde\alpha_{(0,2)}^{2k}\over 3/4-2k(2k+1)}=0\,, \qquad
\sum_{k=1}^6 { 8\,f_{(0,2)}^{2k} + 33\,e_{(0,2)}^{2k}  \over 15/4-2k(2k+1)}=-96\,,\\
&&\sum_{k=1}^6 { f_{(3,0)}^{2k} +e_{(3,0)}^{2k} \over 2k(2k+1)}={1733\over715}\,, \ \qquad\qquad\qquad
\sum_{k=1}^6 { f_{(0,2)}^{2k} +e_{(0,2)}^{2k} \over 2k(2k+1)}={749\over330}\,,\\
&&\sum_{k=1}^6 { f_{(3,0)}^{2k} \over 2k(2k+1)^2}={16000249\over75150075}\,,\ \qquad\qquad\qquad
\sum_{k=1}^6 { f_{(0,2)}^{2k} \over 2k(2k+1)^2}={25658819\over118918800}\, .
\end{eqnarray}

%%%%%%%%%%%%%%%%%%%%%%%%%%%%%%%%%%%%%%%%%%%%%%%%%
\subsubsection{$h_{(3,0)}^0$, $h_{(0,2)}^0$}
%%%%%%%%%%%%%%%%%%%%%%%%%%%%%%%%%%%%%%%%%%%%%%%%%

We now return to the $k=0$  terms, which are determined by the values of the constants $b_{(2,0)}^0$
and $b_{(0,2)}^0$.  In this case we can evaluate the integral
\be
h_{(p,q)}^{0} =\pi^{5}\, b_{(p,q)}^{0}\,
\sum_{(\hat m, \hat n)\in\ZZ^2}\, \int_{\0chi}^{\Vlambda} {dV \over V} \int_{\calF_\Lambda} {d^2 \tau\over \tau_2^2}
\,e^{-\pi\,V\,v\, {|\hat m+\tau \hat n|^2\over\tau_2}}\,
\label{fourconstb}
\ee
for $(p,q)=(3,0)$ and $(p,q)=(0,2)$.
We will write the integral as the sum of two terms,
\be
h_{(p,q)}^0 = h_{(p,q)}^{0\, (1)} - h_{(p,q)}^{0\, (2)}\,
\label{sumint}
\ee
where in $h_{(p,q)}^{0\, (1)}$ the
$\tau$ integral spans the full fundamental domain, $\calF$, whereas $h_{(p,q)}^{0\, (2)}$
subtracts the integral over the range $\Lambda^2/V \le \tau_2 \le \infty$.
In the first contribution we separate the $\hat m=\hat n=0$ term, for which the $\tau$ integral simply gives
the volume of the fundamental domain, $\int d^2\tau/\tau_2^2 = \pi/3$.  The integral over $\tau$ in the
 $(\hat m,\hat n) \ne (0,0)$ piece can be `unfolded' to the infinite strip as in \cite{gkv:twoloop}, giving
\bea
h_{(p,q)}^{0\, (1)} &=&\pi^{5}\, b_{(p,q)}^{0} \int_{\0chi}^{\Vlambda}{dV\over V}\,
\left( \int_{\calF}{d^2\tau\over\tau_{2}^2}
+\int_{0}^\infty {dt\over t^2}\sum_{p\neq0}\exp(-\pi p^2 v V/\tau_{2})\right)\nn\\
&= &{\pi^6\, b_{(p,q)}^{0}\over3}\,\int_{\0chi}^{\Vlambda}{dV\over V} +
{\pi^6\, b_{(p,q)}^0\over3v}\,\int_{\0chi}^{\Vlambda}{dV\over V^2}\, .
\eea

The second term in (\ref{sumint})
may be evaluated by first performing a  Poisson resummation on one of the integers, which
gives  a sum over the
winding number, $\hat n$, and the Kaluza--Klein charge, $m$.
The integral over $\tau_{1}$ projects onto the  terms with
$\hat n\, m =0$, giving
\bea
h_{(p,q)}^{0\, (2)} &=&{2\pi^5\, b_4^{0}\over v^{1\over2}}\,\int_{\0chi}^{\Vlambda}  {dV\over V^{3\over2}}\,
\int_{\Lambda^2/V}^\infty {dt\over t^{3\over2}}\nn\\
&+&{2\pi^5\, b_{(p,q)}^{0}\over v^{1\over2}}\,\int_{\0chi}^{\Vlambda}  {dV\over V^{3\over2}}\,
\int_{\Lambda^2/V}^\infty {dt\over t^{3\over2}}\, \sum_{q\neq0}\left(e^{-\pi v V q^2t}+ e^{-\pi q^2 t/(v V)}\right)
\nn\\
&=&{\pi^6\, b_{(p,q)}^{0}\over3} \, {1\over v}\, \int_{\0chi}^{\Vlambda}{dV\over V^2}
-{\pi^4\, b_{(p,q)}^{0}\over  \,v}\int_{\0chi}^{\Vlambda}{dV\over V^2}\, \sum_{\hat m\neq0}
{1\over \hat m^2}\, e^{-\pi\, \hat m^2\, v\, (V/\Lambda)^2}
\eea
where we have dropped the first term of the second line since it is smaller than $\exp(-\pi v\Lambda^2)$ and
the last line follows by a further Poisson resummation of the second term of the second line.
\bea
h_{(p,q)}^0 &=& h_{(p,q)}^{0\, (1)} -h_{(p,q)}^{0\, (2)}  = {\pi^6\, b_{(p,q)}^{0}\over3}\,
\int_{\0chi}^{\Vlambda}{dV\over V}\, e^{S/V}
 -{\pi^4\, b_{(p,q)}^{0}\over  v}  \,\int_{\0chi}^{\Vlambda}{dV\over V^2}\, e^{S/V}\nn\\
 &=& {\pi^6\, b_{(p,q)}^{0}\over3}\, \log(\chi /\Lambda^2 C'_{(p,q)}) + O(\Lambda^{-1})    \, ,
\label{fullsum}
\eea
where $C'_{(p,q)}$ is an undetermined function of $z$, but is independent of $v$.
After substituting the values of  $b_{(3,0)}^0$ and $b_{(0,2)}^0$
we find
\be
h_{(3,0)}^0 ={386316\over143}\,\zeta(6)\,\log(\chi  /\Lambda^2 \, C_{(3,0)})\,,
\label{h4zerox}
\ee
\be
h_{(0,2)}^0 = {28518\over 11}\,\zeta(6)\,\log(\chi  /\Lambda^2 \, C_{(0,2)})\,.
\label{h4zeroy}
\ee
 Note that in this case
  the scale of the logarithm is $\Lambda$,
 in contrast to the earlier cases, where it was $1/R_{11}^2$
-- there is a new primitive divergence.
 This had to be the case since these terms are the
$\hat m=0 =\hat n$ part of the $L=2$ eleven-dimensional supergravity amplitude, which is the only part that
arises in the limit $R_{11}\to \infty$, where there is a $\log \Lambda$ divergence.
The more conventional dimensional regularization argument that leads to the same coefficient for the
$S^6 \,\log S\,  \calR^4$ term is given in appendix~\ref{sec:elevenreg}.  The pole residue in  (\ref{elevenpole})
matches perfectly with the above coefficient (once the differences in the conventions used for the normalization
are taken into account).

In order to compare with the string result we will write
\bea
h_{(3,0)}^0 &=&{386316\over143}\,\zeta(6)\left(\log(\chi R_{11}^2/C_{(3,0)}) -\log(\Lambda^2R_{11}^2)\right)
\nn\\
 &=&{386316\over143}\,\zeta(6)\left(\log(\chi g_A^2/C_{(3,0)}) - \log(\Lambda^2 v)\right)\,,
\label{h4zeroxnew}
\eea
where the first term on the right-hand side combines nicely with the contributions from $h_{(3,0)}^i$ with $i\ne 0$
to reproduce the correct threshold term.  The left over part is to be subtracted by a new counterterm.

%%%%%%%%%%%%%%%%%%%%%%%%%%%%%%%%%%%%%%%%%%%%%%%%%
\subsubsection{$I_{(3,0)}^{(d=10)}$, $I_{(0,2)}^{(d=10)}$ and counterterm contributions}
%%%%%%%%%%%%%%%%%%%%%%%%%%%%%%%%%%%%%%%%%%%%%%%%%

The values of $h_{(3,0)}^{2k}$ and $h_{(0,2)}^{2k}$ for $k=0,\dots,6$  determine the solutions,
\bea
I_{(3,0)}^{(d=10)}&=& 168 \pi \, \zeta(6)\, v^{3\over2}\, \Lambda^3+{100647\over715}\,
 \zeta(3)\,\zeta(6)
- 3465\, \zeta(6)\, \log(\chi v/\tilde C_{(3,0)})\,
\label{ithreezero}
\eea
and
\bea
I_{(0,2)}^{(d=10)}&=& 168 \pi \, \zeta(6)\, v^{3\over2}\, \Lambda^3+{15827\over110}
\zeta(3)\zeta(6)
-   {6615\over2}\,\zeta(6)\, \log(\chi v/\tilde C_{(0,2)})\, .
\label{izerotwo}
\eea
The $\Lambda^3$ terms are canceled by the counter-term diagram~(\ref{diag:counter}) and replaced by
finite contributions that are interpreted in the IIA string coordinates as genus-four perturbative
contributions\footnote{Since at order $S^6\,\calR^4$
 the diagram regulating the one-loop sub-divergence gives a result proportional to
$\sigma_{6}= \sigma_{2}^3/4+\sigma_{3}^2/3$, it is necessary that the  $\Lambda^3$ coefficients
for  $I_{(3,0)}$ and $I_{(0,2)}$ are the same.}.
There are two distinct terms in (\ref{ithreezero}) and (\ref{izerotwo})
that have no power of $v$ and are independent of $\Lambda$ (they are finite terms).
These correspond to  genus-three IIA string contributions.  The $\log(\chi v)$ term corresponds to the genus-three
part of $E_{5/2}\, s^6\, \log(\chi)$. The genus-one string part of this expression does not arise from two-loop
supergravity diagrams considered in this paper,
but it is easy to see from dimensional arguments that it should be obtained
from the three-loop amplitude of eleven-dimensional supergravity.
The $\Lambda^3$ terms are one-loop sub-divergences regularized by the counter-term diagram of
equation~(\ref{diag:counter})
\bea
\nonumber
I_{(3,0)}+I_{\triangleright\,(3,0)}^{(d=10)} + \delta_{2}I_{(3,0)}&=& 210\, \zeta(8) \,
 v^{3\over2}
+{100647\over715}\,
 \zeta(3)\,\zeta(6)
 - 3465\,\zeta(6)\,  \log(\chi v/C_{(3,0)})\,,\\
\label{counterthree}
\eea
and
\bea
\nonumber
I_{(0,2)}+I_{\triangleright\,(0,2)}^{(d=10)}+ \delta_{2}I_{(0,2)}&=& 210 \, \zeta(8)\,
 v^{3\over2}+{15827\over110}
\zeta(3)\zeta(6)- {6615\over2}\,\zeta(6)\, \log(\chi v/C_{(0,2)})\,.\\
\label{countertwo}
\eea

%%%%%%%%%%%%%%%%%%%%%%%%%%%%%%%%%%%%%%%%%%%%%%
\section{Quasi-zero mode modular functions}
\label{sec:Cste}
%%%%%%%%%%%%%%%%%%%%%%%%%%%%%%%%%%%%%%%%%%%%%%
In this section we will evaluate the coefficient $\calE_{(2,0)}^{(2)\, 0}$ of the
$\sigma_2^2\,\calR^4$ term in (\ref{e:calE})
and the coefficients $\calE_{(3,0)}^{(6)\, 0}$ and $\calE_{(0,2)}^{(6)\, 0}$
of the $S^6\,\calR^4$ terms in (\ref{e:calGx}) and~(\ref{e:calGy}).
These are the cases in which modular function in the integrand,
$B_{(p,q)}(\tau)=b_{(p,q)}$, is a  constant so the eigenvalue
in the inhomogenious Laplace equations is zero and the source term
vanishes.  We will see by direct evaluation that in these cases
the coefficients satisfy Laplace equations of the form
\be
\Delta_\Omega \calE_{(p,q)}^{(r)\, 0} = {D_{(p,q)}\over 2} \,,
\label{zerogen}
\ee
where $D_{(p,q)}$ are constants.

\subsection{ Evaluation of $\calE_{(2,0)}^{(2)\, 0}$}

In the $(2,0)$ case we know that there is a two-loop supergravity threshold (which will be explicitly
evaluated in section~\ref{sec:DimReg}).  This is associated with the zero Kaluza--Klein modes in the
loops, so here we will use the Kaluza--Klein basis for the sums, which means we need to evaluate
\be
\calE_{(2,0)}^{(2)\, 0} = {4\calV_2^2\over \pi^2}\, I_{(2,0)\,0} ={4\calV_2^2\over \pi^2}\,  b_{(2,0)}^0\,
K
\label{kzerodef}
\ee
where $b_{(2,0)}^0 = -13/21$ and
\bea
K &=&{\pi^3\over \calV_2^2}\,
\sum_{\{m_I, n_J\}\in\ZZ^{4}}\, \int_\0chi^{\Vlambda} {dV\over V}\, \int_{\calF_\Lambda} {d^2 \tau\over \tau_2^2}
\,e^{{-\pi{G^{IJ}\over V\tau_2}\left[(m+\tau n)_I( m+\bar\tau n)_J\right]}}\nn\\
&=& {\pi^3\over \calV_2^2}\,
\sum_{\{m_I, n_J\}\in\ZZ^4}\, \int_\0chi^{\Vlambda} {dV\over V}\, \int_{\calF_\Lambda} {d^2 \tau\over \tau_2^2}
\,e^{{-{\pi\over\calV_2 \Omega_2 V\tau_2}\left|m_1+n_1\tau + \Omega(m_2+n_2\tau)\right|^2} - 2\pi{m_1n_2-m_2n_1
\over \calV_2 V} }\,.
\label{e:N2}
\eea
This will be analyzed by separating the integrand into sectors with different patterns of vanishing coefficients,
\be
K \equiv \sum_{m_1,n_1,m_2,n_2}\, \hat K_{(m_1,n_1)(m_2,n_2)} =
\sum_{m_1
,n_1,m_2,n_2}\,\int_{\0chi}^{\Vlambda}\, {dV\over
  V}\int_{\calF_\Lambda} {d^2\tau\over \tau_2^2}\,  J_{(m_1,n_1)(m_2,n_2)}\,.
\label{msects}
\ee
It is convenient to decompose the sums as follows,
\be
\begin{split}
\sum_{m_1,n_1,m_2,n_2}& J_{(m_1,n_1)(m_2,n_2)}\\=&
J_{(0,0)(0,0)}+ \sum_{(m_1,n_1)\ne (0,0)}\, J_{(m_1,n_1)(0,0)}
+ \sum_{(m_2,n_2)\ne(0,0)}\sum_{m_1,n_1}\, J_{(m_1,n_1)(m_2,n_2)}\\
=& J_{(0,0)(0,0)}+ \sum_{(p,q)}\sum_{k_1\ne 0}\, J_{(k_1p,k_1q)(0,0)}
+ \sum_{(p,q)}\sum_{k_2\ne 0}\sum_{m_1,n_1}\, J_{(m_2,n_2)(k_2p,k_2q)}\,,
\end{split}
\label{sumdecom}
\ee
where $p$, $q$ are relatively prime. We may now perform the `unfolding
trick', which replaces the integral of the sum over $p$ and $q$ over $\calF_\Lambda$ by an
integral of only
the $(p,q)=(1,0)$ term over the  rectangle $\calR_\Lambda$: $\{0\le \tau_2\le
\Lambda^2/V$,  $\-1/2\le \tau_1\le 1/2\}$. In principle,  in the presence of the
upper cutoff $\tau_2\le \Lambda^2/V$ this unfolding
leads to a very complicated $\tau_1$ and $V$-dependent lower cutoff on $\tau_2$.  However, as we will see, the
results we need are not sensitive to the lower end of the $\tau_2$ integral and we will set this to zero.
This gives
\be
\begin{split}
\int_{\calF_\Lambda} {d^2\tau\over \tau_2^2}& \sum_{m_1,n_1,m_2,n_2}\, J_{(m_1,n_1)(m_2,n_2)}\\
=&
\int_{\calF_\Lambda} {d^2\tau\over \tau_2^2} \, J_{(0,0)(0,0)}+\int_\calR
    {d^2\tau\over \tau_2^2}
\left( \sum_{k_1\ne 0}\, J_{(k_1,0)(0,0)}
+ \sum_{k_2\ne 0}\sum_{m_1,n_1}\, J_{(m_1,n_1)(k_2,0)}\right)\\
=&\int_{\calF_\Lambda} {d^2\tau\over \tau_2^2} \, J_{(0,0)(0,0)}
 +\int_{\calR_\Lambda}
    {d^2\tau\over \tau_2^2} \left(\sum_{(k_1,k_2)\ne(0,0)}\, J_{(k_1,0)(k_2,0)}
+ \sum_{n_1,k_2\ne 0}\sum_{m_1}\, J_{(m_1,n_1)(k_2,0)}\right)\,,\\
\end{split}
\label{unfold}
\ee
 This is a decomposition into the sum of  singular, degenerate and
 non-degenerate   orbits   of   $SL(2,\ZZ)$   in   the   language   of
 \cite{gkv:twoloop}.

Consider first the $m_I=n_I=0$ term, which contains the $\log(\chi)$ factor.  In this case $J=\pi/3 +O(V/\Lambda^2)$
and the result is
 \be
K_{(0,0)(0,0)}  = {\pi^3\over \calV_2^2}\, {\pi \over 3}\, \log(\chi /C\Lambda^2) +O(\calV_2^{-1})\,,
 \label{zerokk}
\ee
where we have only kept the leading term in the limit $\calV_2 \to 0$, which is the part
that behaves as $\calV_2^{-2}$.

The second term in (\ref{unfold}) leads to
\bea
\sum_{k_1,k_2\ne 0} K_{(k_1,0)(k_2,0)} &=& {\pi^3\over \calV_2^2}\, \int_0^{\Vlambda}  {dV\over V}\,
\int_0^{\Lambda^2/V}{d\tau_2\over \tau_2^2} \, \sum_{k_1,k_2\ne 0}\exp\left(-{\pi\over \calV_2\Omega_2 V \tau_2}
|k_1 + k_2 \Omega_2|^2\right)\nn\\
&=&{\pi^3\over \calV_2^2}\, \int_0^{\Vlambda} dV\,
\int_{\Lambda^{-2}}^\infty d\hat y \, \sum_{k_1,k_2\ne 0}\exp\left(-{\pi \hat y\over \calV_2\Omega_2}
|k_1 + k_2 \Omega_2|^2\right)
\, ,
\label{degenorb}
\eea
where we have defined $\hat y=(V\tau_2)^{-1}$.
It is easy to see that this depends linearly on $\Lambda^2$ and
has an overall power of $\calV_2^{-1}$, so it does not contribute to the term proportional to
$\calV_2^{-2}$ and can be ignored here.

The last term in (\ref{unfold}) leads to
\be
\sum_{k_1,k_2\ne 0}
K_{(m_1,n_1)(k_2,0)} = {2\pi^3\over \calV_2^2} \sum_{n_1>0,k_2\ne 0}\sum_{m_1+0}^{n_1-1} \int_0^{\Vlambda}
 {dV\over V} \int_{\calR_\Lambda} {d^2\tau\over \tau_2^2} \, e^{-{\pi\over \calV_2\Omega_2
 V   \tau_2 }|m_1 + n_1\tau + \Omega k_2|^2 + {2\pi\over V\calV_2}n_1k_2}\, .
    \label{nondegenorb}
 \ee
Integrating over $\tau_1$ gives the expression
\bea
\sum_{k_1,k_2\ne 0}
K_{(m_1,n_1)(k_2,0)} &=& {\pi^3\over \calV_2^2}\, (\calV_2 \Omega_2)^\half \sum_{n_1\neq 0,k_2\neq 0}
\int_0^{\Vlambda}
 {dV\over  V^\half}\int_0^{\Lambda^2/V}  d\tau_2\tau_2^{-\threeh} \, e^{-{\pi\over\calV_2 \Omega_2
    V\tau_2}(n_1^2\tau_2^2 + k_2^2\Omega_2^2)}\nn\\
    &=& {\pi^3\over \calV_2^2}\, (\calV_2 \Omega_2)^\half \sum_{n_1\neq 0,k_2\neq 0}\int_0^{\Vlambda}
 dV  \int_0^{\Lambda^2} dy y^{-\threeh} \, e^{-{\pi\over\calV_2 \Omega_2}(
    {1\over V^2} n_1^2 y+ {1\over y}k_2^2\Omega_2^2)}\, ,
    \label{nondegenresa}
 \eea
where $y=V\tau_2$. Since each term in the sum is dominated by the $V$ cutoff,
we will perform a Poisson resummation of the integer $n_1$ after adding and subtracting the $n_1=0$
term, which is proportional
to $\int dV \sim \Lambda^2$.  Since we are here not keeping terms that are powers of the cutoff (since they will not
have the appropriate power of $\calV_2^{-2}$) we will drop this term.  After the Poisson resummation the result is
(again dropping terms that are positive powers of $\Lambda$ and are therefore  not of order $\calV_2^{-2}$)
\bea
\sum_{k_1,k_2\ne 0}
K_{(m_1,n_1)(k_2,0)} &\sim& {\pi^3\Omega_2\over \calV_2}\, \sum_{n_1\neq0}\sum_{\hat k_2\ne 0}\int_0^{V_\Lambda}
 dV\, V  \int_0^{\Lambda^2} dy y^{-2} \, e^{-{\pi\Omega_2\calV_2 V^2\over y}(\hat n^1)^2 -\pi {\Omega_2\over \calV_2 y}k_2^2 }\nn\\
 &=& {2\pi^2\over \calV_2^2}\,\zeta(2)\,  \sum_{k_2\ne 0}
  \int_0^{\Lambda^2} {dy\over y} \, e^{-\pi {\Omega_2\over \calV_2 y}k_2^2 }
  =
 - 2{\pi^2\over \calV_2^2}\,\zeta(2)\,\log(\Lambda^2\calV_2/\Omega_2 C) \, ,\nn\\
    \label{nondegenresb}
\eea

Therefore, the total contribution to $I^{(2)}_{(2,0)}$ proportional to $\calV_2^{-2}$ (which therefore
does not have a power of $\Lambda$) gives a contribution
\be
K = - 2{\pi^2\over \calV_2^2}\,\zeta(2)\,\log(\chi \calV_2/\Omega_2 C)\, ,
\label{totitwozerb}
\ee
so that from (\ref{kzerodef}) we have
\be
\calE_{(2,0)}^{(2)\, 0} = {104\over 21}\, \zeta(2)\,\log(\chi \calV_2/\Omega_2C)\,,
\label{ezerores}
\ee
so that
\be
\Delta_\Omega\calE_{(2,0)}^{(2)\, 0} = \frac{104}{21}\, \zeta(2)\,.
\label{laplaceoneb}
\ee

\subsection{Evaluation of $\calE_{(3,0)}^{(6)\, 0}$ and $\calE_{(0,2)}^{(6)\, 0}$}

The terms of order $S^6\, \calR^4$ will contribute to a logarithmic
eleven-dimensional threshold term, which means that the zero winding
number sector  $\hat m^I= \hat n^J=0$ possesses the singularity.  In
the winding number basis the expressions we need to evaluate are
\bea
\calE_{(3,0)}^{(6)\, 0} &=& {16\over 3\pi^2}\, I_{(3,0)\, 0} ={16\over \pi^2}\,  b_{(3,0)}^0\,
\hat K\,,\nn\\
\calE_{(0,2)}^{(6)\, 0} &=& {16\over 3\pi^2}\, I_{(0,2)\, 0} ={16\over \pi^2}\,  b_{(2,0)}^0\,
\hat K\,.
\label{khatzerodef}
\eea
where
\be
\hat K =\pi^{5}\,
\sum_{\{\hat m_I, \hat n_J\}}\, \int_\0chi^{\Vlambda} {dV\over V}\, \int_{\calF_\Lambda} {d^2 \tau\over \tau_2^2}
\,e^{{-\pi{VG_{IJ}\over \tau_2}\left[(\hat m+\tau \hat n)^I(\hat m+\bar\tau\hat n)^J\right]}}\,.
\label{fourconst}
\ee
We will now decompose the sums in the same manner as in (\ref{sumdecom}) ,(\ref{unfold}),
writing
\be
\hat K \equiv \sum_{\hat m^1,\hat n^1,\hat m^2,\hat n^2}\, \hat K_{(\hat m^1,\hat n^1)(\hat m^2,\hat n^2)} =
\sum_{\hat m^1,\hat n^1,\hat m^2,\hat n^2}\,\int_{\0chi}^{\Vlambda}\, {dV\over
  V}\int_{\calF_\Lambda} {d^2\tau\over \tau_2^2}\, \hat J_{(\hat m^1,\hat n^1)(\hat m^2,\hat n^2)}\,.
\label{msectswin}
\ee
and
\be
\begin{split}
\int_{\calF_\Lambda} {d^2\tau\over \tau_2^2}& \sum_{\hat m^1,\hat n^1,\hat m^2,\hat n^2}\,
\hat J_{(\hat m^1, \hat n^1)(\hat m^2,\hat n^2)}\nn\\
=&\int_{\calF_\Lambda} {d^2\tau\over \tau_2^2} \, \hat J_{(0,0)(0,0)}
 +\int_{\calR_\Lambda}
    {d^2\tau\over \tau_2^2} \left(\sum_{(\hat k_1,\hat k_2)\ne(0,0)}\,\hat J_{(\hat k_1,0)(\hat k_2,0)}
+ \sum_{\hat n^1,\hat k_2\ne 0}\sum_{\hat m^1}\,\hat J_{(\hat m^1,\hat n^1)(\hat k_2,0)}\right)\,,\\
\end{split}
\label{unfoldwind}
\ee

The zero winding number term is given by
\be
 \hat K_{(0,0)(0,0)}  = \pi^5\, {\pi \over 3}\, \log(\chi /C\Lambda^2) +O(\calV_2{-1})\,,
 \label{zerokkwind}
\ee
The second term in (\ref{unfoldwind}) leads to
\bea
\sum_{k_1,k_2\ne 0}\hat K_{(k_1,0)(k_2,0)} &=& \pi^5\, \int_0^{\Vlambda}  {dV\over V}\,
\int_\0chi^{\Lambda^2/V}{d\tau_2\over \tau_2^2} \, \sum_{\hat k_1,\hat k_2\ne 0}
\exp\left(-\pi{\calV_2 V\over  \Omega_2\tau_2}
|\hat k_1 + \hat k_2 \Omega_2|^2\right)\nn\\
&=&\pi^5\, \int_\0chi^{\Vlambda} dV\,
\int_{\Lambda^{-2}}^\infty d\hat y \, \sum_{k_1,k_2\ne 0}\exp\left(-\pi{\calV_2\over \Omega_2} V^2 \hat y
|\hat  k_1 + \hat k_2 \Omega_2|^2\right)
\, ,
\label{degenorbwind}
\eea
where $\hat y=(V\tau_2)^{-1}$.

The last term in (\ref{unfoldwind}) leads to
\be
\sum_{k_1,k_2\ne 0}
\hat K_{(\hat m^1,\hat n^1)(k_2,0)} = \pi^5\, \sum_{\hat n^1,\hat k_2\ne 0}\sum_{\hat m^1} \int_\0chi^{\Vlambda}
 {dV\over V} \int_{\calR_\Lambda} {d^2\tau\over \tau_2^2} \, e^{-{\pi\calV_2 V\over
 \Omega_2 \tau_2 }|\hat m^1 + \hat n^1\tau + \Omega \hat k_2|^2 + 2\pi V\calV_2\,\hat n^1\hat k_2}\, .
    \label{nondegenorbb}
 \ee
Integrating over $\tau_1$ gives
\bea
\sum_{k_1,k_2\ne 0} \hat K_{(\hat m^1,\hat n^1)(\hat k_2,0)} &=& \pi^5\, \left(\Omega_2\over \calV_2\right)^\half
\sum_{\hat n^1\neq 0,\hat k_2\neq 0}\int_\0chi^{\Vlambda}
 {dV\over  V^\threeh} \int_0^{\Lambda^2/V} d\tau_2\tau_2^{-\threeh} \, e^{-{\pi \calV_2  V\over
    \Omega_2\tau_2}((\hat n^1)^2\tau_2^2 + \hat k_2^2\Omega_2^2)}\nn\\
    &=& \pi^5\,  \left(\Omega_2\over \calV_2\right)^\half  \sum_{\hat
   n_1\neq 0,\hat k_2\neq 0}\int {dx dy\over x y^\threeh}  \,
 e^{- \pi \calV_2 \Omega_2 \hat k_2^2 x  -\pi{\calV_2\over \Omega_2} (\hat n^1)^2 y}\, ,
    \label{nondegenres}
 \eea
where $x=V/\tau_2$ and   $y=V\tau_2$.
The $y$ integral may be performed without worrying about the cutoff
and gives
\be
\int_0^\infty {dy\over y^{\threeh}}\,
\sum_{\hat n^1\ne 0}\, e^{-\pi{\calV_2\over \Omega_2}(\hat n^1)^2 y}=
{\pi \over 3}\left({\calV_2\over \Omega_2}\right)^\half\,,
\label{yintt}
\ee
where we have used the analytic continuation of the Riemann zeta
function to write $\hat\sum |n_1| =-1/6$.
The $x$ integral in (\ref{nondegenres}) gives
\be
\int_{0_\chi}^{\Lambda^2} {dx\over x }  \,
 \sum_{\hat k_2\ne 0}e^{- \pi \calV_2 \Omega_2 \hat k_2^2 x}= \Sigma_1(
 \calV_2 \Omega_2) - \log(\chi /C\,\Lambda^2) = -\log\left(\calV_2\,\Omega_2\,
 \Lambda^2/\hat C\right)- \log(\chi /\hat C\Lambda^2)
\,,
\label{xintt}
\ee
where we have subtracted the $\hat k_2=0$ term (proportional to
$\log(\chi)$) from the sum that
defines $\Sigma_1$ in (\ref{sigone}) and discarded the term
proportional to $S^{-\half}$, which is accompanied by a factor of
$\calV_2^{-\half}$.

Substituting (\ref{yintt}) and (\ref{xintt}) into (\ref{nondegenres})
and combining this with (\ref{zerokkwind}), which is the other  contribution that does not have
a power of $\Lambda$, gives the total contribution
\be
\hat K = \pi^5 {\pi\over 3}\,\log\left(\calV_2\Omega_2\Lambda^2\over \hat C\right)\, ,
\label{totitwozer}
\ee
so that, from (\ref{khatzerodef}),
\bea
\calE_{(3,0)}^{(6)\, 0} &=&{12264\over 715}\, 64\, \zeta(2)^2\,\log\left(\calV_2\Omega_2 \Lambda^2 \over \hat C\right) \,,\nn\\
\calE_{(0,2)}^{(6)\, 0} &=&{2716\over 165}\, 64 \,\zeta(2)^2 \, \log\left(\calV_2\Omega_2\Lambda^2 \over \hat C\right)  \,.
\label{lzerodef}
\eea
Note that, as had to be the case, the $\log(\chi)$ in the zero-winding
sector cancels with the effects of non-zero winding.
The Laplace equations satisfied by these coefficients are
\be
\Delta_\Omega\calE_{(3,0)}^{(6)\, 0} = \frac{12264}{715}\,64\, \zeta(2)^2\,,\qquad
\Delta_\Omega\calE_{(0,2)}^{(6)\, 0} = \frac{2716}{165}\,64\, \zeta(2)^2\,.
\label{laplaceonee}
\ee
%%%%%%%%%%%%%%%%%%%%%%%%%%%%%%%%%%%%%%%%%%%%%%
\section{Weak coupling expansion of the generalized modular functions}\label{sec:Diff}
%%%%%%%%%%%%%%%%%%%%%%%%%%%%%%%%%%%%%%%%%%%%%%

In the main text we found modular functions which are defined by  Poisson equations in the fundamental domain, of the general form
\begin{equation}\label{e:DEsource}
[\Delta_{\Omega}- s(s-1)]\, \mathcal{E}(\Omega)  = S(\Omega)\ ,\qquad s\geq 0\ .
\end{equation}
Here we determine the perturbative part of $\calE$ for a general source term $S$
with a zero mode expansion given by
\begin{equation}\label{e:Szm}
S(\Omega) = \sum_{n=0}^{N} \, \alpha_{n} \Omega_{2}^{n_{0}-n} + S_{cusp}(\Omega_{2})\ .
\end{equation}
We assume that the polynomial part does not contain  $\Omega_{2}^{s}$ or $\Omega_{2}^{1-s}$
 (either because $s>N-1-n_0,\ s>n_0$ or because
 $\alpha_{s}=\alpha_{1-s}=0$). For the relevant cases $n_0$ will be
an integer or half-integer number. $S_{cusp}$ is an
exponentially suppressed  contribution, which nevertheless will contribute to the perturbative
(power-behaved) part of $\calE $.

 The general structure of the zero mode expansion of the solution
 $\mathcal{E}(\Omega)$ of~(\ref{e:DEsource}) is the sum
 of the particular solution with the source term and a solution of the homogeneous equation
 \begin{equation}
 \mathcal{E}(\Omega) = \sum_{n=0}^{N} \, {\alpha_{n}\,
\Omega_{2}^{n_0-n}\over (n_{0}-n)(n_{0}-n-1)- s(s-1)}  + \alpha \Omega_{2}^{s} +\beta \Omega_{2}^{1-s} + \mathcal{O}(\exp(-\Omega_{2}))
 \end{equation}
The parameters $\alpha $ and $\beta $ are integration constants which are fixed by boundary conditions.
For the cases appeared in the main text, one must impose that $\alpha=0$ because $s$ is such that $\Omega_{2}^s$ is more singular than the tree-level contribution in the weak coupling limit.

%%%%%%%%%%%%%%%%%%%%%%%%%%%%%%%%%%%%%%%%%%%%%%
\subsection{General method for determining $\beta$ terms}\label{sec:betaterm}
%%%%%%%%%%%%%%%%%%%%%%%%%%%%%%%%%%%%%%%%%%%%%%

The value of $\beta $ is determined as in section~5.4 of
\cite{gv:D6R4} by integrating over the cutoff fundamental domain for $SL(2,\ZZ)$ the product of
$\calE$ with  the Eisenstein series $E_{s}$ --which is a
solution of the homogeneous equation associated
with~(\ref{e:DEsource}).  Then we perform the partial integrations as
\begin{equation}
0= \int_{\mathcal{F}_{L}} \, {d^2\Omega\over \Omega_{2}^2}\,([\Delta-s(s-1)]\, E_{s})\, \mathcal{E}
 = \int_{\mathcal{F}_{L}} \, {d^2\Omega\over \Omega_{2}^2}\, E_{s}\,S
+\int_{\partial\mathcal{F}_{L}}  \big( \bar\partial E_{s} \mathcal{E}- E_{s}\partial \mathcal{E} \big)
\end{equation}
Computing the boundary term, we find
\begin{equation}\begin{split}
\int_{\mathcal{F}_{L}} {d^2\Omega\over \Omega_{2}^2}\, E_{s}\, S(\Omega)&=2\zeta(2s)\, (1-2s) \,\beta\\
 &+2\zeta(2s)\sum_{n=0}^N \, {\alpha_{n} (n_{0}-n-s)\over (n_{0}-n)(n_{0}-n-1)- s(s-1)} \, L^{s+n_{0}-n-1}\\
&+ \mathcal{O}(L^{-1})
\end{split}
\label{compboun}
\end{equation}
where we have only displayed the terms that do not vanish when the cutoff $L\to\infty$.
Since
\be
E_s(\Omega)   =  \sum_{\gamma\in   \Gamma_\infty   \backslash  \Gamma}
\Im\textrm{m}(\gamma\cdot \Omega)^s\,,
\label{sumrect}
\ee
the integral on the left hand side
can be evaluated by unfolding the Eisenstein series $E_{s}$, giving that in the limit of large $L$
\begin{equation}\begin{split}
\int_{0}^{L} {d\Omega_{2}\over \Omega_{2}^2} \, 2\zeta(2s)\,\Omega_{2}^s\, S(\Omega)&=2\zeta(2s)\, (1-2s) \,\beta\\
&+2\zeta(2s)\sum_{n=0}^N \, {\alpha_{n} (n_{0}-n-s)\over (n_{0}-n)(n_{0}-n-1)- s(s-1)} \, L^{s+n_{0}-n-1}\\
&+ \mathcal{O}(L^{-1})
\end{split}\end{equation}
The power-behaved terms in $L$ in (\ref{compboun}) cancel against the contributions from the power-behaved
terms in $S$,
 so that the value of $\beta$ is
determined by the projection of $S_{cusp}$ on $E_{s}$:
\begin{equation}
(1-2s)\, \beta = \int_{0}^{\infty} \, {d\Omega_{2}\over \Omega_{2}^2}\, \Omega_{2}^s \, S_{cusp}(\Omega_{2})\ .
\label{projcusp}
\end{equation}
%%%%%%%%%%%%%%%%%%%%%%%%%%%%%%%%%%%%%%%%%%%%%%%%
\subsection{ $\beta $ coefficients arising from a source $E_{s_1} E_{s_2}$}\label{sec:Beta}
%%%%%%%%%%%%%%%%%%%%%%%%%%%%%%%%%%%%%%%%%%%%%%%

For the particular case of a source term given by the product of two Eisenstein series,
$E_{s_{1}}E_{s_{2}}$,  the $\beta$-coefficient can be
given in a closed form.
Substituting the well known large $\Omega_2$ expansion of the Eisenstein series (see e.g. \cite{gv:D6R4})
into the right-hand side of (\ref{projcusp}) we find
\begin{eqnarray}
\nonumber
(2s-1)\,\beta^{(s)}_{(s_{1},s_{2})}&=& \int_{0}^{\infty} {dt\over t^{1-s}}\,
 {32\pi^{s_{1}+s_{2}}\over \Gamma(s_{1})\Gamma(s_{2})}\, \sum_{n>0}
  {\sigma_{1-2s_{1}}(n) \sigma_{1-2s_{2}}(n)\over n^{1-s_{1}-s_{2}}}\, K_{s_{1}-{1\over2}}(2\pi n \Omega_{2})K_{s_{2}-{1\over2}}(2\pi n \Omega_{2})\\
&=& {4\pi^{s_{1}+s_{2}-s}}\, \sum_{n>0}  {\sigma_{1-2s_{1}}(n) \sigma_{1-2s_{2}}(n) \over n^{s+1-s_{1}-s_{2}}}\, \\
\nonumber&\times&
{\Gamma\left(s-s_{1}-s_{2}+1\over 2\right)\Gamma\left(s+s_{1}-s_{2}\over 2\right)
\Gamma\left(s-s_{1}+s_{2}\over
2\right)\Gamma\left(s+s_{1}+s_{2}-1\over 2\right)\over
\Gamma(s)\Gamma(s_{1})\Gamma(s_{2})} \, ,
\end{eqnarray}
where we have used the result for the integral of the product of two
Bessel  functions,
\bea
\int_{0}^\infty dt t^{m-1}&& K_{n-{1\over2}}(t) K_{p-{1\over 2}}(t)
={1\over 2^{3-m}\Gamma(m)}\nn\\
&&\Gamma\left(m-n-p+1\over 2\right)\Gamma\left(m+n-p\over2\right)
\Gamma\left(m-n+p\over2\right)\Gamma\left(m+n+p-1\over2\right)
 \, .
\label{betasdefs}
\eea
Using the fact that
$\sigma_{a}(pq)=\sigma_{a}(p)\sigma_{a}(q)$ for $p$ and $q$ prime and the fact that
all integers can be decomposed over a product of primes, one easily
establishes that
\begin{equation}
\sum_{n>0} \, {\sigma_{a}(n)\sigma_{b}(n)\over n^r} =
    {\zeta(r)\zeta(r-a)\zeta(r-b)\zeta(r-a-b)\over \zeta(2r-a-b)} \, ,
\end{equation}
whereby
\begin{equation}
\beta^{(s)}_{(s_{1},s_{2})} = {4 \pi^{s_{1}+s_{2}}\over
  \Gamma(s_{1})\Gamma(s_{2})}\,
{\zeta^*(s-s_{1}-s_{2}+1)\zeta^*(s+s_{1}-s_{2})
\zeta^*(s-s_{1}+s_{2})\zeta^*(s+s_{1}+s_{2}-1)\over (2s-1)\,\zeta^*(2s)}\,,
\end{equation}
 with $\zeta^*(s)=\zeta(s)\Gamma(s/2)/\pi^{s/2}$.

%%%%%%%%%%%%%%%%%%%%%%%%%%%%%%%%%%%%%%%%%%%%%%%%
\section{Two-loop four-graviton supergravity amplitude
in various dimensions}\label{sec:DimReg}
%%%%%%%%%%%%%%%%%%%%%%%%%%%%%%%%%%%%%%%%%%%%%%%%%%%%%%%%%%%%%%%%

In this appendix
 we will consider the two-loop  four-graviton amplitude of maximal supergravity in Minkowski space in nine, ten and
 eleven dimensions using dimensional regularization.   These results make contact at various points with our discussion
 of eleven-dimensional supergravity compactified on a circle and on a two-torus.
 We follow the analysis described  in
 \cite{Smirnov:1999gc,Smirnov:2002kq} and \cite{Tausk:1999vh}
based on a dimensional regularisation
adapted to the ten and nine dimensional case.   Although the
 ten-dimensional and eleven-dimensional results are in
 the literature \cite{BernDunbar} we include  them here for completeness.

\subsection{Ten dimensions}
\label{sec:tenreg}

The amplitude in $D=10-2\epsilon$ dimensions takes the form
\begin{equation} \begin{split}
A_4^{(10-2\epsilon)}& =\calR^4\,
 \Big((-S )^{2}\,\left[
 I^{P\,(10-2\epsilon)}(S,T)+  I^{P\,(10-2\epsilon)}(S,U) +  I^{NP\,(10-2\epsilon)}(S,T) +
 I^{NP\,(10-2\epsilon)}(S,U)\right] \cr
&+(-T)^{2}\,\left[
 I^{P\,(10-2\epsilon)}(T,S)+  I^{P\,(10-2\epsilon)}(T,U) +  I^{NP\,(10-2\epsilon)}(T,S) + I^{NP\,(10-2\epsilon)}(T,U)\right]\cr
&+(-U)^{2}\,\left[
 I^{P\,(10-2\epsilon)}(U,T)+  I^{P\,(10-2\epsilon)}(U,S) +
  I^{NP\,(10-2\epsilon)}(U,T) + I^{NP\,(10-2\epsilon)}(U,S)\right]\Big)\,.
\end{split}\end{equation}
The contributions $I^P$ and $I^{NP}$ are the scalar
field     theory    double-box     diagrams     as    described     in
section~\ref{sec:expand}.

These integrals can be analyzed efficiently using the
{\tt Mathematica} package described in \cite{Czakon:2005rk}.  The integrals can be reduced by a repeated use
 of the first and second Barnes' lemma given in the appendix~D of \cite{Smirnov:1999gc,Smirnov:2004ym}
to reproduce the result of the appendix~C of \cite{BernDunbar}
The planar amplitude $I^P(S,T)$ is given by
\begin{equation}
I^{P\,(10-2\epsilon)}(S,T)={  (-S\mu^2)^{3-2\epsilon}\over  (4\pi)^{10}}  \,
\Big( -{2\over 7!\cdot 5!}{4S+T\over S\,\epsilon^2}
 -\frac{63\,  T^3-252 \,  S\, T^2-55\,S^2  \,T+704\, S^3}{700\cdot
   9!\, S^3\,\epsilon}
+\mathcal{O}(\epsilon^0)\Big)\,,
\end{equation}
and the non-planar amplitude takes to form
\begin{equation}
I^{NP\,(10-2\epsilon)}(S,T)={  (-S\mu^2)^{3-2\epsilon}\over  (4\pi)^{10}} \,
\left({7\over 7!\cdot 5!} {1\over\epsilon^2}+{ 1\over 30\cdot9!}\,
 \frac{917 S^2+ 2 \,T \,U}{S^2\, \epsilon}+\mathcal{O}(\epsilon^0)\right)\,,
\end{equation}
where $\mu$ is an arbitrary scale and we have made use
of the on-shell condition $S+T+U=0$.  The terms of $O(\epsilon^0)$ are functions of the dimensionless
ratio $T/S$.

The $1/\epsilon^2$ pole cancels (on-shell) in the $S$-channel part of the amplitude, $A^{(S)}$, giving
\begin{equation}\begin{split}
&I^{P\,(10-2\epsilon)}(S,T)+I^{P\,(10-2\epsilon)}(S,U)+I^{NP\,(10-2\epsilon)}(S,T)
+I^{NP\,(10-2\epsilon)}(S,U)\cr
&=-{1\over (4\pi)^{10}}\, {  50\, S^3 + 5 S T U\over 6!^2\epsilon }
+ \mathcal{O}(\epsilon^0)\,.
\end{split}\end{equation}
The $\epsilon$ pole contributes to terms proportional to $\log(-S\mu^2)$, which are not symmetric in the
Mandelstam invariants.  However,
summing all the contributions and using the mass-shell constraint the total amplitude takes the symmetric form
\begin{equation}
A_4^{(10-2\epsilon)}= \frac{13}{ (4\pi)^{10}\,466560\, \epsilon}\,\sigma_{2}\,\sigma_{3} \calR^4+ \mathcal{O}(\epsilon^0)\, .
\end{equation}
The striking cancelation of the $1/\epsilon^2$ pole separately in the $S$, $T$ and $U$-channels,
corresponds to the cancelation of the
terms proportional to $\log^2(-S)$, $\log^2(-T)$ and $\log^2(-U)$
in the circle compactification of the term of order $S^5\,\calR^4$
analyzed in appendix~\ref{sec:D10R4circle} and  of the $(\log\Omega_{2})^2$ dependence
in the $\calE_{(1,1)}(\Omega)$ coefficient of equation~(\ref{e:calF}).

Under  a rescaling $(S,T,U) \to \Omega_{2}\, (S,T,U)$ the amplitude behaves as
\begin{equation}
\Omega_{2}^{-5}\,A^{(10-2\epsilon)}_4\to  A^{(10-2\epsilon)}_4+ {13\over (4\pi)^{10}\,233280}\, \sigma_{2}\sigma_{3}\,\log\Omega_{2} \, .
\end{equation}
The $\log\Omega_{2}$ term (properly normalized) should be related to with the $\log\Omega_{2}$ term in (\ref{e:calF}).

\subsubsection{The triangle counterterm diagram}

The ten-dimensional
two-loop amplitude receives an extra contribution from the
triangle diagram with the one-loop counterterm $A_{\triangleright}$ of equation~(\ref{diag:counter}).
In the dimensional regularisation scheme the triangle loop amplitude in $D=10-2\epsilon$ reads
\begin{eqnarray}
\nn I^{(10-2\epsilon)}_{\triangleright}(S)&=&-{1\over (2\pi)^{10}}\int {d^{10-2\epsilon}\ell\over
\ell^2 (\ell-p_{1})^2 (\ell-p_{1}-p_{2})^2}\\
&=&- {\Gamma(-2+\epsilon)\over 2\, (3-\epsilon)^2 (4\pi)^5}\, (-S)^{2-2\epsilon}
\end{eqnarray}
This pole in $\epsilon$
leads to a contribution of order $S^4 \, \log(-S)$.
This is interpreted in the $\calS^1$ compactification to type IIA
string theory as a
genus-two threshold contribution.   Unitarity requires the presence of
this
term since the discontinuity across the threshold is
the product of the genus-one $\calR^4$ term
and the leading contribution from the tree-level amplitude.
To pick out the coefficient we can perform
 a rescaling of the Mandelstam variables  $(S,T,U) \to \Omega_{2}\,
 (S,T,U)$  this amplitude behaves as
\begin{equation}
\Omega_{2}^{-4}\, A^{(10-2\epsilon)}_{\triangleright}\to
 A^{(10-2\epsilon)}_{\triangleright}
+ {1\over (4\pi)^{5}\,72}\, \sigma_{2}^2\,\log\Omega_{2} \,,
\end{equation}
which corresponds to the two-loop  $S^4\, \log\Omega_{2}$ term
that would be contained in a modular function describing the terms at
order $S^4\,\calR^4$ in ten dimensions.

%%%%%%%%%%%%%%%%%%%%%%%%%%%%%%%%%%%%%%%%%%%
\subsection{Eleven dimensions}\label{sec:elevenreg}
%%%%%%%%%%%%%%%%%%%%%%%%%%%%%%%%%%%%%%%%%%%

In similar fashion  the planar and non-planar scalar field theory integrals that  contribute
to the two-loop amplitude in $D=11-2\epsilon$ dimensions
are found to be given by
\begin{eqnarray}
\nn I^{P\,(11-2\epsilon)}(S,T)&=&(-S)^{-2\epsilon}\,{\pi\over(4\pi)^{11}}\,{1\over\epsilon}\, \frac{
2100 \, S^4- 880\, S^3\,T+215 \,S^2\,T^2+30\,S\, T^3+12 \,T^4 }{9451728000}
+ \mathcal{O}(\epsilon^0)\\
\end{eqnarray}
\begin{equation}
I^{NP\,(11-2\epsilon)}(S,T)=(-S)^{-2\epsilon}\,{\pi\over (4\pi)^{11}}\,  {1\over\epsilon}\,
{40383\, S^4 - 1138\ S^2\,T\,U + 144\ U^2\, T^2\over 79394515200}+\mathcal{O}(\epsilon^0)
\end{equation}
The resulting amplitude is
\be
A_{4}^{(11-2\epsilon)}=
{\pi\over(4\pi)^{11}}\,{1\over\epsilon}\,{1971\, \sigma_{2}^3
+2522\, \sigma_{3}^2 \over 5003856000} +\mathcal{O}(\epsilon^0)\ .
\label{elevenpole}
\ee
in agreement with the results of \cite{BernDunbar}.

The $1/\epsilon$ pole in (\ref{elevenpole})
gives a $S^6\, \calR^4$ term in the amplitude in eleven-dimensional Minkowski space
that should correspond to the zero-winding sector of the two-loop amplitude at order
$S^6\, \calR^4$ in the compactified theory.
In section~\ref{sec:hfoureval} we determined the zero-winding coefficients
$h_{(3,0)}^0$ and $h_{(0,2)}^0$ (see (\ref{h4zerox}) and (\ref{h4zeroy})).  Referring back to
the normalizations in equation~(\ref{eberz}), we see that these zero-winding terms agree precisely with
(\ref{elevenpole}) with $1/\epsilon$ replaced by $\log \Lambda^2/C$ (where $C$ is an undetermined constant).

%%%%%%%%%%%%%%%%%%%%%%%%%%%%%%%%%%%%%%%%%%%%%%%%%%%%
\subsection{Nine dimensions}\label{sec:ninereg}
%%%%%%%%%%%%%%%%%%%%%%%%%%%%%%%%%%%%%%%%%%%%%%%%%%%%

The planar and non-planar diagrams in nine dimensions also have logarithmic branch points. These arise from
the coalescence of the square root branch points of the individual one-loop integrals.
For completeness, we note the
result of an analysis of the the planar and non-planar diagrams in $D=9-2\epsilon$ dimensions analogous to the
one of the preceding sub-sections, giving
\begin{equation}
\nn I^{P,(9-2\epsilon)}(S,T)=S^{2-2\epsilon}\,{\pi\over(4\pi)^9}\, \frac{-
45\,S^2 +18\,S\,T + 2\,T^2 }{399168\,\epsilon\,S^2}+\mathcal{O}(\epsilon^0)
\end{equation}
\begin{equation}
I^{NP,(9-2\epsilon)}(S,T,U)=s^{2-2\epsilon}\,{-\pi\over (4\pi)^9\, 332640}\,
{75\,          S^2+          2          \,         T\,          U\over
  S^2\,\epsilon}+\mathcal{O}(\epsilon^0)\ .
\end{equation}
Collecting all the contributions one finds in agreement with \cite{BernDunbar}
\begin{equation}
A^{(9-2\epsilon)}_{4}= -{1\over 8\epsilon}\, {1\over (4\pi)^9}\, {13\pi\over9072}\,
 \sigma_{2}^2\,\calR^4+\mathcal{O}(\epsilon^0) \ .
\end{equation}

%------

\end{document}